\documentclass[sigconf]{acmart}

\AtBeginDocument{%
  \providecommand\BibTeX{{%
    \normalfont B\kern-0.5em{\scshape i\kern-0.25em b}\kern-0.8em\TeX}}}

\copyrightyear{2025}
\acmYear{2025}
\setcopyright{cc}
\setcctype{by}
\acmConference[CHI '25]{CHI Conference on Human Factors in Computing Systems}{April 26-May 1, 2025}{Yokohama, Japan}
\acmBooktitle{CHI Conference on Human Factors in Computing Systems (CHI '25), April 26-May 1, 2025, Yokohama, Japan}\acmDOI{10.1145/3706598.3713187}
\acmISBN{979-8-4007-1394-1/25/04}

\usepackage{cuted}
\usepackage{subcaption} 
\usepackage{tabularx}
\usepackage{url}
\usepackage{xcolor}
\usepackage{booktabs}
\usepackage{verbatim}
\makeatletter
\g@addto@macro{\UrlBreaks}{\UrlOrds}
\makeatother
\usepackage{multirow} 
\usepackage{color} 
\usepackage{enumitem} 

\usepackage{longtable}
\usepackage{xspace}
\usepackage{hyperref}

\newcommand{\nuserstudyrecuruted}{460\xspace}
\newcommand{\nuserstudyregret}{316\xspace}
\newcommand{\userstudyfinal}{72\xspace}
\newcommand{\chisq}{$\chi^2$}

\newcommand{\notextrefrq}[1]{\hyperref[rq:#1]{\highlight{\textbf{\texttt{\textcolor{gray}{#1}}}}}}

\begin{document}

\title[Analysing Contextual Influences on Intervention Effectiveness during Infinite Scrolling]{Scrolling in the Deep: Analysing Contextual Influences on Intervention Effectiveness during Infinite Scrolling on Social Media}

\author{Luca-Maxim Meinhardt}
\email{luca.meinhardt@uni-ulm.de}
\orcid{0000-0002-9524-4926}
\affiliation{%
  \institution{Institute of Media Informatics, Ulm University}
  \city{Ulm}
  \country{Germany}
}

\author{Maryam Elhaidary}
\email{maryam.elhaidary@uni-ulm.de}
\orcid{0009-0002-7658-0788}
\affiliation{%
  \institution{Institute of Media Informatics, Ulm University}
  \city{Ulm}
  \country{Germany}
}

\author{Mark Colley}
\email{m.colley@ucl.ac.uk}
\orcid{0000-0001-5207-5029}
\affiliation{%
  \institution{Institute of Media Informatics}
  \city{Ulm}
  \country{Germany}
}
\affiliation{%
  \institution{UCL Interaction Centre}
  \city{London}
  \country{United Kingdom}
}

\author{Michael Rietzler}
\email{michael.rietzler@uni-ulm.de}
\orcid{0000-0003-2599-8308}
\affiliation{%
  \institution{Institute of Media Informatics, Ulm University}
  \city{Ulm}
  \country{Germany}
}

\author{Jan Ole Rixen}
\email{jan.rixen@uni-ulm.de}
\orcid{0000-0002-6003-7900}
\affiliation{%
  \institution{Institute of Media Informatics}
  \city{Ulm}
  \country{Germany}
}
\affiliation{%
  \institution{Karlsruhe Institute of Technology}
  \city{Karlsruhe}
  \country{Germany}
}

\author{Aditya Kumar Purohit}
\email{aditya.purohit@cais-research.de}
\orcid{0000-0002-9766-6575}
\affiliation{%
  \institution{Center for Advanced Internet Studies (CAIS) gGmbH}
  \city{Bochum}
  \country{Germany}
}

\author{Enrico Rukzio}
\email{enrico.rukzio@uni-ulm.de}
\orcid{0000-0002-4213-2226}
\affiliation{%
  \institution{Institute of Media Informatics, Ulm University}
  \city{Ulm}
  \country{Germany}
}

\renewcommand{\shortauthors}{Meinhardt et al.}

\begin{abstract}
Infinite scrolling on social media platforms is designed to encourage prolonged engagement, leading users to spend more time than desired, which can provoke negative emotions. 
Interventions to mitigate infinite scrolling have shown initial success, yet users become desensitized due to the lack of contextual relevance. 
Understanding how contextual factors influence intervention effectiveness remains underexplored.
We conducted a 7-day user study (N=72) investigating how these contextual factors affect users' \textit{reactance} and \textit{responsiveness} to interventions during infinite scrolling. 
Our study revealed an interplay, with contextual factors such as being at home, sleepiness, and valence playing significant roles in the intervention's effectiveness. Low valence coupled with being at home slows down the \textit{responsiveness} to interventions, and sleepiness lowers \textit{reactance} towards interventions, increasing user acceptance of the intervention.
Overall, our work contributes to a deeper understanding of user responses toward interventions and paves the way for developing more effective interventions during infinite scrolling.

\end{abstract}



\maketitle

\section{Introduction}
In the era of social media (SoMe), platforms such as TikTok and Instagram changed how we consume digital content by employing interaction mechanisms such as infinite scrolling. This mechanism, where content endlessly loads as users swipe or scroll, can lead to prolonged screen time~\cite{Mildner.2021}, leading users to a feeling of being caught in a loop of unconscious and habitual use~\cite{RixenIS.2023} and post-usage regret~\cite{Cho.2017}. It is, therefore, classified as an attention-capturing dark pattern~\cite{mongeroffarello2022towards} designed to manipulate users into actions contrary to their interests~\cite{Gray.2018}.
Infinite scrolling is particularly prevalent in SoMe, as this mechanism is designed to capture users' attention and increase engagement with the presented content. The interaction design of TikTok is a prominent example of infinite scrolling, which was recently suspected of violating the Digital Services Act by the European Commission~\cite{DSA}, as the design of TikTok's system ``\textit{[...] may stimulate behavioral addictions and/or create so-called `rabbit hole effects'}\,''~\cite{ECagainstTiktok}. 
In support of this, \citet{Mildner.2021} highlighted that 25\% of their participants expressed regret over the excessive duration spent infinitely scrolling through Facebook's newsfeed. This interaction is categorized as a passive form of interaction~\cite{Frison.2020} and is therefore often perceived as lacking in meaningfulness, reducing users' sense of control~\cite{Lukoff_2018} and their affective well-being~\cite{Verduyn.2015}. Despite users' awareness of the issue and their intentions to limit digital media consumption~\cite{ko2015nugu} (e.g., with the aid of digital well-being applications for Android~\cite{Android.Wellbeing} and iOS~\cite{iOS.Wellbeing}), users often encounter resistance to reminders and self-imposed limitations~\cite{hiniker2016mytime}. This resistance often stems from a deficiency in self-control regarding digital media consumption~\cite{Delaney.2017}. 
Hence, there have been several attempts to develop interventions to reduce SoMe usage driven by dark patterns such as infinite scrolling (e.g., time limits~\cite{hiniker2016mytime}, mindful intention prompts~\cite{Terzimehic.2022}, virbrations~\cite{okeke2018good}, or lockout task interventions~\cite{Kim.2019b}). However, these refer to SoMe usage as an isolated interaction without considering the users' context during usage. For instance, whether the user is at work or relaxing during leisure time, their emotional state (e.g., stressed or content) or their social situation (e.g., alone, with friends, or in a public setting) might affect how they respond to interventions. In fact, evidence hints that the context plays a crucial role in how users respond to interventions~\cite{Pinder.2018, Ding.2016}.

Nevertheless, while the influence of context in behavior change~\cite{Ding.2016, Metcalfe2012Behavioural, Pinder.2018} and mobile phone interactions~\cite{Akpinar.2023, Barnard.2007} has been well studied, these findings are often based on active interactions with mobile devices (e.g., typing performance~\cite{Akpinar.2023}). \citet{meske2017dinu} have highlighted the importance of optimal timing for digital nudges, and \citet{Purohit.2019} advocated for context-aware intervention timings to enhance user receptivity and foster healthier digital habits. However, these approaches do not fully address passive interaction like infinite scrolling, where users are prone to normative dissociation, meaning that ``[...] users’ volition is not accessible to them, which may prevent them from disengaging''~\cite[p. 11]{baughan_i_2022}. Infinite scrolling can place users in a "trance-like" state driven by the need to pass time~\cite{Lukoff_2018}, resulting in an absorption that is difficult to break effectively through interventions. Thus, the contextual influence on interventions for infinite scrolling is likely to be different, as users may not respond in the same way as they would in more active phone interactions.

Recognizing the research gap in context-aware interventions in infinite scrolling, we explored how contextual factors influence intervention effectiveness during scrolling.
To assess effectiveness, we defined it based on two dimensions: \textit{responsiveness} and \textit{reactance}.
We measured users' \textit{responsiveness} as the objective effect of an intervention, defined as the duration it took for users to stop infinite scrolling after an intervention occurred. 
However, while some interventions objectively reduce SoMe usage, they could elicit subjective negative reactions, causing participants to revert to their initial habits, as stated by \citet{Okeke.2018}. Thus, subjective evaluations of the effects of these interventions also have to be taken into account~\cite{MILLS.2023}. 
Therefore, we also measured the \textit{reactance} toward the intervention. \textit{Reactance} within the HCI context is adopted from \citet{Ehrenbrink.2020}, who defined it as the resistance individuals feel when their freedom of choice is perceived to be under threat. This resistance is rooted in psychological models~\cite{Dillard.2005}, implying that ``\textit{messages [interventions] designed with the objective of behavior change must necessarily (implicitly or explicitly) limit an audience’s freedom}''~\cite[p. 67]{Rains.2013}. Thus, an intervention during infinite scrolling may also be perceived as threatening the individual's freedom to continue scrolling, thus creating \textit{reactance}. Hence, we defined the following research question: 

\smallskip
\begin{quote}
      \textbf{How does the user's context influence the \textit{\textit{reactance}} and \textit{\textit{responsiveness}} towards interventions during infinite scrolling?}
\end{quote}
\smallskip

\noindent We conducted a longitudinal field study with N=\userstudyfinal participants over 7 days who installed \textit{InfiniteScape}, a native Android application that tracks users' infinite scrolling behavior. Once prolonged ($>15min$) infinite scrolling was detected, participants were shown an intervention overlay stating that it is time to take a break from scrolling. Participants were then prompted with a questionnaire asking for their current context, including their valence, social situation, current activity, location (being at home or not), multitasking behavior, and level of sleepiness, as well as their \textit{reactance} toward the intervention.

Our findings suggest an interplay between multiple contextual factors. Hence, different contextual elements are closely linked and influence each other.
We found that users tend to accept interventions more when they are tired, possibly due to an awareness of the negative impacts of bedtime procrastination. However, this awareness does not translate into action, as users did not disengage from scrolling after an intervention. 
In addition, the familiar and comfortable environment of their home may not provide enough distraction from negative emotions, leading users to ignore interventions and continue scrolling. Further, multitasking, particularly during moments of these negative emotions, emerged as a factor that encouraged users to stop scrolling sooner. This suggests that additional activities can serve as an effective distraction from infinite scrolling and coping with negative emotions.

\smallskip
\medskip

\noindent\fcolorbox{orange}{orange!30}{\textbf{Contribution Statement}~\cite{Wobbrock.2016}}


\medskip

\noindent\textbf{Empirical study that tells us about people.} Through our longitudinal, 7-days-long study (N=72), we provide empirical evidence that the effectiveness of interventions during infinite scrolling is contextually influenced. Our analysis revealed that multiple interconnected contextual factors, such as location, sleepiness, and valence, significantly influence users' \textit{responsiveness} and \textit{reactance} to these interventions. 

\section{Related Work}
This section outlines proposed digital interventions designed to mitigate SoMe overuse, highlighting the potential benefits of reduced phone usage for individuals' digital well-being. Further, we discuss previous research that investigated the contextual influence on behavior change, including smartphone usage. Problematic smartphone use has been widely researched~\cite{Thomee.2018, bashir2017effects, Panova.2018, Lee.2014, Sha.2021, Verduyn.2015}, with two main perspectives defining it. Firstly, research has examined whether users show addictive behaviors towards their phones~\cite{lin2015time}. This approach focuses on the patterns and frequency of phone usage that resemble addictive characteristics. Secondly, it considers whether specific designs or usage patterns are problematic~\cite{Mildner.2021, Mildner.2023}. Hence, there have been several attempts in academia to develop interventions to reduce SoMe or smartphone usage, which we briefly describe. 

\subsection{Interventions for Limiting Social Media Use}

There are two main types of interventions~\cite{Purohit.2023, kai_internal}: external and internal. On the one hand, internal interventions involve making changes inside the application itself. For example, removing the newsfeed of SoMe applications~\cite{Purohit.2023}. On the other hand, external interventions do not change the functionality of an application but intervene on a higher phone level, meaning they affect the overall smartphone system rather than individual apps. Within these external interventions, four distinct features exist~\cite{mongeroffarello2019race}, varying in level of severeness. Firstly, phone timers merely provide users with data regarding their smartphone usage, aiding in awareness and potential habit alteration~\cite{hiniker2016mytime}. Secondly, persuasive interventions involve sending reminders and notifications to users, prompting them to reconsider their smartphone habits and fostering a more conscious phone usage~\cite{Purohit2021,2023cocreation}. Thirdly, take-a-break prompts remind users to take breaks from smartphone usage, for instance, to engage in other more meaningful activities~\cite{Terzimehic.2022b}. Here, design frictions such as breathing exercises are mainly used to limit SoMe use~\cite{haliburton_longitudinal_2024}. Finally, phone blockers increase the difficulty of phone use, benefiting individuals who experience difficulty with self-regulation~\cite{Kim.2019b}. 

There remains a notable research gap in interventions tailored to the user's specific context. Yet, many researchers emphasize the need for interventions that are not one-size-fits-all but adaptable to each user’s unique circumstances and environment~\cite{alberto_2021,alberto_2023, RixenIS.2023, Purohit.2019, Okeke.2018framework, Sobolev.2021}. This approach recognizes that the impact and effectiveness of interventions can be enhanced when tailored to the users' context, considering the unique behaviors, needs, and challenges users face daily~\cite{VandenAbeele.2021b}. Therefore the next section discusses contextual influences on behavior change and smartphone usage. 


\subsection{Contextual Influence on Digital Behavior Change}

In behavior change, context plays a significant role~\cite{thomas2021systematic, Purohit.2019, karppinen2018opportunities}. 
\citet{Ding.2016} emphasized the crucial role of time and location when it comes to setting reminders to change behavior. They argue that using time and location effectively can make reminders more helpful and less bothersome as ``\textit{[...] context information plays a very important role in increasing the effectiveness and reducing the annoyingness of reminders}''~\cite[p. 7]{Ding.2016}. Further, \citet{Pinder.2018} points out that various factors, such as one's location, the time of day, current mood, and even physiological states like hunger, can affect how people react toward digital behavior change interventions. This highlights the importance of delivering interventions in the right context to be effective~\cite{Purohit.2019}. \citet{orzikulova_time2stop_2024} demonstrated this in their field study, showing that just-in-time interventions resulted in significantly lower smartphone overuse than static interventions.
Additionally, \citet{Akpinar.2023} asserted that context shapes user interactions with their devices, identifying environment, mobility, social interaction, multitasking, and distractions as key factors. While these studies underscore the importance of context on digital behavior change, much of this research centers on general smartphone interactions without distinguishing specific interaction types or states of engagement. In the context of infinite scrolling, however, the user's sense of control~\cite{Lukoff_2018} may diminish, leading to normative dissociation~\cite{baughan_i_2022} and reduced affective well-being compared to active interaction, such as direct exchanges with others~\cite{Verduyn.2015}. In this state, users are more habitual and may respond differently to interventions than in other smartphone interactions.

\citet{RixenIS.2023} specifically explored why users might disengage from SoMe with infinite scrolling, finding that users often express regret for time spent using SoMe to cope with negative emotions or procrastinate. They highlight the potential benefits of interventions that react to the users' context, which is defined as device-specific, real-world related, and internal context. \citet{Purohit.2019} demanded similar by proposing a model to determine the best timing for digital nudges, categorizing context into five areas: location, social setting, internal state, current situation, and individual behavior patterns. However, their model is in contrast to \citet{mongeroffarello2019race}. They allowed users to add contextual conditions such as location to their personalized interventions. However, only a few users used this feature. Therefore, they imply that ``\textit{[...] users consider their behaviors problematic independently of their contextual situation}''~\cite[p. 11]{mongeroffarello2019race}. 

Although previous research suggested that contextual interventions can increase the effectiveness of digital interventions, there is a lack of evidence. Thus, our study investigated this claim.
Therefore, in the next section, we identified contextual factors from related work to examine their influences on intervention effectiveness.

\section{Contextual Factors for the User Study}\label{sec:hypotheses}

Recognizing the wide range of possible contextual influences, we used existing research to identify six key factors most likely to influence the effectiveness of interventions during infinite scrolling. These factors were investigated in the subsequent user study. We assumed that the following contextual factors are closely linked and influence each other, as hinted by \citet{Purohit.2019}. Hence, \textbf{we refrained from formulating specific hypotheses} but rather explored how they influence the effectiveness of interventions during infinite scrolling. 


\textbf{Current Activity.}
\citet{RixenIS.2023} hinted that SoMe sessions are shorter when the declared breakout reason is due to work activity compared to leisure activity. Therefore, we assume that the current activity (work or leisure activity) influences the effectiveness of interventions during infinite scrolling. 



\textbf{Social Situation.}
When people are in social situations, like sitting in a coffee shop surrounded by strangers or having a meal with friends, using the phone is often perceived as impolite~\cite{Miller-Ott.2017, FORGAYS2014314}. Studies show that checking the phone during social situations can interrupt conversations and reduce people's connectedness and empathy to each other~\cite{Przybylski.2013, Misra.2016, Lutz.2020}. However, when people are eating alone, they tend to use their phones more, usually for fun or to pass the time~\cite{Weber.2020}. As social norms create pressure not to use the phone during social gatherings, we assume that this is an important factor in influencing users' \textit{reactance} and \textit{responsiveness} towards an intervention. 



\textbf{At Home.} \citet{Hintze.2017} found that mobile phone session durations were twice as long when users were at home compared to other locations. This difference in usage patterns could be influenced by the absence of social norms around phone use in private spaces like the home, potentially affecting how users respond to interventions in these environments.



\textbf{Multitasking.} \citet{Akpinar.2023} found that multitasking during phone use leads to more typing errors as users are distracted. Although typing is not directly related to infinite scrolling, we believe that when users are multitasking—like eating, cooking, or watching TV—an intervention during infinite scrolling could redirect their focus back to the main activity. This shift in attention could potentially affect the intervention's effectiveness.



\textbf{Valence and Sleepiness.}
\citet{RixenIS.2023} found ``\textit{[...] that participants reported significantly higher levels of valence on sessions that were not only composed of scrolling activity}''~\cite[p. 15]{RixenIS.2023}. Valence refers to the positive or negative emotions that individuals experience~\cite{Frijda1986-FRITE, Vazard.2022}. This suggests that infinite scrolling has a negative impact on users' valence. Additionally, \citet{Diefenbach.2019} noted that people often turn to their smartphones as a way to cope with negative emotions. This raises the possibility that the emotional impact of infinite scrolling might influence how users respond to interventions. 
Further, \citet{YANG2020112686} found that excessive smartphone usage is significantly related to poor sleep quality, also influencing daytime sleepiness~\cite{Nathan.2013}. In particular, ``\textit{longer average screentimes during bedtime and the sleeping period were associated with poor sleep quality}''\cite[p. 2]{Christensen.2016}. We, therefore, assume that sleepiness might be related to the effectiveness of an intervention during infinite scrolling.

\section{User Study}
To investigate how contextual factors influence users' \textit{responsiveness} and \textit{reactance} towards interventions during infinite scrolling, we conducted a 7-day-long field study with N=\userstudyfinal participants.
For the user study, we selected six of the most used SoMe applications in the United States in 2023~\cite{StatistaSNS}, which were also investigated in related work~\cite{RixenIS.2023}. These applications are Facebook, Instagram, X (former Twitter), Reddit, TikTok, and YouTube (specifically their Shorts feature).

\subsection{Apparatus}\label{sec:apparatus}

\begin{figure}[ht!]
\centering
\small
    \begin{subfigure}[c]{0.32\linewidth}
    \centering
        \includegraphics[width=\linewidth]{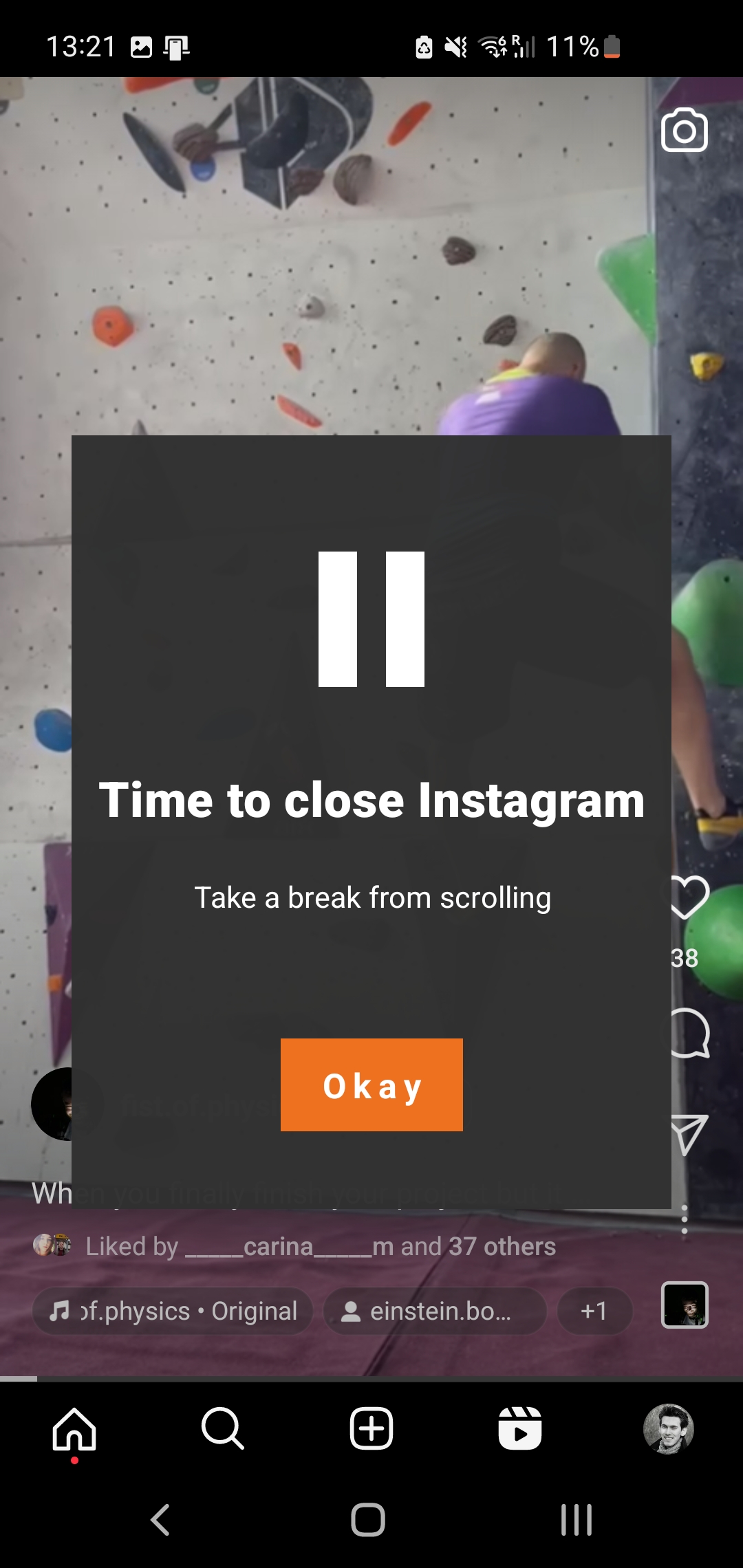}
        \caption{Intervention}\label{fig:intervention}
        \Description{This screenshot shows the pop-up window as an intervention over Instagram. The pop-up shows a pause button with the text ``Time to close Instagram. Take a break from scrolling.'' Below it is an orange button with the caption ``Okay''}
    \end{subfigure}
    \hspace{.3mm} 
    \begin{subfigure}[c]{0.32\linewidth}
        \includegraphics[width=\linewidth]{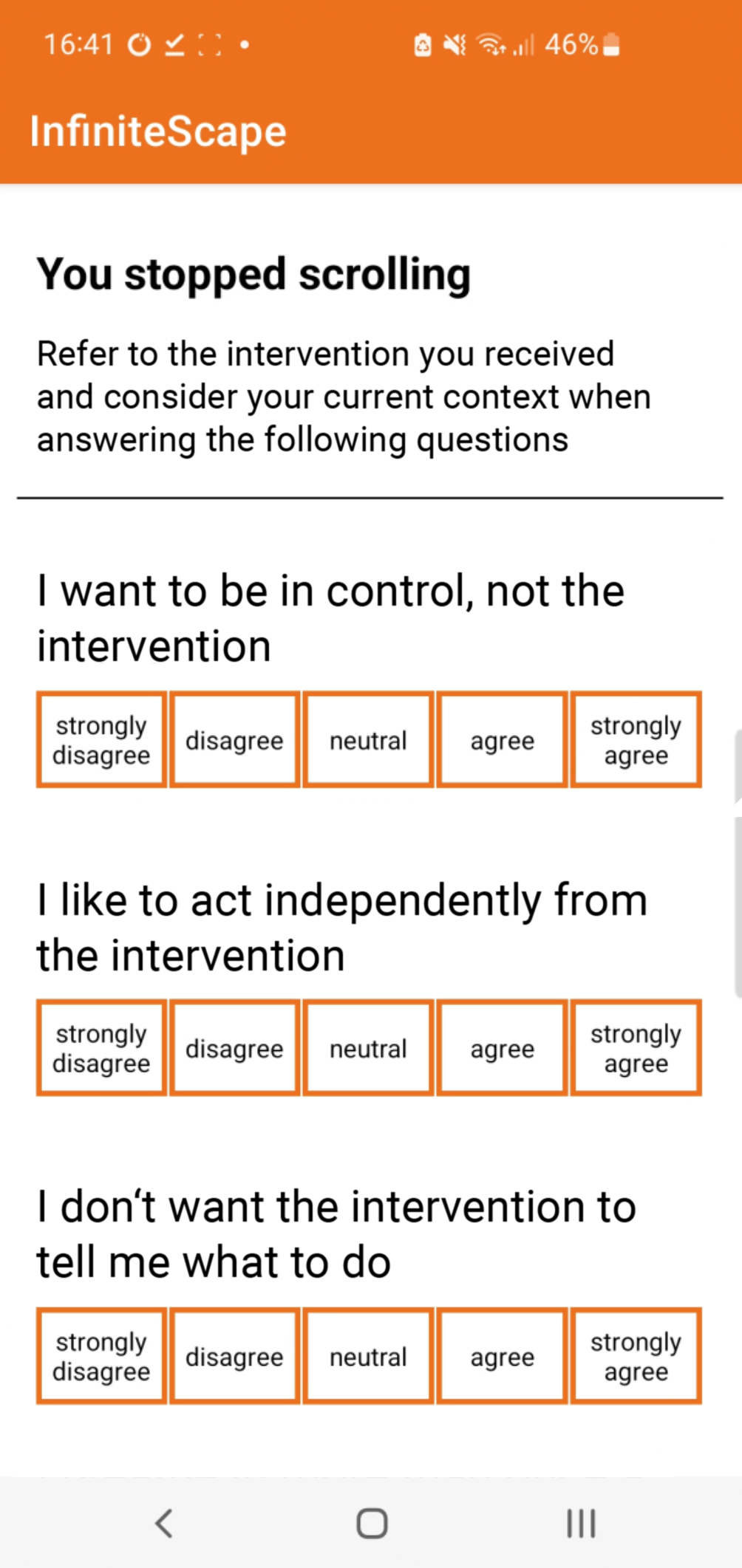}
        \caption{Questionnaire pt. 1 }
        \Description{This screenshot shows the first part of the questionnaire used in the user study}
    \end{subfigure}
    \hspace{.3mm} 
    \begin{subfigure}[c]{0.32\linewidth}
        \includegraphics[width=\linewidth]{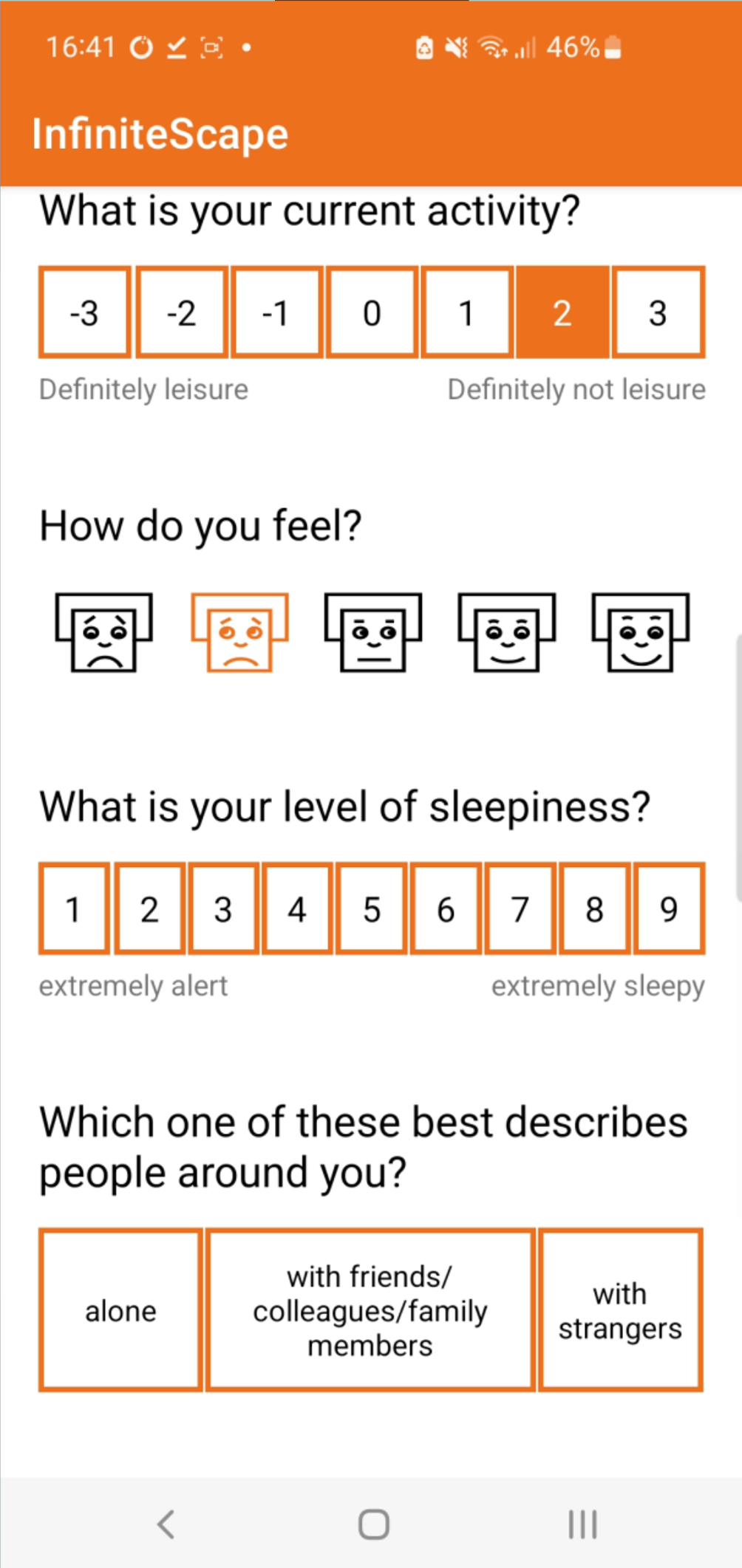}
        \caption{Questionnaire pt. 2}
        \Description{This screenshot shows the second part of the questionnaire used in the user study}
    \end{subfigure}
   \caption{InfiniteScape. Including the intervention overlay and the questionnaire making for the participants' \textit{reactance} and current context. The questionnaire is only partly visible (see \autoref{sec:questionaire_design} for more details)}~\label{fig:infiniteScape}
   \Description{This figure shows three screenshots of InfiniteScape, including the intervention and the questionnaires}
\end{figure}

To answer our research question, we developed \textit{InfiniteScape}, 
a native Android application that monitors users' infinite scrolling behaviors across the six SoMe platforms. To achieve this, we implemented Android's Accessibility Service,\footnote{\url{https://developer.android.com/reference/android/accessibilityservice/AccessibilityService}, accessed: March 11, 2024}, which enables access to the content variable of each application’s window tree. This setup allowed \textit{InfiniteScape} to detect the active tab or section within each SoMe app and determine whether the user was engaged in infinite scrolling.
Our monitoring focused solely on infinite scrolling, deliberately excluding other interactions such as direct messaging or content creation within these platforms. For instance, if a user switched from the "Reels" tab in Instagram to direct messaging, the content variable would update from "Reels" to "Messages," which \textit{InfiniteScape} interpreted as an interruption in infinite scrolling. A similar logic was applied across other SoMe platforms; for example, navigating outside of YouTube’s Shorts section within YouTube would also be interpreted as an interruption in scrolling. Closing the application while engaged in infinite scrolling was likewise logged as an interruption of continuous infinite scrolling. This approach ensured that only continuous scrolling sessions, without any interruptions, were categorized as infinite scrolling.

Upon detecting uninterrupted, continuous infinite scrolling for 15~minutes, an intervention overlay (see \autoref{fig:intervention}) appears on the smartphone's screen, modeled after the screen-time reminders from TikTok\footnote{\url{https://support.tiktok.com/en/account-and-privacy/account-information/screen-time}, accessed: March 11, 2024} and Instagram\footnote{\url{https://help.instagram.com/2049425491975359/?cms_platform=android-app&helpref=platform_switcher}, accessed: March 11, 2024}. The 15-minute latency to start the intervention after scrolling was informed by \citet{Terzimehic.2022b}. They found that after approximately 10--20 minutes, most participants stated negative feelings toward smartphone usage. Further, \citet{RixenIS.2023} reported that in sessions exceeding 10 minutes, infinite scrolling was the predominant activity during SoMe sessions.
Thus, the intervention overlay always appears when users scroll continuously for 15 minutes without interruption. Users could remove this intervention by tapping “okay”. This allowed users to continue infinite scrolling. If users ultimately stopped infinite scrolling after dismissing the intervention—such as by closing the application or switching to a non-infinite-scrolling activity inside the same app—they were shown a questionnaire. This questionnaire captured their current context (see \autoref{sec:questionaire_design}) and assessed the \textit{reactance} they experienced towards the intervention.
Additionally, we measured the time between the intervention overlay occurred until the users eventually stopped infinite scrolling, defining this duration as the \textit{responsiveness} towards the intervention. We also logged the time of day (hh:mm:ss) of the intervention.

This real-time feedback collection uses the Experience Sampling Method (ESM), which has been validated by existing research (e.g.,~\cite{beyens2020effect, kross2013facebook, RixenIS.2023}). 
Unlike traditional ESM implementations that trigger questionnaires at pre-determined (periodic) times~\cite{beyens2020effect, reissmann2018role, kross2013facebook, verduyn2015passive}, our method triggers questionnaires when participants stop infinite scrolling. This event-based approach received higher response rates than the traditional, periodic approach~\cite{vanBerkel.2019}. However, this method also has its limitations, as it only captures the participant's context at the moment they decide to stop scrolling. Hence, we missed data from those who continued scrolling, as the questionnaire would not be triggered. Despite this, we chose event-based ESM because of its effective usage in other SoMe studies (e.g.,~\cite{Cho.2021, RixenIS.2023, Chang.2015, Bayer.2018}).

\subsection{Procedure}
Prior to participating in the longitudinal 7-day study, participants were guided through a short registration survey. Here, they were provided with an in-depth explanation of the study's objectives and procedure to explore contextual influences towards interventions during infinite scrolling. 
We pre-screened participants from the United States via \href{https://www.prolific.com}{Prolific}. Further, we excluded participants from the study who did not own an Android phone with version 10 or higher as the required permissions were optimized for these versions, and more than 85\% of American Android users had version 10 or higher during the study period~\cite{AndroidVersionUSA}. Further, only participants who expressed regret during infinite scrolling (``\textit{Do you ever regret scrolling too much on social media apps?}'') were invited to take part in the 7-day study by downloading the application. 
This decision aligns with Self-Determination Theory~\cite{ryan_self-determination_2000}, which emphasizes that intrinsic motivation, driven by autonomy and alignment with personal values, is essential for behavior change. Regret signals participants' recognition of excessive scrolling as problematic, thus fostering a willingness to engage in interventions. In contrast, those without regret may lack intrinsic drive, reducing the study's ability to evaluate intervention effectiveness.
Regardless of whether they downloaded the application, participants in the registration study were compensated 0.19£ for their median effort of 1:40~minutes.

Those participants who proceeded with the longitudinal study received an instructional video detailing the \textit{InfiniteScape} application's download and installation process. Due to the use of Android's Accessibility Service for detecting infinite scrolling, the application could not be hosted on the Google Play Store~\cite{GooglePlayStore.AccesabilityService}. Thus, an anonymized repository was available for the download of the \textit{.apk} file. This was accessible either via a QR code for desktop-based registrants or a direct download link for mobile participants. Upon installing \textit{InfiniteScape}, participants were shown the terms-of-consent form, which they were encouraged to read carefully before agreeing to participate. 
Both the study and consent form received approval from the university's Ethics Committee, ensuring that all privacy protocols and ethical standards, such as anonymization of the data, were upheld.
After the form, the application guided them to grant the necessary permissions. Finally, the application prompted participants to enter their age and specify the gender with which they most closely identify (male, female, non-binary, prefer not to answer). After the demographic survey, the application started its service, indicated by a continuously displayed icon in the phone's top bar—a standard requirement for Android foreground services.\footnote{\url{https://developer.android.com/develop/background-work/services/foreground-services}, accessed: November 13, 2024}. This icon was present during the entire duration of the study. Given its constant presence, we expect participants to become habituated to it, minimizing any influence on their natural scrolling behavior.

During the 7-day user study, the participant's infinite scrolling behavior was tracked, intervening with an overlay after 15 minutes of continuous scrolling. After the participants stopped infinite scrolling by closing the application, they were provided with a questionnaire asking them about their perceived \textit{reactance} and current context. For each completed questionnaire, participants were compensated with a bonus payment of 0.5£. This results in an average bonus payment of 5.20£ per participant who completed the study.
After the 7-day study, the application notified participants that the study had finished and that they could delete the application.

\subsection{Questionnaire Design}\label{sec:questionaire_design}
This section outlines the specific questions used to measure the dependent variables and contextual factors collected during the study. 
Recognizing the importance of participant engagement and the potential for survey fatigue, we predominantly utilized concise, single-item measures. Although single-item measures can produce measurement error~\cite{Dejonckheere.2022}, their use is well-established and validated in the field of SoMe research, offering a balance between data quality and respondent burden~\cite{beyens2020effect, RixenIS.2023, Bayer.2018, Chang.2015}. Nonetheless, to mitigate potential measurement errors and maintain data integrity, we incorporated random attention checks within the questionnaire. A detailed list of the concrete items used during the user study can be found in \autoref{app:usedquestionaires}.

\paragraph{Dependent Variables}
According to \citet{Rains.2013}, interventions aiming for behavior change reduce user's sense of control~\cite{KaiLukoff.2022.Dissertation}. Hence, we assume this is also true for interventions during infinite scrolling. Thus, intervening can create \textit{reactance} towards the intervention. This phenomenon in HCI is defined by \citet{Ehrenbrink.2020}, who refers to \textit{\textit{reactance}} as the resistance individuals feel when their freedom of choice is perceived to be under threat. Hence, increased \textit{reactance} can reduce the effectiveness of the intervention by affecting the user's acceptance of the guidance or constraints imposed. Consequently, we used \textit{reactance} as our first dependent variable, as suggested by \citet{meinhardt_balancing_2023}. We measured \textit{reactance} using the subscale Threat of the Reactance Scale for Human-Computer Interaction (RSHCI)~\cite{Ehrenbrink.2020}. This subscale included five question items that were rated on a five-point Likert scale ranging from 1=``strongly disagree'' to 5=``strongly agree.'' For each observation, we calculated the \textit{reactance} score by averaging across these five items.
Additionally, we used \textit{responsiveness} towards an intervention, defined as the time span in seconds from displaying the intervention and the moment when participants eventually stopped infinite scrolling.

\paragraph{Contextual Factors}
The specific questionaires that we used for surveying the contextual factors (see \autoref{sec:hypotheses}) are described in the following:

\begin{itemize}[leftmargin=*, itemsep=0pt]
    \item For the \textbf{Current Activity}, we used the interval scale proposed by \citet{Samdahl.1991}, which extends from (-3),``definitely leisure'', to (+3), ``definitely not leisure'' context.
    \item For assessing the \textbf{Social Situation}, we employed a question inspired by \citet{Akpinar.2023}, who defined social context during smartphone usage as ``Which one of these best describes people around you?''. We gave three possible responses: alone, with acquaintances (friends, family, colleagues), or with strangers.
    \item To assess whether participants were \textbf{At Home}, they were asked, ``Are you currently at home?'' and were given a yes or no answer.
    \item We added the contextual factor of \textbf{Multitasking} by asking, ``Did you do anything else besides being on [App Name]?''. This question could be answered either with yes or no.
    \item To define the internal context, we asked for the participants' \textbf{Valence} using the self-assessment Manikin scale (SAM)~\cite{Bradley.1994} as already employed by \citet{RixenIS.2023}. The scale contains five images of manikin. However, we only used the dimension for valence. Further, we simplified the scale by only using the faces of the images used in the SAM, as this is the only part changing for valence in the SAM.
    \item To measure \textbf{Sleepiness}, we used the Karolinska Sleepiness Scale (KSS)~\cite{Shahid.2012b} ranging from 1=``extremely alert'' to 9=``extremely sleepy''.
\end{itemize}

\subsection{Participants}
We recruited participants over approximately one month to recruit a total of \nuserstudyrecuruted participants who completed the registration phase of the study. As mentioned above, participants who did not experience regret during infinite scrolling were excluded from the user study, resulting in \nuserstudyregret eligible participants. A total of N=160 participants successfully downloaded \textit{InfiniteScape} and enrolled in the longitudinal study. The participants who refrained from downloading the application mentioned reasons such as privacy concerns, difficulties with the download process, or the perceived burden of a 7-day commitment to the study. In addition, we believe this drop-out range can largely be attributed to the low initial effort required for registration. This may have led participants to claim the initial reward without full commitment to completing the longitudinal study.
While the drop-out rate may appear substantial, it is consistent with \citet{RixenIS.2023}, who reported comparable rates. 
To ensure consistent exposure duration for each participant, we excluded data from 88 participants who did not complete the full 7-days during the study. 
This resulted in a final sample size of N=\userstudyfinal participants, with a mean age of $MD=35.50$, $SD=10.01$ years (33 male, 29 female, 10 non-binary, 0 prefer not to answer). These participants provided a total of 946 data points, which were used in our subsequent analysis.

\subsection{Results}

\begin{figure*}[ht!]
\centering
\small
    \begin{subfigure}[c]{0.1595\linewidth}
        \includegraphics[width=\linewidth]{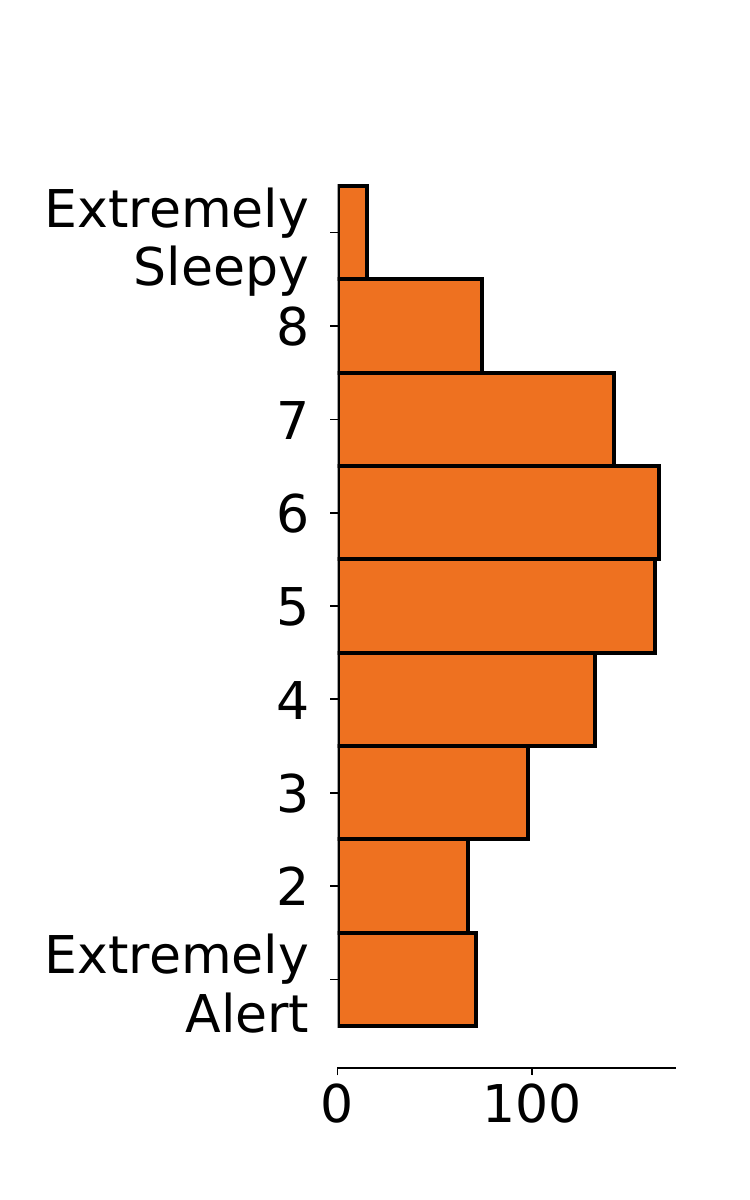}
        \caption{Sleepiness~\cite{Shahid.2012b}}\label{fig:distribution_kss}
        \Description{Distribution of Sleepiness}
    \end{subfigure}
    \begin{subfigure}[c]{0.1595\linewidth}
        \includegraphics[width=\linewidth]{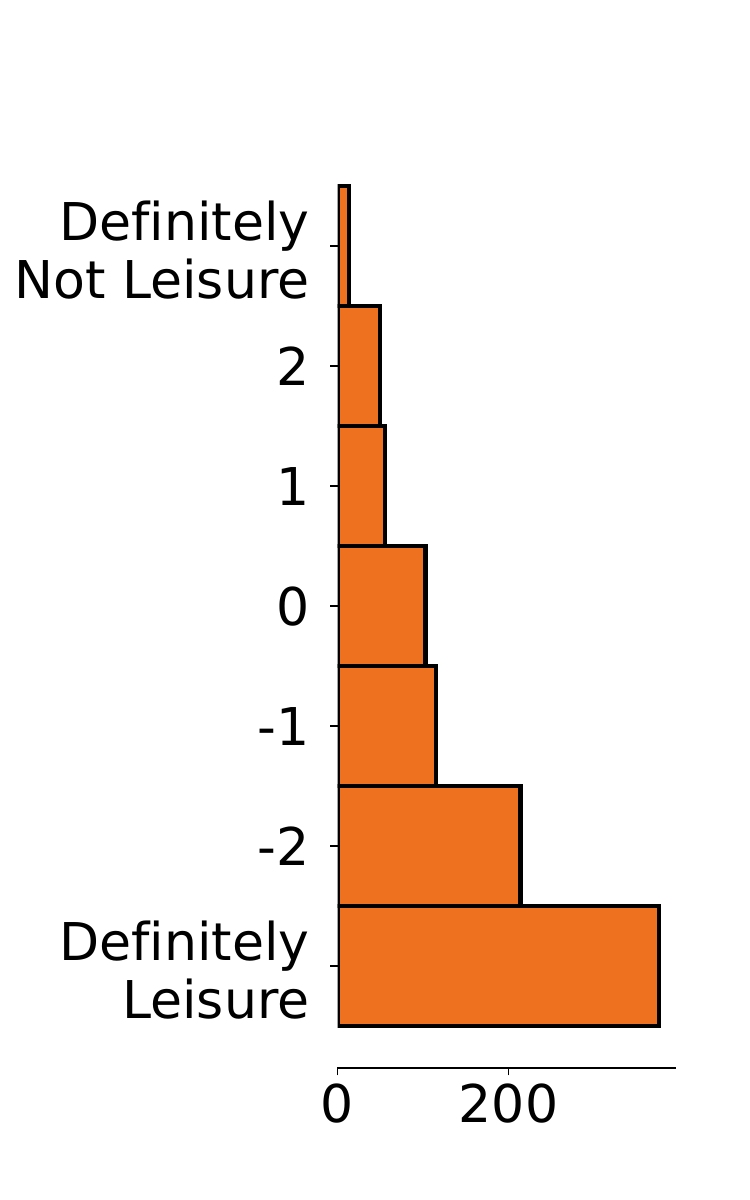}
        \caption{Current Act.~\cite{Samdahl.1991}}\label{fig:distribution_current_activity}
        \Description{Distribution of the current activity}
    \end{subfigure}
    \begin{subfigure}[c]{0.1595\linewidth}
        \includegraphics[width=\linewidth]{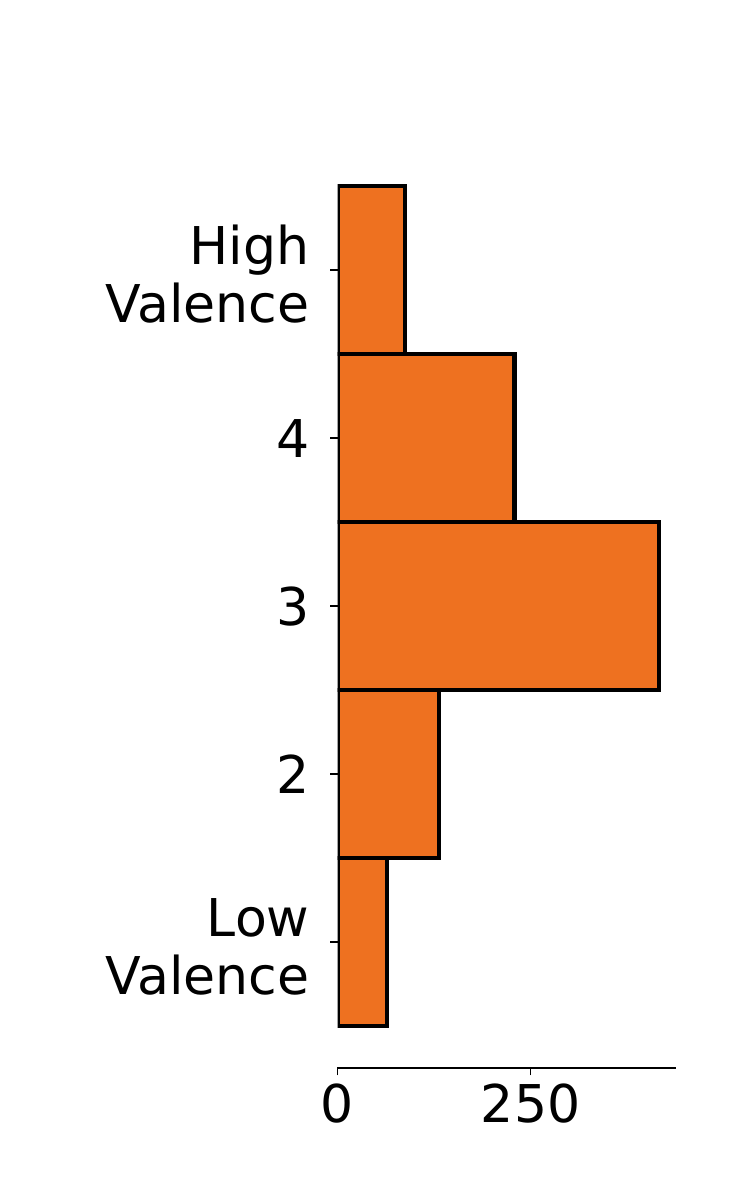}
        \caption{Valence~\cite{Bradley.1994}}\label{fig:distribution_sam}
        \Description{Distribution of the Valence}
    \end{subfigure}
    \begin{subfigure}[c]{0.1595\linewidth}
        \includegraphics[width=\linewidth]{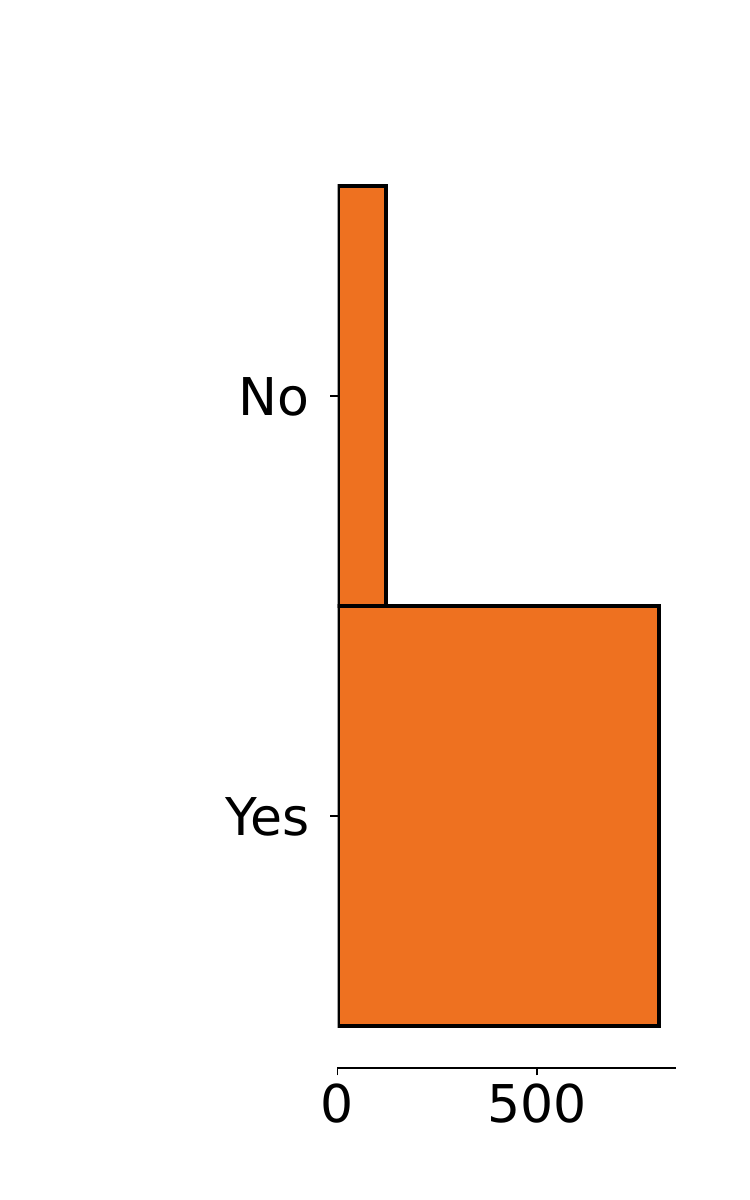}
        \caption{At Home}\label{fig:distribution_at_home}
        \Description{Distribution of contextual factor at home}
    \end{subfigure}
    \begin{subfigure}[c]{0.1595\linewidth}
        \includegraphics[width=\linewidth]{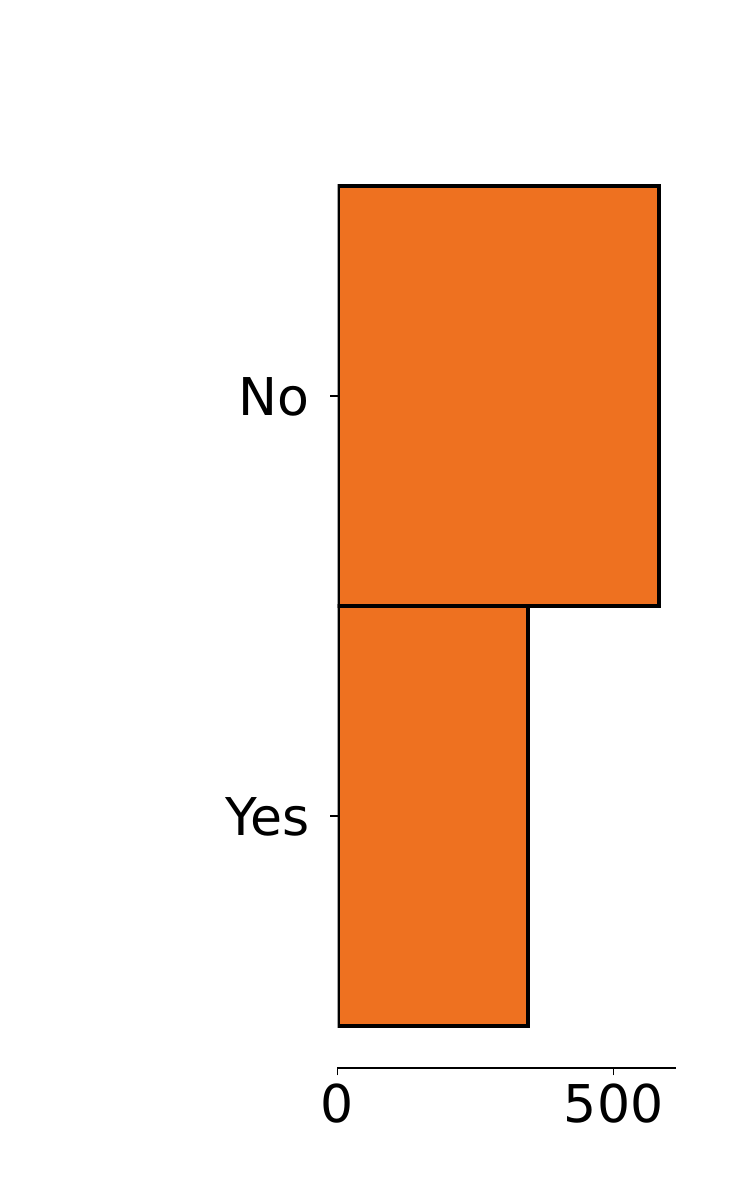}
        \caption{Multitasking}\label{fig:distribution_side_activity}
        \Description{Distribution of contextual factor multitasking}
    \end{subfigure}
    \begin{subfigure}[c]{0.1595\linewidth}
        \includegraphics[width=\linewidth]{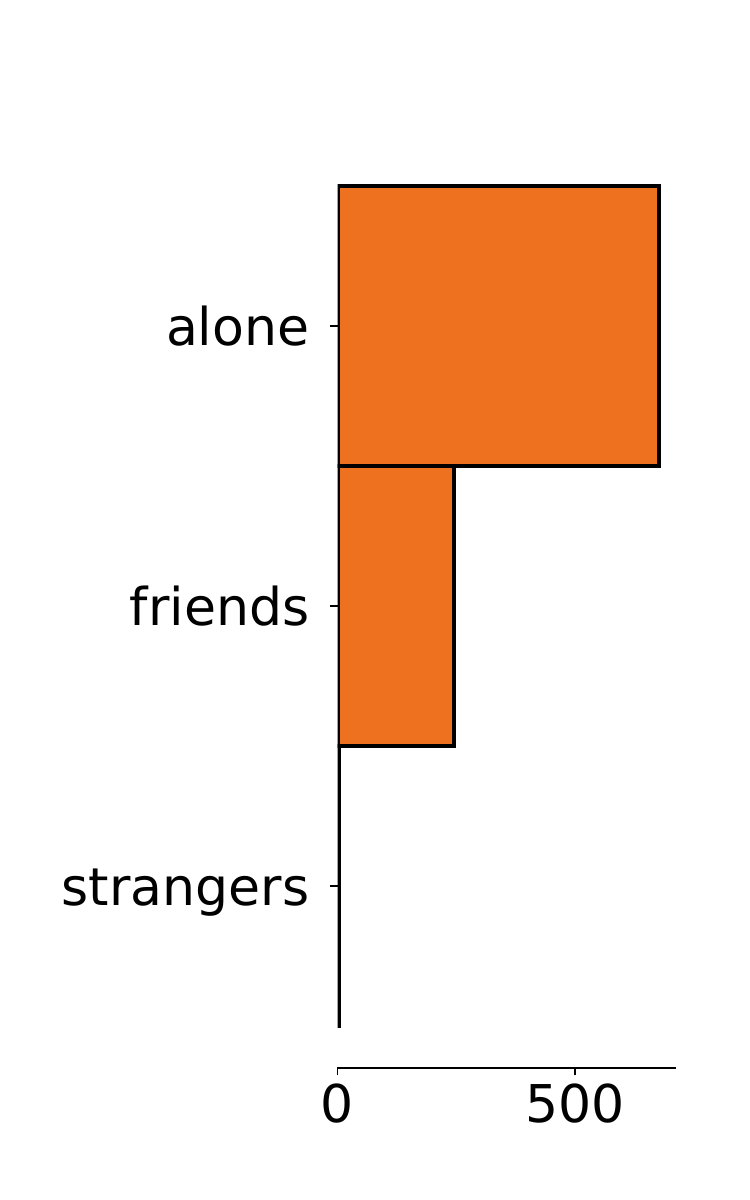}
        \caption{Social Situation}\label{fig:distribution_social_situation}
        \Description{Distribution of contextual factor social situation}
    \end{subfigure}
   \caption{Distribution of the investigated contextual factors during infinite scrolling interventions}~\label{fig:distribution}
   \Description{These plots show the distributions of the contextual factors assessed during the user study}
\end{figure*}

To address our research question of how context affects users' \textit{reactance} and \textit{responsiveness} towards interventions during infinite scrolling, we fitted linear mixed models (LMM) for each dependent variable to explore the main effects and interactions. This enabled us to include a random intercept for each participant, as the models account for the correlation between repeated measures of \textit{reactance} and \textit{responsiveness} within the same participant. \\
We used R version 4.3.1 and RStudio version 2023.12.1 with up-to-date packages as of September 2024 for analysis and Python version 3.10.4 for plotting. 

\subsubsection{Data Pre-Processing}
Initially, we removed 11 data points of participants who failed the attention checks. We then used the z-score method to identify and remove outliers in the dependent variables, setting the threshold at a z-score of 3. Thus, data points that were not within three standard deviations of the mean are considered statistically rare and were removed from the data set. Accordingly, 8 data points were removed as they exceeded the z-score threshold for \textit{responsiveness}. Looking at these data points, the time to stop infinite scrolling after the intervention exceeds 3 hours. Hence, we assume that there were technical issues during these sessions. 
After preprocessing, 927 data points from 72 participants (with an average of $12.88$ (SD=$13.02$) data points per participant) remained for subsequent analysis (see \autoref{app:distribution} for detailed frequency of data points per participant). Subsequently, we evaluated the distribution of our dependent variables using the Shapiro-Wilk test~\cite{Shapiro-Wilk}. The results indicated that both \textit{reactance} (W = 0.95, p < 0.001) and \textit{responsiveness} (W = 0.48, p < 0.001) are not normally distributed. Despite this non-normality, according to \citet{Arnau.2012}, deviations from normality have only minimal impact on the standard errors of estimation methods for longitudinal studies. Consequently, we assessed skewness and found that while \textit{reactance} was nearly symmetric (-0.36), \textit{responsiveness} was strongly right-skewed (3.62). To correct this and gain more robust estimates, as suggested by \citet{draper1998applied}, we applied a logarithmic transformation to reduce the skewness of \textit{responsiveness} to 0.782. After this correction, the Shapiro-Wilk test showed W = 0.88, p < 0.001, indicating an improvement toward a normal distribution.

\subsubsection{Descriptive Data} 

\begin{figure}[ht!]
\centering
\small
    \begin{subfigure}[c]{0.55\linewidth}
        \includegraphics[width=\linewidth]{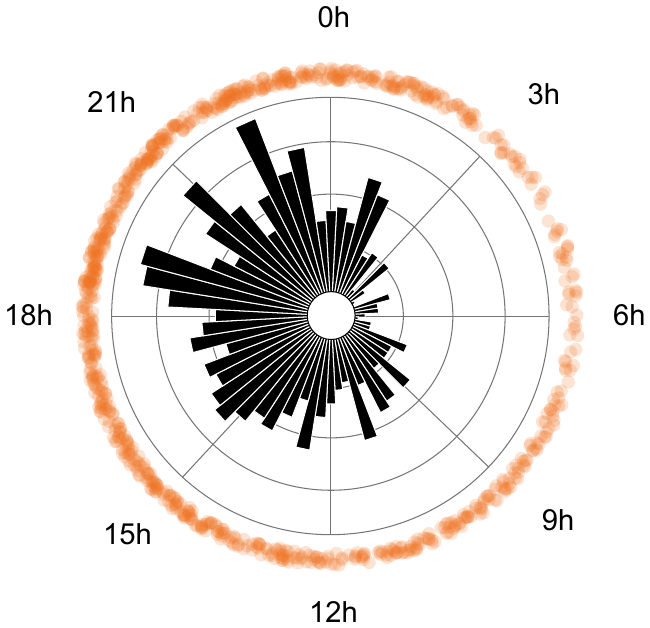}
        \caption{Distribution of intervention data points over time of day}\label{fig:time_distribution}
        \Description{This circular bar plot depicts the distribution of interventions during the time, indicating that most interventions occurred between 6pm and 11pm}
    
    \end{subfigure}
        \hspace{1.5cm}
    \begin{subfigure}[c]{0.55\linewidth}
        \includegraphics[width=\linewidth]{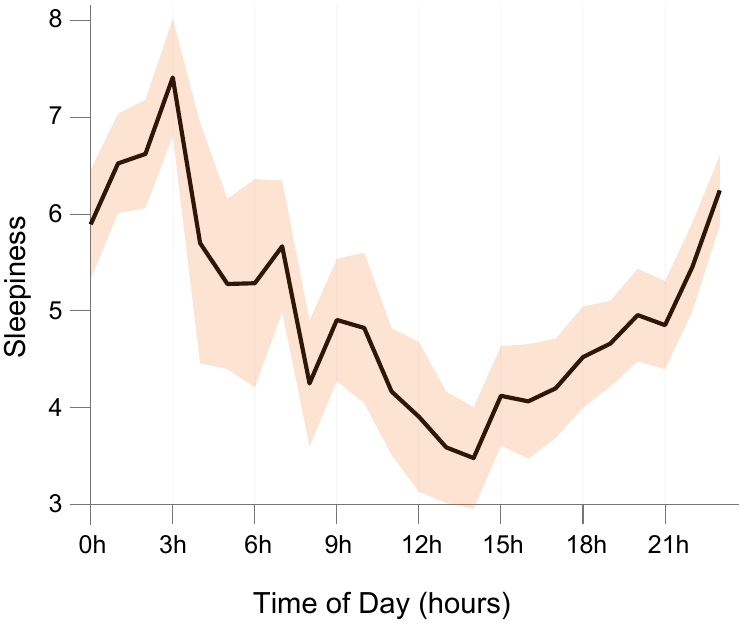}
        \caption{Relation between sleepiness~\cite{Shahid.2012b} and time of day in hours with 95\% CI}\label{fig:sleepiness_timeOfDay}
        \Description{This line plot shows the relation between the sleepiness rating and the time of day, indicating an increase of sleepiness from 1pm to 3am and a decrease between 4am and 12am}
    \end{subfigure}
   \caption{Distribution of the interventions over time of day and the relation between sleepiness and time of day}~\label{fig:descriptive_data_time}
   \Description{These plots show the distribution of the interventions over time of day and the relation between sleepiness and time of day}
\end{figure}

The descriptive data indicate that the overall level of \textit{reactance} was rated as medium on a range from 1 to 5, with a mean of $3.55$ (SD=$1.03$). In terms of \textit{responsiveness}, the duration users continued infinite scrolling after an intervention varied widely, from 0 seconds to 67 minutes and 20 seconds.
On average, users stopped scrolling after 3 minutes and 31 seconds (SD=8 minutes and 21 seconds). The interventions of the six SoMe applications were distributed as follows: TikTok was the most used at 40.30\%, followed by Reddit (26.48\%), Facebook (13.49\%), Instagram (11.13\%), X (5.56\%), and YouTube Shorts (3.03\%). The distribution of the contextual factors is depicted in \autoref{fig:distribution}. Further details regarding the contextual factors are summarized in \autoref{app:descriptive}. 
The majority of interventions occurred late afternoon and evening (see \autoref{fig:time_distribution}), with a peak between 18h and 23h (40.99\% of the data points). During the night (between 0h and 6h), a minimal of interventions occurred (14.78\%), suggesting minimal engagement in infinite scrolling during these hours.

\subsubsection{Linear Mixed Models}

For our analysis, we fitted two LMMs for each of our independent variables (see \autoref{table:lmm}): \textit{reactance} and \textit{responsiveness}. For Models 1 and 3, we investigated the main effects, employing the formula: \textit{Reactance}/\textit{Responsiveness} $\sim$ At Home + Current Activity + Sleepiness + Valence + Side Activity + Social Situation. Conversely, Models 2 and 4 examined interaction effects, using the formula: \textit{Reactance}/\textit{Responsiveness} $\sim$ At Home * Current Activity * Sleepiness * Valence * Multitasking * Social Situation. In all models, participants were included as a random effect, denoted as $\sim$ 1 | ProlificID.
To control for the increased risk of Type I errors due to multiple comparisons and the exploratory nature of this study, we adjusted the alpha level using the Bonferroni correction to $\alpha = 0.025$, ensuring that the results are robust against the possibility of finding false positives.

In Models 1 and 3, which focused on the main effects, one significant main effect was found. In Models 2 and 4, which assessed interaction effects, four significant main effects were also observed. Notably, while the main effects for Models 1 and 3 are valid for interpretation, the main effects in Models 2 and 4 should not be interpreted due to the presence of interaction effects as noted by the guidelines from \citet{Nelder.1977}. This is due to the relationship between independent and dependent variables being altered, making it inappropriate to interpret main effects in isolation~\cite{venables1998exegeses}. Thus, in \autoref{table:lmm}, these main effects in Models 2 and 4 are grayed out, and only their interaction effects are considered for the subsequent interpretation.
Detailed results for all four models (two for each independent variable) are presented in \autoref{table:lmm}.

Although the time of intervention could also potentially influence how users respond to an intervention, we refrained from including this factor in the LMMs. The rationale behind this exclusion is that the periodic nature of daytime does not fit well with the linear analysis used in LMMs. Furthermore, individuals' daily schedules vary widely (e.g., a shift worker might wake up at 1 am compared to a student waking up at 9 am), making it difficult to generalize the effect of daytime on participants' reactions toward interventions. Instead, we argue that sleepiness is a more appropriate variable. As shown in \autoref{fig:sleepiness_timeOfDay}, sleepiness increases during the night and decreases during the day. This pattern not only represents the individual physiological rhythms common to all participants but also serves as a linear factor for our models.

\begin{table*}[ht!] \centering
\small
  \caption{Linear Mixed Models predicting \textit{reactance} and \textit{responsiveness}. Coefficient (Standard Error)} 
  \Description{This table shows the results of all calculated Linear mixed models for \textit{reactance} and \textit{responsiveness}}
\renewcommand{\arraystretch}{0.47}
\begin{tabular}{lp{1.65cm}p{1.65cm}p{1.65cm}p{1.65cm}} 
\\[-1.8ex]\hline 
\hline \\[-1.3ex]
\addlinespace[5pt] 
 & \multicolumn{4}{c}{\textit{Dependent variable:}} \\ 
\addlinespace[5pt] 
\cline{2-5} 
 \addlinespace[2pt] 
\\[-0.5ex] & \multicolumn{2}{c}{Reactance} & \multicolumn{2}{c}{Responsiveness} \\ 
\\[-0.5ex] & Model 1 & Model 2 & Model 3 & Model 4\\
\addlinespace[5pt] 
\hline \\[-1.8ex] 
\addlinespace[5pt] 
\textbf{\textit{Main Effects}} \\ [0.4ex] 
\addlinespace[5pt] 
Social Sit. [Strangers] (Ref. alone) & $-$0.04 & \textcolor{lightgray}{8.76} & 0.76 & \textcolor{lightgray}{93.39$^{*}$} \\ 
  & (0.40) & \textcolor{lightgray}{(15.90)} & (0.93) & \textcolor{lightgray}{(36.99)} \\ 
  & & & & \\ 
 Social Sit. [Friends] (Ref. alone) & $-$0.09 & \textcolor{lightgray}{40.46} & $-$0.06 & \textcolor{lightgray}{74.43} \\ 
  & (0.08) & \textcolor{lightgray}{(26.73)} & (0.18) & \textcolor{lightgray}{(61.99)} \\ 
  & & & & \\ 
 Multitasking [True] & $-$0.05 & \textcolor{lightgray}{3.56} & $-$0.26 & \textcolor{lightgray}{24.36$^{*}$} \\ 
  & (0.07) & \textcolor{lightgray}{(4.58)} & (0.16) & \textcolor{lightgray}{(10.64)} \\ 
  & & & & \\ 
 Valence & $-$0.06 & \textcolor{lightgray}{0.76} & $-$0.06 & \textcolor{lightgray}{7.11$^{*}$} \\ 
  & (0.04) & \textcolor{lightgray}{(1.10)} & (0.08) & \textcolor{lightgray}{(2.56)} \\ 
  & & & & \\ 
   Sleepiness & \textbf{$-$0.05$^{**}$} & \textcolor{lightgray}{0.40} & 0.01 & \textcolor{lightgray}{2.70} \\ 
  & (0.02) & \textcolor{lightgray}{(0.96)} & (0.04) & \textcolor{lightgray}{(2.22)} \\ 
  & & & & \\ 
  Current Activity & $-$0.03 & \textcolor{lightgray}{1.77} & $-$0.01 & \textcolor{lightgray}{$-$6.86} \\ 
  & (0.02) & \textcolor{lightgray}{(1.66)} & (0.05) & \textcolor{lightgray}{(3.88)} \\ 
  & & & & \\ 
  At Home [True] & $-$0.11 & \textcolor{lightgray}{2.77} & 0.19 & \textcolor{lightgray}{30.76$^{**}$} \\ 
  & (0.11) & \textcolor{lightgray}{(4.13)} & (0.25) & \textcolor{lightgray}{(9.63)} \\ 
  & & & & \\ 
\textbf{\textit{2-way Interaction Effects}} \\[0.3ex]
\addlinespace[5pt] 
  Valence x Social Sit. [Strangers] && $-$2.58 && \textbf{$-$20.69$^{*}$} \\ 
  && (3.54) && (8.23) \\ 
  & & \\  
   Sleepiness x Social Situation [Strangers] && 0.16 && \textbf{$-$5.14$^{*}$} \\ 
  && (0.92) && (2.13) \\ 
  & & \\ 
   Current Activity x Social Sit. [Strangers] && 0.84 && 0.16 \\ 
  && (0.43) && (0.99) \\ 
  & & \\ 
   Valence x Multitasking [True] && $-$0.90 && \textbf{$-$7.27$^{*}$} \\ 
  && (1.26) && (2.93) \\ 
  & & \\ 
   At Home [True] x Multitasking [True] && $-$3.27 && \textbf{$-$29.21$^{**}$} \\ 
  && (4.70) && (10.94) \\ 
  & & \\ 
  At Home [True] x Valence && $-$0.78 && \textbf{$-$8.25$^{**}$} \\ 
  && (1.12) && (2.61) \\ 
  & & \\
   At Home [True] x Current Activity && $-$1.80 && 8.53 \\ 
  && (1.69) && (3.93) \\ 
  & & \\
  \textbf{\textit{3-way Interaction Effect}} \\[0.3ex] 
  \addlinespace[5pt] 
  At Home [True] x Valence x Multitasking [True] && 0.87 && \textbf{8.33$^{*}$} \\ 
  && (1.29) && (3.00) \\ 
  & & \\ 
  \hline 
  \\
$AIC$  & 2271.9 & 2308.1 & 3816.1 & 3957.9 \\ 
$BIC$  & 2320.2 & 2646.3 & 3864.4 & 4196.1 \\ 
$log-likelihood$  & -1126 & -1084 & -1898.1 & -1858.9 \\ 
$deviance$  & 2251.9 & 2168.1 & 3796.1 & 3717.9 \\ 
$R^{2} conditional  $ & 0.42& 0.41 & 0.34 &  0.36\\ 
$R^{2} marginal  $ &0.02 & 0.02 & 0.0072 & 0.07 \\ 
  \addlinespace[5pt] 
\hline 
\hline \\
\textit{Significance Codes:}  & \multicolumn{2}{r}{$^{*}$p$<$0.025; $^{**}$p$<$0.005; $^{***}$p$<$0.0005} \\ \\
\end{tabular} 
\Description{This table displays the results of a Linear Mixed Models Predicting \comf}
  \label{table:lmm} 
\end{table*}

\subsubsection{Main Effects}

\begin{figure}[ht!]
\centering
\small
    \begin{subfigure}[c]{0.7\linewidth}
        \includegraphics[width=\linewidth]{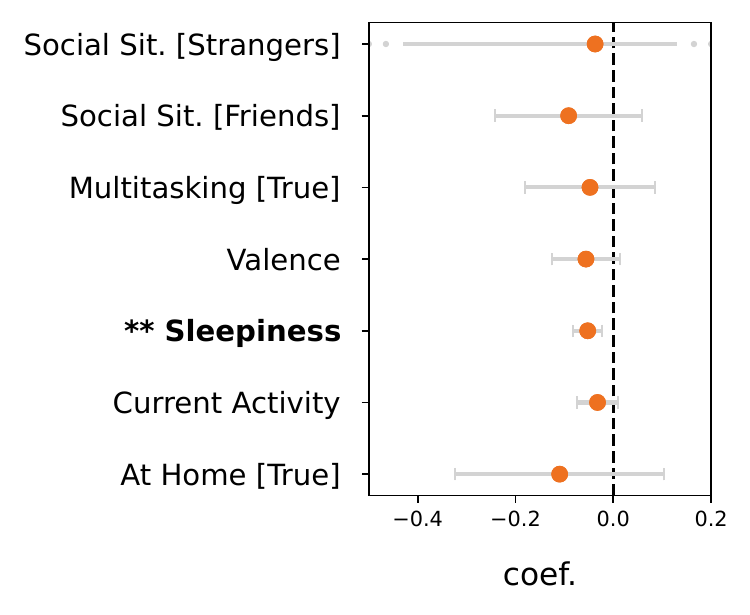}
        \caption{Main effects for \textit{reactance}  }\label{fig:main_effect_reactance}
        \Description{Main effects of reactance}
    \end{subfigure}
      \hspace{1cm}
    \begin{subfigure}[c]{0.7\linewidth}
        \includegraphics[width=\linewidth]{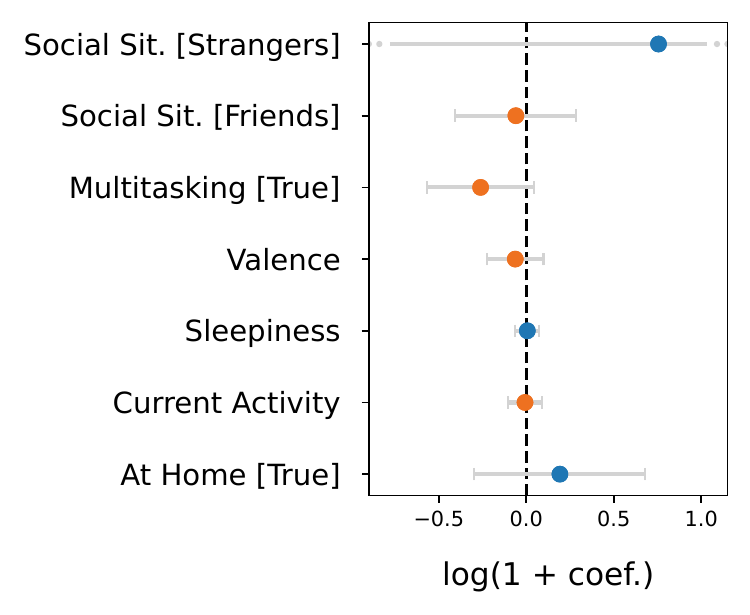}
        \caption{Main effects for \textit{responsiveness} }\label{fig:main_effect_responsiveness}
        \Description{Main effects of responsiveness}
    \end{subfigure}

   \caption{Coefficients of the main effects for \textit{reactance} and \textit{responsiveness}. As the \textit{responsiveness} was logarithmically transformed, the coefficients must be considered as log(1+ coef.). Blue dots indicate a positive coefficient, orange dots negative ones.}~\label{fig:lmm_main_effects}
   \Description{This plot shows the influence of each contextual factor for the two dependent variables. The sleepiness significantly influences the reactance towards an intervention}
\end{figure}

This section will report the significant main effect for \textit{reactance} and \textit{responsiveness} depicted in \autoref{fig:lmm_main_effects}. 
We found that sleepiness negatively affects \textit{reactance} (t(917) = -3.40, p < .001). This indicates that users experience less \textit{reactance} towards an intervention as they become more sleepy. However, our analysis did not reveal a significant main effect of sleepiness on users' \textit{responsiveness} to interventions. Additionally, we found no other significant main effects impacting user's \textit{reactance} or \textit{responsiveness}. However, we found multiple interaction effects on our dependent variables. 

\subsubsection{Interaction Effects}

\begin{figure*}[h!]
\centering
\small
    \begin{subfigure}[c]{0.26\linewidth}
        \includegraphics[width=\linewidth]{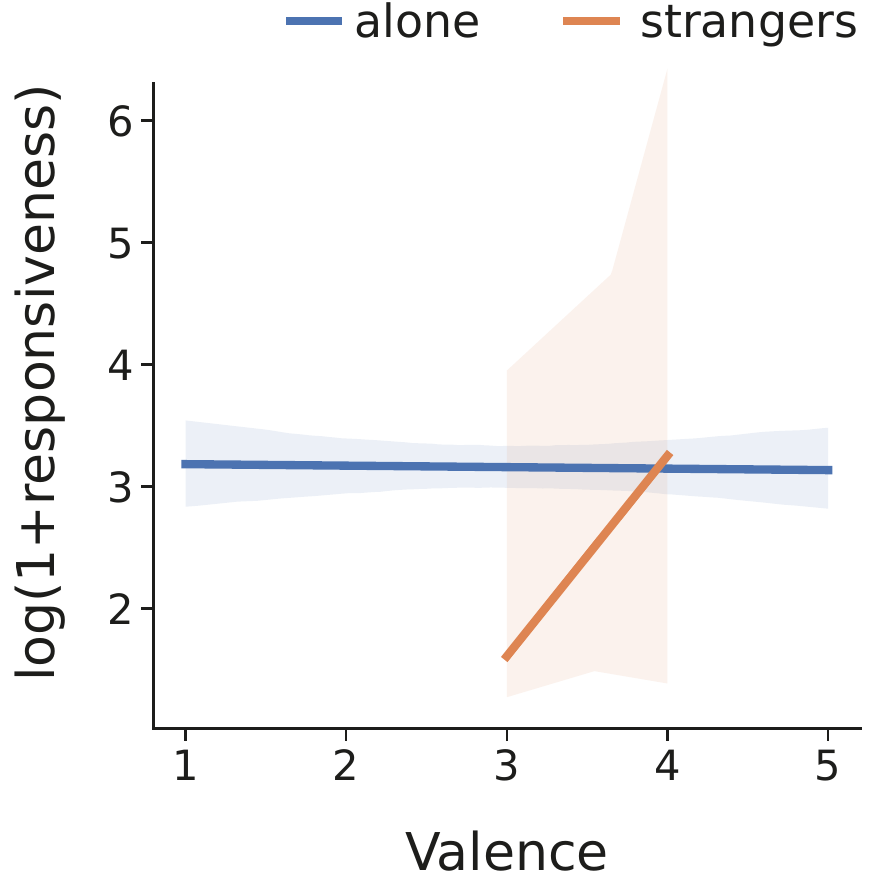}
        \caption{Valence × Social Situation}\label{fig:intEffect_valence_socialSit}
        \Description{Interaction effect between valence and Social Situation}
    \end{subfigure}
    \hspace{5mm}
        \vspace{5mm}
    \begin{subfigure}[c]{0.26\linewidth}
        \includegraphics[width=\linewidth]{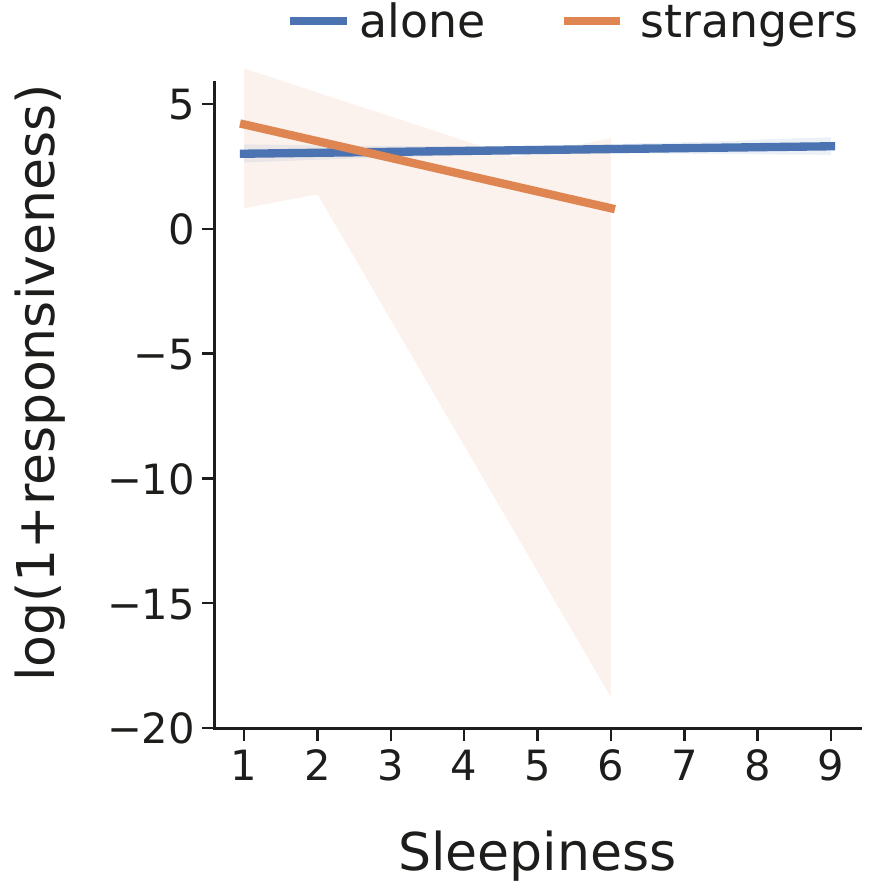}
        \caption{Sleepiness × Social Situation}\label{fig:intEffect_kss_socialSit}
        \Description{Interaction effect between sleepiness and Social situation}
    \end{subfigure}
    \hspace{5mm}
    \begin{subfigure}[c]{0.26\linewidth}
        \includegraphics[width=\linewidth]{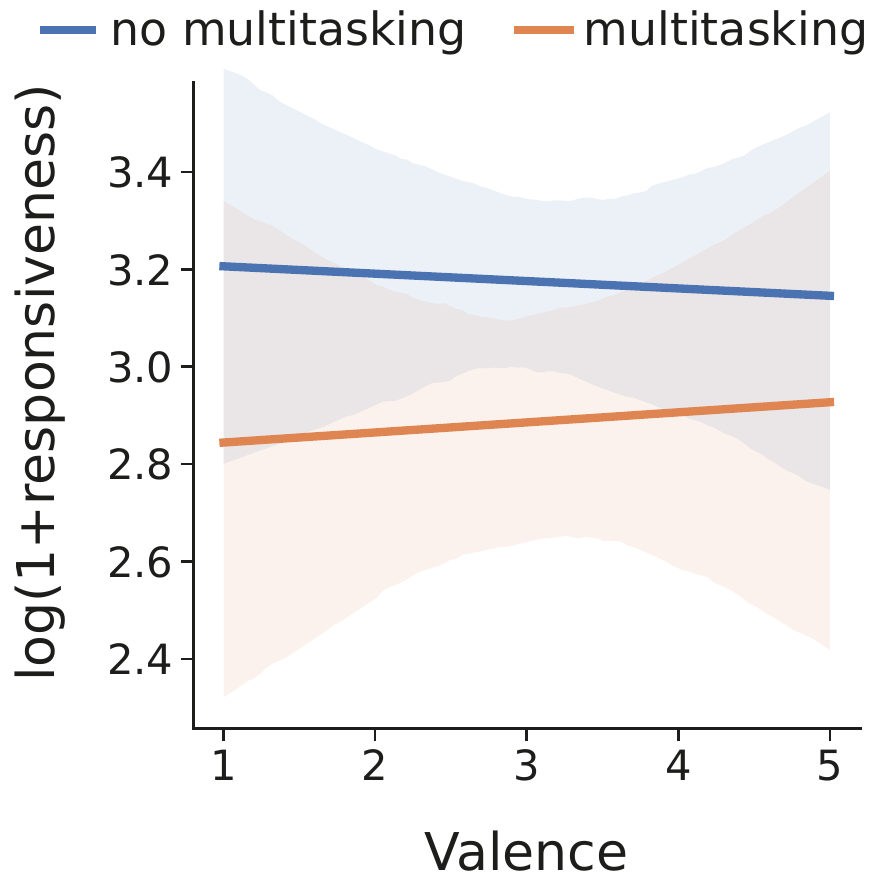}
        \caption{Valence × Multitasking}\label{fig:intEffect_sam_multitasking}
        \Description{Interaction effect between valence and multitasking}
    \end{subfigure}
    \hspace{5mm}

    \begin{subfigure}[c]{0.26\linewidth}
        \includegraphics[width=\linewidth]{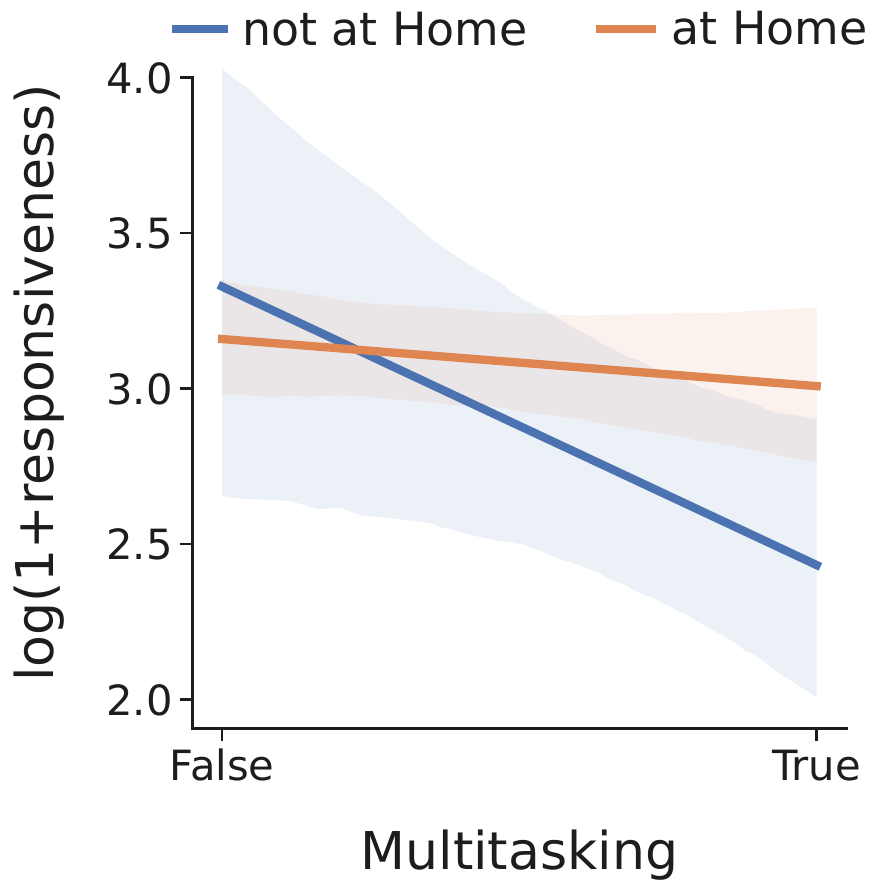}
        \caption{At Home × Multitasking}\label{fig:intEffect_multitasking_atHome}
        \Description{Interaction effect between multitasking and being at home or not}
    \end{subfigure}
    \hspace{5mm}
    \begin{subfigure}[c]{0.26\linewidth}
        \includegraphics[width=\linewidth]{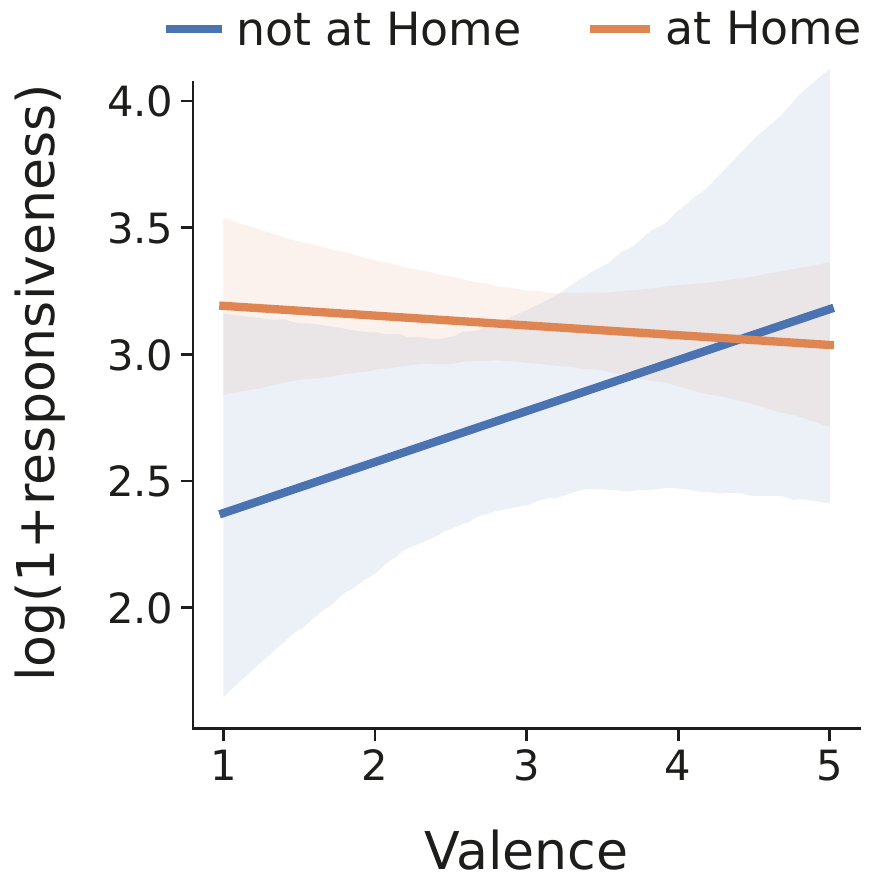}
        \caption{At Home × Valence}\label{fig:intEffect_sam_atHome}
        \Description{Interaction effect between valence and being at home}
    \end{subfigure}
   
   \caption{Interaction effects of \textit{responsiveness} with 95\% CI}~\label{fig:interaction_effects_responsiveness}
   \Description{These plots show the significant interaction effects for the responsiveness}
\end{figure*}



We found no significant interaction effects on \textit{reactance}. However, for the dependent variable \textit{responsiveness}, we found a negative interaction effect between Valence × Social Situation [Strangers] (t(857) = -2.51, p = 0.012), which was significant; see \autoref{fig:intEffect_valence_socialSit}. This indicates that while being alone, the \textit{responsiveness} to the intervention is almost unaffected, while it strongly increases with increasing valence when participants are in a social situation with strangers. However, one should note that no data points were obtained for low valence (1 and 2) and high valence(5). 

There is another significant negative interaction effect for Valence × Multitasking [True] (t(857) = -2.48, p = 0.013); see \autoref{fig:intEffect_sam_multitasking}. Hence, while having a side activity besides infinite scrolling, the \textit{responsiveness} increases with increasing valence. However, when participants have no side activity, the effect reverses, and the \textit{responsiveness} decreases with increasing valence. 

Another significant negative interaction effect was identified between being at home and valence (t(857) = -3.17, p = 0.002), as detailed in \autoref{fig:intEffect_sam_atHome}. This finding suggests that when participants are at home, their \textit{responsiveness} moderately decreases as valence increases. In contrast, when participants are not at home, there is a notable increase in \textit{responsiveness} corresponding with an increase in valence.
Further, the interaction between Sleepiness × Social Situation [Strangers] is statistically significant and negative (t(857) = -2.41, p = 0.016); see \autoref{fig:intEffect_kss_socialSit}. This interaction indicates that when alone, an increase in sleepiness leads to a moderate increase in \textit{responsiveness} to interventions. However, when in the presence of strangers, an increase in sleepiness results in a much stronger decrease in \textit{responsiveness}. It is important to note that during instances of extreme sleepiness while being with strangers, the intervention did not occur.
There was an interaction between being At Home [True] × Multitasking [True], which is significantly negative (t(857) = -2.67, p = 0.008; see \autoref{fig:intEffect_multitasking_atHome}). This result suggests that when users are at home, engaging in a side activity alongside infinite scrolling does not impact their \textit{responsiveness} much. In contrast, when users are not at home, multitasking alongside scrolling has a pronounced negative effect on their \textit{responsiveness}.


Lastly, we found a significant positive three-way interaction between At Home [True] x Valence x Multitasking [True] (t(857) = 2.77, p = 0.006).

\subsubsection{Model Comparison}\label{sec:model_comparision}
To identify whether interaction effects or main effects are better to explain the influences of the context factors, we conducted likelihood-ratio tests to compare the models, including main effects (n = 10 parameters), with the models including interaction effects (n= 70 parameters). 
On the one hand, for \textit{reactance}, the main effects model (Model 1) yielded an AIC of 2271.9 and log-likelihood of -1126, while the interaction effects model (Model 2) produced an AIC of 2308.1 and a log-likelihood of -1084. Comparing the two models indicated a significant improvement in fit with the inclusion of interactions (\chisq(60) = 83.848, p = .023). Hence, despite the increased complexity of the interaction model, the significant p-value suggests that the interactions between contextual factors provide an improvement in explaining the contextual influences for \textit{reactance}.
On the other hand, for the \textit{responsiveness}, the model with main effects (Model 3) produced an AIC of 3816.1 and a log-likelihood of -1898.1. In contrast, the interaction effects model (Model 4) resulted in an AIC of 3857.9 and a log-likelihood of -1858.9. The likelihood ratio test indicated a \chisq(60) = 78.265, p = .057. Thus, although the interaction model showed a lower deviance (3717.9) compared to the main effects model (3796.1), suggesting a better fit to the data, the increase in model complexity and the p-value slightly above the alpha level of .05 suggest that the improvement in fit may not justify the additional complexity introduced by the interaction terms.


\section{Discussion}
This work explored how contextual factors influence users' \textit{reactance} and \textit{responsiveness} towards an intervention during infinite scrolling on SoMe. 
We conducted a longitudinal user study for 7 days with N=\userstudyfinal participants, who installed our self-developed \textit{InfiniteScape}, a native Android application tracking their infinite scrolling behavior. Upon detecting continuous scrolling (e.g., in Instagram or TikTok) for more than 15 minutes, participants were prompted with an intervention overlay nudging them to stop scrolling. 
We gave participants the option to dismiss this intervention and continue scrolling. Once they stopped infinite scrolling, such as by closing the SoMe application, we asked participants about their \textit{reactance} to the intervention and their current context, including valence, social situation, current activity, being at home or not, multitasking behavior, and level of sleepiness. These six contextual factors were based on previous research~\cite{RixenIS.2023, Purohit.2019}. Additionally, we recorded the time span between the intervention and when participants stopped scrolling to measure their \textit{responsiveness}.
In this section, we will explore the implications of our findings, discuss how the identified contextual factors play a role in user behavior, and offer practical implications for designing context-aware interventions during infinite scrolling on SoMe.

\subsection{Contextual Influences}
Our analysis revealed only one significant main effect, while five significant interaction effects between the contextual factors were found. 
However, when comparing models (see \autoref{sec:model_comparision}), the addition of interaction effects showed only slight improvements over the models that considered main effects alone. In particular, the model for \textit{reactance} improved significantly with interaction effects, but the enhancement for the \textit{responsiveness} model was not statistically significant (p = .057). This observation implies that the interaction effects should be interpreted with caution. However, \citet{Jameson.2001} suggests the importance of incorporating multiple contextual factors to assess the user's context accurately. This indicates that context cannot be considered individually. The interplay of these factors points to a complex network of interrelated contextual influences, each intricately connected to and affecting the others. This complexity highlights the need to view context as an integrated system, with various components interacting to impact interventions' effectiveness during infinite scrolling.


\subsubsection{Bedtime Procrastination}
We observed a main effect that increased sleepiness led to a decrease in \textit{reactance} towards interventions during infinite scrolling. This implies that \textbf{people more likely accept interventions when tired}. Supporting this, \citet{Christensen.2016} found that phone usage at bedtime is linked to poor sleep quality, which might be an issue users are aware of, thus reducing their \textit{reactance} towards interventions. Interestingly, our results did not show that sleepiness led to a faster response to the intervention.
Instead, being alone--in a private context--actually increased the time users spent on infinite scrolling after the intervention. In contrast, being surrounded by strangers reduced this time. However, we did not record any data for situations where users were extremely sleepy and in the company of strangers. Therefore, we assume that high sleepiness levels are more likely in private contexts, such as in bed.
The increased \textit{responsiveness} when alone, coupled with the reduced \textit{reactance} due to sleepiness, suggests that users recognize the adverse effects of poor sleep quality and, therefore, accept the intervention when in bed. However, they do not necessarily react to it by stopping their scrolling behavior. We attribute this contradicting behavior to bedtime procrastination, a tendency to delay going to sleep in favor of more engaging activities such as watching TV~\cite{Kroese.2014}. Related to smartphone usage, it has been noted that ``\textit{individuals with smartphone addiction are inclined to postpone their bedtime}''~\cite[p. 1]{Geng.2021}. We infer that while people may be \textbf{aware of the negative effects of bedtime procrastination} and thus more receptive to interventions, they \textbf{still find it challenging to disengage} from infinite scrolling when tired. This suggests an internal conflict between awareness of habits and the difficulty in altering them, particularly in the context of infinite scrolling at bedtime.

\subsubsection{Infinite Scrolling as Coping Strategy for Negative Emotions}
Smartphone usage has been found to be a coping mechanism for negative emotions~\cite{Diefenbach.2019}. In particular, the consumption of SoMe is often used as a way to procrastinate on undesirable tasks~\cite{reinecke_slacking_2016}, providing a short-term mood boost~\cite{sirois_procrastination_2013}. However, this temporary relief often leads to negative feelings such as guilt or regret~\cite{hofmann_spoiled_2013, Cho.2017}. However, we could not find any main effect on \textit{reactance} or \textit{responsiveness} with regard to the participant's valence. Instead, the interaction with valence, being at home, and multitasking revealed more nuanced insights. Particularly, the decision to stop infinite scrolling after the intervention was influenced not only by the users' valence but also by whether they are at home and if they perform an activity alongside infinite scrolling. Our findings show that when users are at home, their response time to an intervention remains relatively long, regardless of other factors like multitasking (see \autoref{fig:intEffect_multitasking_atHome}) or their valence (see \autoref{fig:intEffect_sam_atHome}). In this context, \textbf{being at home seems to act as a stabilizing factor}, reducing the influence of other variables on responsiveness. 

In detail, we found that high valance does not alter \textit{responsiveness} to interventions when users are at home or elsewhere. In contrast, low valence tends to slow down the response time to interventions when users are at home (see \autoref{fig:intEffect_sam_atHome}). This suggests that \textbf{the familiar environment} of beginning at home may \textbf{not provide enough distractions from negative emotions}, leading users to ignore interventions and continue scrolling. Conversely, when users are in different settings, external stimuli offer more distractions from their negative emotions, resulting in a faster reaction to stop infinite scrolling after an intervention, as shown by the interaction effect between multitasking and being at home (see \autoref{fig:intEffect_multitasking_atHome}).
This observation aligns with the interaction effect between valence and multitasking (see \autoref{fig:intEffect_sam_atHome}). When users are engaged in \textbf{multitasking during moments of negative emotions}, they tend to \textbf{disengage in infinite scrolling faster} compared to when their focus is solely on scrolling. This effect can be explained by the Multiple Resource Theory~\cite{wickens_multiple_2002}, which posits that interference between tasks increases when they compete for the same cognitive resources, such as modality or type of attention (e.g., focal vs. ambient). When multitasking, activities that draw from overlapping resource pools increase cognitive demand. In this context, multitasking alongside infinite scrolling likely increases interference, compelling users to free cognitive capacity by responding to the intervention faster. Hence, performing a peripheral activity while infinite scrolling may demand sufficient shared resources to nudge users toward disengaging from infinite scrolling when prompted by an intervention.

\subsection{Does Context Truly Matter?}
Although we found multiple significant effects for certain contextual factors, the overall influence of context on the intervention's effectiveness appears limited. For instance, while some factors, such as sleepiness, were found to impact intervention effectiveness significantly, other contextual factors, like the current activity (whether the participant was engaged in leisure or working activities), did not occur in any significant main or interaction effects. This raises the question of the true importance of context in designing effective interventions for infinite scrolling.
Interestingly, prior research offers mixed insights into this question. The user study by \citet{mongeroffarello2019race} found that participants rarely used the personalization feature of interventions, suggesting that they did not perceive their context to be crucial in interventing their smartphone use. However, this contrasts with several studies that emphasize the importance of context in behavior change. For example, research argues that timely, context-aware interventions are more effective because they align with the user's immediate environment, mood, or task~\cite{Ding.2016, Pinder.2018}. Similarly, \citet{Purohit.2019} highlights the need for interventions to be aware of location, time, and social settings to optimize behavior change, particularly in digital well-being.
This disparity between our findings and existing research suggests that the role of context may be more nuanced than previously understood. It is possible that some contexts, such as sleepiness, directly influence the user’s valence. In contrast, other contextual factors, like current activity, may not have had a strong enough or immediate impact to show significant effects in this study. Another possibility is that the design of the intervention itself plays a role in how much context matters. For example, more immersive or intrusive interventions could override the need for context awareness by being effective regardless of those factors. Additionally, device-specific contexts, such as whether users scroll over old or new content in their feed, might affect intervention effectiveness~\cite{RixenIS.2023}. Despite this, our study provides statistical evidence that certain contextual factors—such as sleepiness, valence, being at home or not, and multitasking—significantly influence interventions' effectiveness.

\subsection{Practical Implications for Designing Context-Aware Interventions}
Although previous work indicated that digital interventions should be context-aware, they lacked empirical investigation. The findings of our study emphasize the nuanced and interconnected influences of contextual factors in shaping the effectiveness of interventions during infinite scrolling. This highlights the need for context-aware interventions that consider being at home, social situations, valance, multitasking, and sleepiness as the main factors of an integrated system. For example, the reduced \textit{reactance} observed during increased sleepiness suggests an opportunity for bedtime interventions to increase acceptance of it. Interventions such as promoting calming activities such as mindfulness prompts~\cite{terzimehic_implicit_2023} or journal writing~\cite{sakel_social_2024} might subtly encourage disengagement during bedtime. However, the lack of effect on \textit{responsiveness} suggests that a multi-step approach may be necessary, with gradual intensification of interventions during bedtime, which could help elicit faster responses without initially overwhelming the user.
Building on the recommendations of \citet{ruiz_design_2024}, integrating design friction interventions during infinite scrolling could be effective. Their study showed that requiring users to rate each post before accessing the next increased frustration and effectively reduced engagement. Adapting this approach to bedtime procrastination by progressively increasing interaction friction could strike a balance by maintaining low \textit{responsiveness} at the beginning and gradually provoking faster \textit{responsiveness} as the intervention intensifies.

We further found that being at home acts as a stabilizing factor, diminishing the impact of other variables like valence or multitasking on \textit{responsiveness} to interventions. When users are not at home, their response time varies depending on their valence or whether they are multitasking. However, when users are at home, their response time remains consistently high. This emphasizes the need to focus on tailoring interventions specifically for when users are at home, e.g. by synchronizing with smart home devices. While we could not find significant effects on \textit{reactance} associated with being at home, we suggest using more severe interventions when users are at home compared to interventions when they are elsewhere to enhance \textit{responsiveness}. \citet{Terzimehic.2022b} found that users often wished they had engaged in more meaningful activities, such as physical exercise or social interaction, instead of regretful smartphone use. Hence, interventions could build on this insight by promoting outdoor activities that align with these preferences, such as suggesting nearby parks, fitness classes, or social meetups. A similar approach was already taken by \citet{consolvo_goal-setting_2009} by setting goals to encourage physical activities.
By leveraging these insights, SoMe platforms can move beyond one-size-fits-all approaches to foster meaningful, sustainable changes in infinite scrolling behavior, aligning with their promise of reducing excessive screen time.


\subsection{Detecting Contextual Factors}
For the practical implications discussed earlier to be effective, it is crucial to detect the users' context while they engage in infinite scrolling. 
While detecting the users' location using GPS data to determine if they are at home or elsewhere is relatively straightforward, identifying other contextual factors presents a greater challenge. Factors such as the user's current activity, valence, social situation, or whether they are multitasking require more sophisticated approaches for detection.
However, recent advancements in sensing technology, particularly in machine learning, have significantly improved our ability to detect specific aspects of the users' context. For instance, \citet{Dawei.2023} demonstrated the use of smartphone recordings to detect face-to-face conversations, providing valuable information about the user's social situation. Further, \citet{Mandi.2023} developed a framework capable of assessing a user's valence and arousal through facial image analysis using a smartphone camera.
In addition, the detection of sleepiness~\cite{Huda.2020} and multitasking~\cite{Kashevnik.2021} has primarily been explored within the context of driving, utilizing eye-tracking technology. Transferring these methods to the domain of smartphone usage, particularly in the context of infinite scrolling, could offer novel ways to tailor context-aware interventions more effectively. While these technologies can provide valuable context data, they also raise privacy concerns. Thus, any implementation of context-aware interventions must prioritize user privacy and ensure that such technology respects individual boundaries and maintains ethical data handling.

While those contextual factors could be detected with current and future technology, our study required using Android's Accessibility Service to monitor infinite scrolling behavior. However, apps utilizing this service face restrictions on the Google Play Store, as they are not permitted for non-accessible purposes~\cite{GooglePlayStore.AccesabilityService}. This presents a challenge for the practical application of apps capable of tracking infinite scrolling and intervening in such behavior. Nonetheless, ensuring user privacy while effectively tracking digital behaviors is crucial. Future developments in this area must balance the technical capabilities for tracking infinite scrolling with privacy standards and marketplace regulations to make these tools available to a broader user base.

\subsection{Limitations and Future Work}
Looking ahead, future research should extend this research by investigating various interventions to determine the most effective ones for specific contexts. While our study employed a simple pop-up intervention adopted from current state-of-the-art interventions in SoMe applications (see \autoref{sec:apparatus}), it is plausible that alternative types of interventions may perform better or worse depending on the context. 

In reflecting on the limitations of our study, it is important to acknowledge certain aspects that could influence the interpretation of our findings. First, our participant pool was limited to Android users, which inherently excludes a substantial number of smartphone users, particularly those using iOS devices. This restriction potentially limits the diversity of our study sample and may impact the applicability of our findings across different technological platforms. Further, our study's 7-day duration may not capture the full scope of longer-term effects of contextual influences on infinite scrolling. While there are longer-term studies on general smartphone overuse (e.g., approximately 13 weeks~\cite{haliburton_longitudinal_2024}), future research should examine the extended impacts specifically related to infinite scrolling behavior.
Additionally, our approach to assessing participants' current context after the intervention relied on self-reporting, not objective detection~\cite{Dawei.2023, Mandi.2023, Huda.2020}. While our work gave first insights into the complexity of contextual influences, future work should take those objective detection approaches to investigate whether context detection matches the outcomes of our study. 
Another limitation is that, due to the event-based ESM, only contextual information was collected from participants, who eventually stopped infinite scrolling after the intervention occured. Therefore, we are missing data from those who continued scrolling and, therefore, ignored the interventions and did not answer the questionnaire. In this study, interventions were triggered after 15~minutes of continuous infinite scrolling. While a baseline condition, in which no intervention would be triggered, could have provided further insights into contextual factors on participants' unaffected reasons for stopping infinitive scrolling (such as already hinted by \citet{RixenIS.2023}), it was not included in the current study due to the primary focus on contextual factors on intervention effectiveness. Further, we only included participants who expressed regret during infinite scrolling, ensuring intrinsic motivation to engage with interventions, as supported by Self-Determination Theory~\cite{ryan_self-determination_2000}. However, individuals who unconsciously scroll without regret may require different interventions, such as increasing awareness or breaking habits through external triggers. Future work should address this group to broaden intervention applicability.

Our study examined specific contextual factors identified in prior research~\cite{RixenIS.2023, Weber.2020, Hintze.2017, Akpinar.2023, Purohit.2019}, but these represent only a subset of potential influences on user behavior. Future research could expand on this by exploring a broader range of contextual elements. This expansion could reveal additional layers of complexity in user behavior on interventions during infinite scrolling. Besides contextual factors, \citet{VandenAbeele.2021b} hints that the content consumed during infinite scrolling might also influence reactions towards an intervention (e.g., engaging content might cause higher \textit{reactance} than boring content). Hence, future work should look into the influence of the consumed content, e.g., via screenshots~\cite{Chen.2023, orzikulova_time2stop_2024}.

Concerning statistical power, our approach shows the inherent challenges in estimating power for LMMs~\cite{Kumle.2021}. Proper power analysis requires simulations based on data from prior studies, which may introduce variability in the estimated power depending on the prior study's sample. This lack of precise power calculation means that we cannot fully assess the risk of Type II errors - missing true effects due to insufficient sample size. As a result, there may be significant effects that we have not detected. Nevertheless, the fact that significant interactions were found is already an indication of sufficient power. Nevertheless, the low R² marginal values (see \autoref{table:lmm}) in the LMMs indicate that the variance in user responses explained by our models is subtle. This suggests that individual differences between users may have a more pronounced impact than the specific contextual factors identified. This insight is interesting for future research because it highlights the importance of personalization in intervention design, recognizing that individual user characteristics may play a key role in determining intervention effectiveness.

\section{Conclusion}
This paper explored the impact of contextual factors on the effectiveness of interventions during infinite scrolling on SoMe, defined by the \textit{reactance} and \textit{responsiveness} towards the intervention. To achieve this, we developed \textit{InfiniteScape}, designed to monitor users' infinite scrolling behaviors and present an overlay intervention after 15 minutes of continuous activity. Once participants stopped scrolling, a follow-up questionnaire captured their \textit{reactance} toward the intervention and their prevailing contextual factors. Furthermore, the duration between the intervention and the moment participants stopped scrolling was recorded as their \textit{responsiveness}. 

Our study spanned 7 days and involved N=72 participants who installed \textit{InfiniteScape}. The findings reveal that multiple contextual factors are interlinked, showing that they should not be considered in isolation. In particular, we found interaction effects on the \textit{responsiveness} for the users' valence with their social situation, whether they are at home or elsewhere, and whether they were multitasking while scrolling. Specifically, low valence combined with being at home tended to slow users' \textit{responsiveness} to interventions, whereas multitasking during low valence resulted in users responding to the intervention more quickly. Further, we observed a main effect indicating that increased sleepiness reduces users' \textit{reactance} towards interventions, suggesting that users are more likely to accept an intervention when they feel tired.
These findings underscore the complexity of intervention effectiveness and emphasize the need to design context-aware strategies to mitigate excessive infinite scrolling on SoMe platforms. Examining how these different contextual aspects interact together within an overall system is important.

Our research contributes to our understanding of how contextual factors impact the effectiveness of digital interventions and provides evidence to support the development of more effective, context-aware interventions to address infinite scrolling.

\section*{Open Science}
The source code of the native Android application \textit{InfiniteScape}, and the RScript for analysis are available under the following link: \\
\url{https://github.com/luca-maxim/scrollingInTheDeep}. \\
The study data is provided in an anonymized format. Hence, to ensure privacy, we replaced each participant's Prolific ID with an unique sequential Participant ID.

\begin{acks}
This research was conducted in the context of the DFG project \textit{"Beyond Screen Time: Context- and Content-tailored Interventions to Social Media Usage to Enhance Digital Well-being"}.

We would like to thank Johannes Schöning for his valuable mental support during the \textit{CHITogether 2024~\cite{scott2024doing}} in St. Gallen.
\end{acks}

\bibliographystyle{ACM-Reference-Format}
\bibliography{sample-base}


\begin{thebibliography}{108}


\ifx \showCODEN    \undefined \def \showCODEN     #1{\unskip}     \fi
\ifx \showISBNx    \undefined \def \showISBNx     #1{\unskip}     \fi
\ifx \showISBNxiii \undefined \def \showISBNxiii  #1{\unskip}     \fi
\ifx \showISSN     \undefined \def \showISSN      #1{\unskip}     \fi
\ifx \showLCCN     \undefined \def \showLCCN      #1{\unskip}     \fi
\ifx \shownote     \undefined \def \shownote      #1{#1}          \fi
\ifx \showarticletitle \undefined \def \showarticletitle #1{#1}   \fi
\ifx \showURL      \undefined \def \showURL       {\relax}        \fi
\providecommand\bibfield[2]{#2}
\providecommand\bibinfo[2]{#2}
\providecommand\natexlab[1]{#1}
\providecommand\showeprint[2][]{arXiv:#2}

\bibitem[Akpinar et~al\mbox{.}(2023)]%
        {Akpinar.2023}
\bibfield{author}{\bibinfo{person}{Elgin Akpinar}, \bibinfo{person}{Yeliz Ye{\c{s}}ilada}, {and} \bibinfo{person}{Pinar Karag{\"o}z}.} \bibinfo{year}{2023}\natexlab{}.
\newblock \showarticletitle{Effect of Context on Smartphone Users' Typing Performance in the Wild}.
\newblock \bibinfo{journal}{\emph{ACM Transactions on Computer-Human Interaction}} \bibinfo{volume}{30}, \bibinfo{number}{3} (\bibinfo{year}{2023}), \bibinfo{pages}{1--44}.
\newblock
\showISSN{1073-0516}
\href{https://doi.org/10.1145/3577013}{doi:\nolinkurl{10.1145/3577013}}


\bibitem[Android(2023)]%
        {Android.Wellbeing}
\bibfield{author}{\bibinfo{person}{Android}.} \bibinfo{year}{2023}\natexlab{}.
\newblock \bibinfo{booktitle}{\emph{Digital wellbeing | Android}}.
\newblock
\urldef\tempurl%
\url{https://www.android.com/digital-wellbeing/}
\showURL{%
Retrieved April 8, 2024 from \tempurl}


\bibitem[Apple(2023)]%
        {iOS.Wellbeing}
\bibfield{author}{\bibinfo{person}{Apple}.} \bibinfo{year}{2023}\natexlab{}.
\newblock \bibinfo{booktitle}{\emph{Use Screen Time on your iPhone, iPad, or iPod touch}}.
\newblock
\urldef\tempurl%
\url{https://support.apple.com/en-us/HT208982}
\showURL{%
Retrieved April 8, 2024 from \tempurl}


\bibitem[Arnau et~al\mbox{.}(2012)]%
        {Arnau.2012}
\bibfield{author}{\bibinfo{person}{Jaume Arnau}, \bibinfo{person}{Roser Bono}, \bibinfo{person}{Mar{\'i}a~J. Blanca}, {and} \bibinfo{person}{Rebecca Bendayan}.} \bibinfo{year}{2012}\natexlab{}.
\newblock \showarticletitle{Using the linear mixed model to analyze nonnormal data distributions in longitudinal designs}.
\newblock \bibinfo{journal}{\emph{Behavior research methods}} \bibinfo{volume}{44}, \bibinfo{number}{4} (\bibinfo{year}{2012}), \bibinfo{pages}{1224--1238}.
\newblock
\href{https://doi.org/10.3758/s13428-012-0196-y}{doi:\nolinkurl{10.3758/s13428-012-0196-y}}


\bibitem[Barnard et~al\mbox{.}(2007)]%
        {Barnard.2007}
\bibfield{author}{\bibinfo{person}{Leon Barnard}, \bibinfo{person}{Ji~Soo Yi}, \bibinfo{person}{Julie~A. Jacko}, {and} \bibinfo{person}{Andrew Sears}.} \bibinfo{year}{2007}\natexlab{}.
\newblock \showarticletitle{Capturing the effects of context on human performance in mobile computing systems}.
\newblock \bibinfo{journal}{\emph{Personal and Ubiquitous Computing}} \bibinfo{volume}{11}, \bibinfo{number}{2} (\bibinfo{year}{2007}), \bibinfo{pages}{81--96}.
\newblock
\showISSN{1617-4909}
\href{https://doi.org/10.1007/s00779-006-0063-x}{doi:\nolinkurl{10.1007/s00779-006-0063-x}}


\bibitem[Bashir and Bhat(2017)]%
        {bashir2017effects}
\bibfield{author}{\bibinfo{person}{Hilal Bashir} {and} \bibinfo{person}{Shabir~Ahmad Bhat}.} \bibinfo{year}{2017}\natexlab{}.
\newblock \showarticletitle{Effects of social media on mental health: A review}.
\newblock \bibinfo{journal}{\emph{International Journal of Indian Psychology}} \bibinfo{volume}{4}, \bibinfo{number}{3} (\bibinfo{year}{2017}), \bibinfo{pages}{125--131}.
\newblock
\href{https://doi.org/10.25215/0403.134}{doi:\nolinkurl{10.25215/0403.134}}


\bibitem[Baughan et~al\mbox{.}(2022)]%
        {baughan_i_2022}
\bibfield{author}{\bibinfo{person}{Amanda Baughan}, \bibinfo{person}{Mingrui~Ray Zhang}, \bibinfo{person}{Raveena Rao}, \bibinfo{person}{Kai Lukoff}, \bibinfo{person}{Anastasia Schaadhardt}, \bibinfo{person}{Lisa~D. Butler}, {and} \bibinfo{person}{Alexis Hiniker}.} \bibinfo{year}{2022}\natexlab{}.
\newblock \showarticletitle{“{I} {Don}’t {Even} {Remember} {What} {I} {Read}”: {How} {Design} {Influences} {Dissociation} on {Social} {Media}}. In \bibinfo{booktitle}{\emph{{CHI} {Conference} on {Human} {Factors} in {Computing} {Systems}}}. \bibinfo{publisher}{ACM}, \bibinfo{address}{New Orleans LA USA}, \bibinfo{pages}{1--13}.
\newblock
\showISBNx{978-1-4503-9157-3}
\href{https://doi.org/10.1145/3491102.3501899}{doi:\nolinkurl{10.1145/3491102.3501899}}


\bibitem[Bayer et~al\mbox{.}(2018)]%
        {Bayer.2018}
\bibfield{author}{\bibinfo{person}{Joseph Bayer}, \bibinfo{person}{Nicole Ellison}, \bibinfo{person}{Sarita Schoenebeck}, \bibinfo{person}{Erin Brady}, {and} \bibinfo{person}{Emily~B. Falk}.} \bibinfo{year}{2018}\natexlab{}.
\newblock \showarticletitle{Facebook in context(s): Measuring emotional responses across time and space}.
\newblock \bibinfo{journal}{\emph{New Media {\&} Society}} \bibinfo{volume}{20}, \bibinfo{number}{3} (\bibinfo{year}{2018}), \bibinfo{pages}{1047--1067}.
\newblock
\showISSN{1461-4448}
\href{https://doi.org/10.1177/1461444816681522}{doi:\nolinkurl{10.1177/1461444816681522}}


\bibitem[Beyens et~al\mbox{.}(2020)]%
        {beyens2020effect}
\bibfield{author}{\bibinfo{person}{Ine Beyens}, \bibinfo{person}{J~Loes Pouwels}, \bibinfo{person}{Irene~I van Driel}, \bibinfo{person}{Loes Keijsers}, {and} \bibinfo{person}{Patti~M Valkenburg}.} \bibinfo{year}{2020}\natexlab{}.
\newblock \showarticletitle{The effect of social media on well-being differs from adolescent to adolescent}.
\newblock \bibinfo{journal}{\emph{Scientific Reports}} \bibinfo{volume}{10}, \bibinfo{number}{1} (\bibinfo{year}{2020}), \bibinfo{pages}{1--11}.
\newblock
\href{https://doi.org/10.1038/s41598-020-67727-7}{doi:\nolinkurl{10.1038/s41598-020-67727-7}}


\bibitem[Bradley and Lang(1994)]%
        {Bradley.1994}
\bibfield{author}{\bibinfo{person}{M.~M. Bradley} {and} \bibinfo{person}{P.~J. Lang}.} \bibinfo{year}{1994}\natexlab{}.
\newblock \showarticletitle{Measuring emotion: the Self-Assessment Manikin and the Semantic Differential}.
\newblock \bibinfo{journal}{\emph{Journal of behavior therapy and experimental psychiatry}} \bibinfo{volume}{25}, \bibinfo{number}{1} (\bibinfo{year}{1994}), \bibinfo{pages}{49--59}.
\newblock
\showISSN{0005-7916}
\href{https://doi.org/10.1016/0005-7916(94)90063-9}{doi:\nolinkurl{10.1016/0005-7916(94)90063-9}}


\bibitem[Chang et~al\mbox{.}(2015)]%
        {Chang.2015}
\bibfield{author}{\bibinfo{person}{Yung-Ju Chang}, \bibinfo{person}{Gaurav Paruthi}, {and} \bibinfo{person}{Mark~W. Newman}.} \bibinfo{year}{2015}\natexlab{}.
\newblock \showarticletitle{A Field Study Comparing Approaches to Collecting Annotated Activity Data in Real-World Settings}. In \bibinfo{booktitle}{\emph{Proceedings of the 2015 ACM International Joint Conference on Pervasive and Ubiquitous Computing}} (Osaka, Japan) \emph{(\bibinfo{series}{UbiComp '15})}. \bibinfo{publisher}{Association for Computing Machinery}, \bibinfo{address}{New York, NY, USA}, \bibinfo{pages}{671–682}.
\newblock
\showISBNx{9781450335744}
\href{https://doi.org/10.1145/2750858.2807524}{doi:\nolinkurl{10.1145/2750858.2807524}}


\bibitem[Chen et~al\mbox{.}(2023)]%
        {Chen.2023}
\bibfield{author}{\bibinfo{person}{Yu-Chun Chen}, \bibinfo{person}{Yu-Jen Lee}, \bibinfo{person}{Kuei-Chun Kao}, \bibinfo{person}{Jie Tsai}, \bibinfo{person}{En-Chi Liang}, \bibinfo{person}{Wei-Chen Chiu}, \bibinfo{person}{Faye Shih}, {and} \bibinfo{person}{Yung-Ju Chang}.} \bibinfo{year}{2023}\natexlab{}.
\newblock \showarticletitle{Are You Killing Time? Predicting Smartphone Users' Time-killing Moments via Fusion of Smartphone Sensor Data and Screenshots}. In \bibinfo{booktitle}{\emph{CHI}}, \bibfield{editor}{\bibinfo{person}{Albrecht Schmidt}, \bibinfo{person}{Kaisa V{\"a}{\"a}n{\"a}nen}, \bibinfo{person}{Tesh Goyal}, \bibinfo{person}{Per~Ola Kristensson}, \bibinfo{person}{Anicia Peters}, \bibinfo{person}{Stefanie Mueller}, \bibinfo{person}{Julie~R. Williamson}, {and} \bibinfo{person}{Max~L. Wilson}} (Eds.). \bibinfo{publisher}{ACM}, \bibinfo{address}{New York, NY, USA}, \bibinfo{pages}{1--19}.
\newblock
\showISBNx{9781450394215}
\href{https://doi.org/10.1145/3544548.3580689}{doi:\nolinkurl{10.1145/3544548.3580689}}


\bibitem[Cho et~al\mbox{.}(2021)]%
        {Cho.2021}
\bibfield{author}{\bibinfo{person}{Hyunsung Cho}, \bibinfo{person}{DaEun Choi}, \bibinfo{person}{Donghwi Kim}, \bibinfo{person}{Wan~Ju Kang}, \bibinfo{person}{Eun~Kyoung Choe}, {and} \bibinfo{person}{Sung-Ju Lee}.} \bibinfo{year}{2021}\natexlab{}.
\newblock \showarticletitle{Reflect, not Regret: Understanding Regretful Smartphone Use with App Feature-Level Analysis}.
\newblock \bibinfo{journal}{\emph{Proceedings of the ACM on Human-Computer Interaction}} \bibinfo{volume}{5}, \bibinfo{number}{CSCW2} (\bibinfo{year}{2021}), \bibinfo{pages}{1--36}.
\newblock
\href{https://doi.org/10.1145/3479600}{doi:\nolinkurl{10.1145/3479600}}


\bibitem[Cho and Saakes(2017)]%
        {Cho.2017}
\bibfield{author}{\bibinfo{person}{Minjoo Cho} {and} \bibinfo{person}{Daniel Saakes}.} \bibinfo{year}{2017}\natexlab{}.
\newblock \showarticletitle{Calm Automaton}. In \bibinfo{booktitle}{\emph{Proceedings of the 2017 CHI Conference Extended Abstracts on Human Factors in Computing Systems}}, \bibfield{editor}{\bibinfo{person}{Gloria Mark}, \bibinfo{person}{Susan Fussell}, \bibinfo{person}{Cliff Lampe}, \bibinfo{person}{m.c. schraefel}, \bibinfo{person}{Juan~Pablo Hourcade}, \bibinfo{person}{Caroline Appert}, {and} \bibinfo{person}{Daniel Wigdor}} (Eds.). \bibinfo{publisher}{ACM}, \bibinfo{address}{New York, NY, USA}, \bibinfo{pages}{393--396}.
\newblock
\showISBNx{9781450346566}
\href{https://doi.org/10.1145/3027063.3052968}{doi:\nolinkurl{10.1145/3027063.3052968}}


\bibitem[Christensen et~al\mbox{.}(2016)]%
        {Christensen.2016}
\bibfield{author}{\bibinfo{person}{Matthew~A. Christensen}, \bibinfo{person}{Laura Bettencourt}, \bibinfo{person}{Leanne Kaye}, \bibinfo{person}{Sai~T. Moturu}, \bibinfo{person}{Kaylin~T. Nguyen}, \bibinfo{person}{Jeffrey~E. Olgin}, \bibinfo{person}{Mark~J. Pletcher}, {and} \bibinfo{person}{Gregory~M. Marcus}.} \bibinfo{year}{2016}\natexlab{}.
\newblock \showarticletitle{Direct Measurements of Smartphone Screen-Time: Relationships with Demographics and Sleep}.
\newblock \bibinfo{journal}{\emph{PloS one}} \bibinfo{volume}{11}, \bibinfo{number}{11} (\bibinfo{year}{2016}), \bibinfo{pages}{e0165331}.
\newblock
\href{https://doi.org/10.1371/journal.pone.0165331}{doi:\nolinkurl{10.1371/journal.pone.0165331}}


\bibitem[Consolvo et~al\mbox{.}(2009)]%
        {consolvo_goal-setting_2009}
\bibfield{author}{\bibinfo{person}{Sunny Consolvo}, \bibinfo{person}{Predrag Klasnja}, \bibinfo{person}{David~W. McDonald}, {and} \bibinfo{person}{James~A. Landay}.} \bibinfo{year}{2009}\natexlab{}.
\newblock \showarticletitle{Goal-setting considerations for persuasive technologies that encourage physical activity}. In \bibinfo{booktitle}{\emph{Proceedings of the 4th {International} {Conference} on {Persuasive} {Technology}}}. \bibinfo{publisher}{ACM}, \bibinfo{address}{Claremont California USA}, \bibinfo{pages}{1--8}.
\newblock
\showISBNx{978-1-60558-376-1}
\href{https://doi.org/10.1145/1541948.1541960}{doi:\nolinkurl{10.1145/1541948.1541960}}


\bibitem[Dejonckheere et~al\mbox{.}(2022)]%
        {Dejonckheere.2022}
\bibfield{author}{\bibinfo{person}{Egon Dejonckheere}, \bibinfo{person}{Febe Demeyer}, \bibinfo{person}{Birte Geusens}, \bibinfo{person}{Maarten Piot}, \bibinfo{person}{Francis Tuerlinckx}, \bibinfo{person}{Stijn Verdonck}, {and} \bibinfo{person}{Merijn Mestdagh}.} \bibinfo{year}{2022}\natexlab{}.
\newblock \showarticletitle{Assessing the reliability of single-item momentary affective measurements in experience sampling}.
\newblock \bibinfo{journal}{\emph{Psychological assessment}} \bibinfo{volume}{34}, \bibinfo{number}{12} (\bibinfo{year}{2022}), \bibinfo{pages}{1138--1154}.
\newblock
\href{https://doi.org/10.1037/pas0001178}{doi:\nolinkurl{10.1037/pas0001178}}


\bibitem[Delaney and Lades(2017)]%
        {Delaney.2017}
\bibfield{author}{\bibinfo{person}{Liam Delaney} {and} \bibinfo{person}{Leonhard~K. Lades}.} \bibinfo{year}{2017}\natexlab{}.
\newblock \showarticletitle{Present Bias and Everyday Self-Control Failures: A Day Reconstruction Study}.
\newblock \bibinfo{journal}{\emph{Journal of Behavioral Decision Making}} \bibinfo{volume}{30}, \bibinfo{number}{5} (\bibinfo{year}{2017}), \bibinfo{pages}{1157--1167}.
\newblock
\showISSN{08943257}
\href{https://doi.org/10.1002/bdm.2031}{doi:\nolinkurl{10.1002/bdm.2031}}


\bibitem[Diefenbach and Borrmann(2019)]%
        {Diefenbach.2019}
\bibfield{author}{\bibinfo{person}{Sarah Diefenbach} {and} \bibinfo{person}{Kim Borrmann}.} \bibinfo{year}{2019}\natexlab{}.
\newblock \showarticletitle{The Smartphone as a Pacifier and Its Consequences: Young Adults' Smartphone Usage in Moments of Solitude and Correlations to Self-Reflection}. In \bibinfo{booktitle}{\emph{Proceedings of the 2019 CHI Conference on Human Factors in Computing Systems}} (Glasgow, Scotland Uk) \emph{(\bibinfo{series}{CHI '19})}. \bibinfo{publisher}{Association for Computing Machinery}, \bibinfo{address}{New York, NY, USA}, \bibinfo{pages}{1–14}.
\newblock
\showISBNx{9781450359702}
\href{https://doi.org/10.1145/3290605.3300536}{doi:\nolinkurl{10.1145/3290605.3300536}}


\bibitem[Dillard and Shen(2005)]%
        {Dillard.2005}
\bibfield{author}{\bibinfo{person}{James~Price Dillard} {and} \bibinfo{person}{Lijiang Shen}.} \bibinfo{year}{2005}\natexlab{}.
\newblock \showarticletitle{On the Nature of Reactance and its Role in Persuasive Health Communication}.
\newblock \bibinfo{journal}{\emph{Communication Monographs}} \bibinfo{volume}{72}, \bibinfo{number}{2} (\bibinfo{year}{2005}), \bibinfo{pages}{144--168}.
\newblock
\showISSN{0363-7751}
\href{https://doi.org/10.1080/03637750500111815}{doi:\nolinkurl{10.1080/03637750500111815}}


\bibitem[Ding et~al\mbox{.}(2016)]%
        {Ding.2016}
\bibfield{author}{\bibinfo{person}{Xiang Ding}, \bibinfo{person}{Jing Xu}, \bibinfo{person}{Honghao Wang}, \bibinfo{person}{Guanling Chen}, \bibinfo{person}{Herpreet Thind}, {and} \bibinfo{person}{Yuan Zhang}.} \bibinfo{year}{2016}\natexlab{}.
\newblock \showarticletitle{WalkMore: promoting walking with just-in-time context-aware prompts}. In \bibinfo{booktitle}{\emph{2016 IEEE Wireless Health (WH)}}. \bibinfo{pages}{1--8}.
\newblock
\href{https://doi.org/10.1109/WH.2016.7764558}{doi:\nolinkurl{10.1109/WH.2016.7764558}}


\bibitem[Draper and Smith(1998)]%
        {draper1998applied}
\bibfield{author}{\bibinfo{person}{Norman~R Draper} {and} \bibinfo{person}{Harry Smith}.} \bibinfo{year}{1998}\natexlab{}.
\newblock \bibinfo{booktitle}{\emph{Applied regression analysis}}. Vol.~\bibinfo{volume}{326}.
\newblock \bibinfo{publisher}{John Wiley \& Sons}.
\newblock
\href{https://doi.org/10.1002/9781118625590}{doi:\nolinkurl{10.1002/9781118625590}}


\bibitem[Ehrenbrink(2020)]%
        {Ehrenbrink.2020}
\bibfield{author}{\bibinfo{person}{Patrick Ehrenbrink}.} \bibinfo{year}{2020}\natexlab{}.
\newblock \showarticletitle{Reactance Scale for Human--Computer Interaction}.
\newblock In \bibinfo{booktitle}{\emph{The Role of Psychological Reactance in Human--Computer Interaction}}, \bibfield{editor}{\bibinfo{person}{Patrick Ehrenbrink}} (Ed.). \bibinfo{publisher}{{Springer International Publishing}}, \bibinfo{address}{Cham}, \bibinfo{pages}{71--81}.
\newblock
\showISBNx{978-3-030-30309-9}
\href{https://doi.org/10.1007/978-3-030-30310-5 {\_}8}{doi:\nolinkurl{10.1007/978-3-030-30310-5 {\_}8}}


\bibitem[{European Commission}(2022)]%
        {DSA}
\bibfield{author}{\bibinfo{person}{{European Commission}}.} \bibinfo{year}{2022}\natexlab{}.
\newblock \bibinfo{booktitle}{\emph{Regulation (EU) 2022/2065 of the European Parliament and of the Council of 19 October 2022 on a Single Market For Digital Services and amending Directive 2000/31/EC (Digital Services Act)}}.
\newblock
\urldef\tempurl%
\url{http://data.europa.eu/eli/reg/2022/2065/oj}
\showURL{%
\tempurl}


\bibitem[{European Commission}(2024)]%
        {ECagainstTiktok}
\bibfield{author}{\bibinfo{person}{{European Commission}}.} \bibinfo{year}{2024}\natexlab{}.
\newblock \bibinfo{booktitle}{\emph{Commission opens formal proceedings against TikTok under the Digital Services Act}}.
\newblock
\urldef\tempurl%
\url{https://ec.europa.eu/commission/presscorner/detail/en/IP_24_926}
\showURL{%
\tempurl}
\newblock
\shownote{, 19 February 2024}.


\bibitem[Forgays et~al\mbox{.}(2014)]%
        {FORGAYS2014314}
\bibfield{author}{\bibinfo{person}{Deborah~Kirby Forgays}, \bibinfo{person}{Ira Hyman}, {and} \bibinfo{person}{Jessie Schreiber}.} \bibinfo{year}{2014}\natexlab{}.
\newblock \showarticletitle{Texting everywhere for everything: Gender and age differences in cell phone etiquette and use}.
\newblock \bibinfo{journal}{\emph{Computers in Human Behavior}}  \bibinfo{volume}{31} (\bibinfo{year}{2014}), \bibinfo{pages}{314--321}.
\newblock
\showISSN{0747-5632}
\href{https://doi.org/10.1016/j.chb.2013.10.053}{doi:\nolinkurl{10.1016/j.chb.2013.10.053}}


\bibitem[Frijda(2001)]%
        {Frijda1986-FRITE}
\bibfield{author}{\bibinfo{person}{Nico~H. Frijda}.} \bibinfo{year}{2001}\natexlab{}.
\newblock \bibinfo{booktitle}{\emph{The emotions}}.
\newblock \bibinfo{publisher}{{Cambridge Univ. Press}}, \bibinfo{address}{Cambridge}.
\newblock
\showISBNx{9780521316002}
\newblock
\shownote{{ISBN:} 978-05-213-1600-2}.


\bibitem[Frison and Eggermont(2020)]%
        {Frison.2020}
\bibfield{author}{\bibinfo{person}{Eline Frison} {and} \bibinfo{person}{Steven Eggermont}.} \bibinfo{year}{2020}\natexlab{}.
\newblock \showarticletitle{Toward an Integrated and Differential Approach to the Relationships Between Loneliness, Different Types of Facebook Use, and Adolescents' Depressed Mood}.
\newblock \bibinfo{journal}{\emph{Communication Research}} \bibinfo{volume}{47}, \bibinfo{number}{5} (\bibinfo{year}{2020}), \bibinfo{pages}{701--728}.
\newblock
\showISSN{0093-6502}
\href{https://doi.org/10.1177/0093650215617506}{doi:\nolinkurl{10.1177/0093650215617506}}


\bibitem[Geng et~al\mbox{.}(2021)]%
        {Geng.2021}
\bibfield{author}{\bibinfo{person}{Yaoguo Geng}, \bibinfo{person}{Jingjing Gu}, \bibinfo{person}{Jing Wang}, {and} \bibinfo{person}{Ruiping Zhang}.} \bibinfo{year}{2021}\natexlab{}.
\newblock \showarticletitle{Smartphone addiction and depression, anxiety: The role of bedtime procrastination and self-control}.
\newblock \bibinfo{journal}{\emph{Journal of affective disorders}}  \bibinfo{volume}{293} (\bibinfo{year}{2021}), \bibinfo{pages}{415--421}.
\newblock
\href{https://doi.org/10.1016/j.jad.2021.06.062}{doi:\nolinkurl{10.1016/j.jad.2021.06.062}}


\bibitem[{Google Play Store}({[n.\,d.]})]%
        {GooglePlayStore.AccesabilityService}
\bibfield{author}{\bibinfo{person}{{Google Play Store}}.} \bibinfo{year}{[n.\,d.]}\natexlab{}.
\newblock \bibinfo{booktitle}{\emph{Use of the AccessibilityService API}}.
\newblock
\urldef\tempurl%
\url{https://support.google.com/googleplay/android-developer/answer/10964491?hl=en}
\showURL{%
Retrieved Aug 10, 2024 from \tempurl}


\bibitem[Gray et~al\mbox{.}(2018)]%
        {Gray.2018}
\bibfield{author}{\bibinfo{person}{Colin~M. Gray}, \bibinfo{person}{Yubo Kou}, \bibinfo{person}{Bryan Battles}, \bibinfo{person}{Joseph Hoggatt}, {and} \bibinfo{person}{Austin~L. Toombs}.} \bibinfo{year}{2018}\natexlab{}.
\newblock \showarticletitle{The Dark (Patterns) Side of UX Design}. In \bibinfo{booktitle}{\emph{Proceedings of the 2018 CHI Conference on Human Factors in Computing Systems}}, \bibfield{editor}{\bibinfo{person}{Regan Mandryk}, \bibinfo{person}{Mark Hancock}, \bibinfo{person}{Mark Perry}, {and} \bibinfo{person}{Anna Cox}} (Eds.). \bibinfo{publisher}{ACM}, \bibinfo{address}{New York, NY, USA}, \bibinfo{pages}{1--14}.
\newblock
\showISBNx{9781450356206}
\href{https://doi.org/10.1145/3173574.3174108}{doi:\nolinkurl{10.1145/3173574.3174108}}


\bibitem[Haliburton et~al\mbox{.}(2024)]%
        {haliburton_longitudinal_2024}
\bibfield{author}{\bibinfo{person}{Luke Haliburton}, \bibinfo{person}{David~Joachim Grüning}, \bibinfo{person}{Frederik Riedel}, \bibinfo{person}{Albrecht Schmidt}, {and} \bibinfo{person}{Nađa Terzimehić}.} \bibinfo{year}{2024}\natexlab{}.
\newblock \showarticletitle{A {Longitudinal} {In}-the-{Wild} {Investigation} of {Design} {Frictions} to {Prevent} {Smartphone} {Overuse}}. In \bibinfo{booktitle}{\emph{Proceedings of the {CHI} {Conference} on {Human} {Factors} in {Computing} {Systems}}}. \bibinfo{publisher}{ACM}, \bibinfo{address}{Honolulu HI USA}, \bibinfo{pages}{1--16}.
\newblock
\showISBNx{9798400703300}
\href{https://doi.org/10.1145/3613904.3642370}{doi:\nolinkurl{10.1145/3613904.3642370}}


\bibitem[Hiniker et~al\mbox{.}(2016)]%
        {hiniker2016mytime}
\bibfield{author}{\bibinfo{person}{Alexis Hiniker}, \bibinfo{person}{Sungsoo Hong}, \bibinfo{person}{Tadayoshi Kohno}, {and} \bibinfo{person}{Julie~A. Kientz}.} \bibinfo{year}{2016}\natexlab{}.
\newblock \showarticletitle{MyTime: Designing and Evaluating an Intervention for Smartphone Non-Use}. In \bibinfo{booktitle}{\emph{Proceedings of the 2016 CHI Conference on Human Factors in Computing Systems}} \emph{(\bibinfo{series}{CHI '16})}. \bibinfo{publisher}{{Association for Computing Machinery}}, \bibinfo{address}{New York, NY, USA}, \bibinfo{pages}{4746--4757}.
\newblock
\showISBNx{9781450333627}
\href{https://doi.org/10.1145/2858036.2858403}{doi:\nolinkurl{10.1145/2858036.2858403}}


\bibitem[Hintze et~al\mbox{.}(2017)]%
        {Hintze.2017}
\bibfield{author}{\bibinfo{person}{Daniel Hintze}, \bibinfo{person}{Philipp Hintze}, \bibinfo{person}{Rainhard~D. Findling}, {and} \bibinfo{person}{Ren\'{e} Mayrhofer}.} \bibinfo{year}{2017}\natexlab{}.
\newblock \showarticletitle{A Large-Scale, Long-Term Analysis of Mobile Device Usage Characteristics}.
\newblock \bibinfo{journal}{\emph{Proc. ACM Interact. Mob. Wearable Ubiquitous Technol.}} \bibinfo{volume}{1}, \bibinfo{number}{2}, Article \bibinfo{articleno}{13} (\bibinfo{date}{jun} \bibinfo{year}{2017}), \bibinfo{numpages}{21}~pages.
\newblock
\href{https://doi.org/10.1145/3090078}{doi:\nolinkurl{10.1145/3090078}}


\bibitem[Hofmann et~al\mbox{.}(2013)]%
        {hofmann_spoiled_2013}
\bibfield{author}{\bibinfo{person}{Wilhelm Hofmann}, \bibinfo{person}{Hiroki Kotabe}, {and} \bibinfo{person}{Maike Luhmann}.} \bibinfo{year}{2013}\natexlab{}.
\newblock \showarticletitle{The spoiled pleasure of giving in to temptation}.
\newblock \bibinfo{journal}{\emph{Motivation and Emotion}} \bibinfo{volume}{37}, \bibinfo{number}{4} (\bibinfo{date}{Dec.} \bibinfo{year}{2013}), \bibinfo{pages}{733--742}.
\newblock
\showISSN{0146-7239, 1573-6644}
\href{https://doi.org/10.1007/s11031-013-9355-4}{doi:\nolinkurl{10.1007/s11031-013-9355-4}}


\bibitem[Huda et~al\mbox{.}(2020)]%
        {Huda.2020}
\bibfield{author}{\bibinfo{person}{Choirul Huda}, \bibinfo{person}{Herman Tolle}, {and} \bibinfo{person}{Fitri Utaminingrum}.} \bibinfo{year}{2020}\natexlab{}.
\newblock \showarticletitle{Mobile-Based Driver Sleepiness Detection Using Facial Landmarks and Analysis of EAR Values}.
\newblock \bibinfo{journal}{\emph{International Journal of Interactive Mobile Technologies (iJIM)}} \bibinfo{volume}{14}, \bibinfo{number}{14} (\bibinfo{year}{2020}), \bibinfo{pages}{16}.
\newblock
\href{https://doi.org/10.3991/ijim.v14i14.14105}{doi:\nolinkurl{10.3991/ijim.v14i14.14105}}


\bibitem[Jameson(2001)]%
        {Jameson.2001}
\bibfield{author}{\bibinfo{person}{Anthony Jameson}.} \bibinfo{year}{2001}\natexlab{}.
\newblock \showarticletitle{Modelling both the Context and the User}.
\newblock \bibinfo{journal}{\emph{Personal and Ubiquitous Computing}} \bibinfo{volume}{5}, \bibinfo{number}{1} (\bibinfo{year}{2001}), \bibinfo{pages}{29--33}.
\newblock
\showISSN{1617-4909}
\href{https://doi.org/10.1007/s007790170025}{doi:\nolinkurl{10.1007/s007790170025}}


\bibitem[Karppinen et~al\mbox{.}(2018)]%
        {karppinen2018opportunities}
\bibfield{author}{\bibinfo{person}{Pasi Karppinen}, \bibinfo{person}{Harri Oinas-Kukkonen}, \bibinfo{person}{Tuomas Alah{\"a}iv{\"a}l{\"a}}, \bibinfo{person}{Terhi Jokelainen}, \bibinfo{person}{Anna-Maria Teeriniemi}, \bibinfo{person}{Tuire Salonurmi}, {and} \bibinfo{person}{Markku~J Savolainen}.} \bibinfo{year}{2018}\natexlab{}.
\newblock \showarticletitle{Opportunities and challenges of behavior change support systems for enhancing habit formation: A qualitative study}.
\newblock \bibinfo{journal}{\emph{Journal of biomedical informatics}}  \bibinfo{volume}{84} (\bibinfo{year}{2018}), \bibinfo{pages}{82--92}.
\newblock
\href{https://doi.org/10.1016/j.jbi.2018.06.012}{doi:\nolinkurl{10.1016/j.jbi.2018.06.012}}


\bibitem[Kashevnik et~al\mbox{.}(2021)]%
        {Kashevnik.2021}
\bibfield{author}{\bibinfo{person}{Alexey Kashevnik}, \bibinfo{person}{Roman Shchedrin}, \bibinfo{person}{Christian Kaiser}, {and} \bibinfo{person}{Alexander Stocker}.} \bibinfo{year}{2021}\natexlab{}.
\newblock \showarticletitle{Driver Distraction Detection Methods: A Literature Review and Framework}.
\newblock \bibinfo{journal}{\emph{IEEE Access}}  \bibinfo{volume}{9} (\bibinfo{year}{2021}), \bibinfo{pages}{60063--60076}.
\newblock
\href{https://doi.org/10.1109/ACCESS.2021.3073599}{doi:\nolinkurl{10.1109/ACCESS.2021.3073599}}


\bibitem[Kim et~al\mbox{.}(2019)]%
        {Kim.2019b}
\bibfield{author}{\bibinfo{person}{Jaejeung Kim}, \bibinfo{person}{Joonyoung Park}, \bibinfo{person}{Hyunsoo Lee}, \bibinfo{person}{Minsam Ko}, {and} \bibinfo{person}{Uichin Lee}.} \bibinfo{year}{2019}\natexlab{}.
\newblock \showarticletitle{LocknType}. In \bibinfo{booktitle}{\emph{Proceedings of the 2019 CHI Conference on Human Factors in Computing Systems}}, \bibfield{editor}{\bibinfo{person}{Stephen Brewster}, \bibinfo{person}{Geraldine Fitzpatrick}, \bibinfo{person}{Anna Cox}, {and} \bibinfo{person}{Vassilis Kostakos}} (Eds.). \bibinfo{publisher}{ACM}, \bibinfo{address}{New York, NY, USA}, \bibinfo{pages}{1--12}.
\newblock
\showISBNx{9781450359702}
\href{https://doi.org/10.1145/3290605.3300927}{doi:\nolinkurl{10.1145/3290605.3300927}}


\bibitem[Ko et~al\mbox{.}(2015)]%
        {ko2015nugu}
\bibfield{author}{\bibinfo{person}{Minsam Ko}, \bibinfo{person}{Subin Yang}, \bibinfo{person}{Joonwon Lee}, \bibinfo{person}{Christian Heizmann}, \bibinfo{person}{Jinyoung Jeong}, \bibinfo{person}{Uichin Lee}, \bibinfo{person}{Daehee Shin}, \bibinfo{person}{Koji Yatani}, \bibinfo{person}{Junehwa Song}, {and} \bibinfo{person}{Kyong-Mee Chung}.} \bibinfo{year}{2015}\natexlab{}.
\newblock \showarticletitle{NUGU: A Group-based Intervention App for Improving Self-Regulation of Limiting Smartphone Use}.
\newblock  (\bibinfo{year}{2015}), \bibinfo{pages}{1235--1245}.
\newblock
\showISBNx{9781450329224}
\href{https://doi.org/10.1145/2675133.2675244}{doi:\nolinkurl{10.1145/2675133.2675244}}


\bibitem[Kroese et~al\mbox{.}(2014)]%
        {Kroese.2014}
\bibfield{author}{\bibinfo{person}{Floor~M. Kroese}, \bibinfo{person}{Denise T.~D. de Ridder}, \bibinfo{person}{Catharine Evers}, {and} \bibinfo{person}{Marieke~A. Adriaanse}.} \bibinfo{year}{2014}\natexlab{}.
\newblock \showarticletitle{Bedtime procrastination: introducing a new area of procrastination}.
\newblock \bibinfo{journal}{\emph{Frontiers in psychology}}  \bibinfo{volume}{5} (\bibinfo{year}{2014}), \bibinfo{pages}{611}.
\newblock
\showISSN{1664-1078}
\href{https://doi.org/10.3389/fpsyg.2014.00611}{doi:\nolinkurl{10.3389/fpsyg.2014.00611}}


\bibitem[Kross et~al\mbox{.}(2013)]%
        {kross2013facebook}
\bibfield{author}{\bibinfo{person}{Ethan Kross}, \bibinfo{person}{Philippe Verduyn}, \bibinfo{person}{Emre Demiralp}, \bibinfo{person}{Jiyoung Park}, \bibinfo{person}{David~Seungjae Lee}, \bibinfo{person}{Natalie Lin}, \bibinfo{person}{Holly Shablack}, \bibinfo{person}{John Jonides}, {and} \bibinfo{person}{Oscar Ybarra}.} \bibinfo{year}{2013}\natexlab{}.
\newblock \showarticletitle{Facebook use predicts declines in subjective well-being in young adults}.
\newblock \bibinfo{journal}{\emph{PloS one}} \bibinfo{volume}{8}, \bibinfo{number}{8} (\bibinfo{year}{2013}), \bibinfo{pages}{e69841}.
\newblock
\href{https://doi.org/10.1371/journal.pone.0069841}{doi:\nolinkurl{10.1371/journal.pone.0069841}}


\bibitem[Kumle et~al\mbox{.}(2021)]%
        {Kumle.2021}
\bibfield{author}{\bibinfo{person}{Levi Kumle}, \bibinfo{person}{Melissa L-H V{\~o}}, {and} \bibinfo{person}{Dejan Draschkow}.} \bibinfo{year}{2021}\natexlab{}.
\newblock \showarticletitle{Estimating power in (generalized) linear mixed models: An open introduction and tutorial in R}.
\newblock \bibinfo{journal}{\emph{Behavior research methods}} \bibinfo{volume}{53}, \bibinfo{number}{6} (\bibinfo{year}{2021}), \bibinfo{pages}{2528--2543}.
\newblock
\href{https://doi.org/10.3758/s13428-021-01546-0}{doi:\nolinkurl{10.3758/s13428-021-01546-0}}


\bibitem[Lee et~al\mbox{.}(2014)]%
        {Lee.2014}
\bibfield{author}{\bibinfo{person}{Uichin Lee}, \bibinfo{person}{Joonwon Lee}, \bibinfo{person}{Minsam Ko}, \bibinfo{person}{Changhun Lee}, \bibinfo{person}{Yuhwan Kim}, \bibinfo{person}{Subin Yang}, \bibinfo{person}{Koji Yatani}, \bibinfo{person}{Gahgene Gweon}, \bibinfo{person}{Kyong-Mee Chung}, {and} \bibinfo{person}{Junehwa Song}.} \bibinfo{year}{2014}\natexlab{}.
\newblock \showarticletitle{Hooked on smartphones}. In \bibinfo{booktitle}{\emph{Proceedings of the SIGCHI Conference on Human Factors in Computing Systems}}, \bibfield{editor}{\bibinfo{person}{Matt Jones}, \bibinfo{person}{Philippe Palanque}, \bibinfo{person}{Albrecht Schmidt}, {and} \bibinfo{person}{Tovi Grossman}} (Eds.). \bibinfo{publisher}{ACM}, \bibinfo{address}{New York, NY, USA}, \bibinfo{pages}{2327--2336}.
\newblock
\showISBNx{9781450324731}
\href{https://doi.org/10.1145/2556288.2557366}{doi:\nolinkurl{10.1145/2556288.2557366}}


\bibitem[Liang et~al\mbox{.}(2023)]%
        {Dawei.2023}
\bibfield{author}{\bibinfo{person}{Dawei Liang}, \bibinfo{person}{Alice Zhang}, {and} \bibinfo{person}{Edison Thomaz}.} \bibinfo{year}{2023}\natexlab{}.
\newblock \showarticletitle{Automated Face-To-Face Conversation Detection on a Commodity Smartwatch with Acoustic Sensing}.
\newblock \bibinfo{journal}{\emph{Proc. ACM Interact. Mob. Wearable Ubiquitous Technol.}} \bibinfo{volume}{7}, \bibinfo{number}{3}, Article \bibinfo{articleno}{109} (\bibinfo{date}{sep} \bibinfo{year}{2023}), \bibinfo{numpages}{29}~pages.
\newblock
\href{https://doi.org/10.1145/3610882}{doi:\nolinkurl{10.1145/3610882}}


\bibitem[Lin et~al\mbox{.}(2015)]%
        {lin2015time}
\bibfield{author}{\bibinfo{person}{Yu-Hsuan Lin}, \bibinfo{person}{Yu-Cheng Lin}, \bibinfo{person}{Yang-Han Lee}, \bibinfo{person}{Po-Hsien Lin}, \bibinfo{person}{Sheng-Hsuan Lin}, \bibinfo{person}{Li-Ren Chang}, \bibinfo{person}{Hsien-Wei Tseng}, \bibinfo{person}{Liang-Yu Yen}, \bibinfo{person}{Cheryl~CH Yang}, {and} \bibinfo{person}{Terry~BJ Kuo}.} \bibinfo{year}{2015}\natexlab{}.
\newblock \showarticletitle{Time distortion associated with smartphone addiction: Identifying smartphone addiction via a mobile application (App)}.
\newblock \bibinfo{journal}{\emph{Journal of psychiatric research}}  \bibinfo{volume}{65} (\bibinfo{year}{2015}), \bibinfo{pages}{139--145}.
\newblock
\href{https://doi.org/10.1016/j.jpsychires.2015.04.003}{doi:\nolinkurl{10.1016/j.jpsychires.2015.04.003}}


\bibitem[Lukoff(2022)]%
        {KaiLukoff.2022.Dissertation}
\bibfield{author}{\bibinfo{person}{Kai Lukoff}.} \bibinfo{year}{2022}\natexlab{}.
\newblock \emph{\bibinfo{title}{Designing to Support Sense of Agency for Time Spent on Digital Interfaces}}.
\newblock Dissertation. \bibinfo{school}{{University of Washington}}.
\newblock
\urldef\tempurl%
\url{http://hdl.handle.net/1773/49196}
\showURL{%
\tempurl}


\bibitem[Lukoff et~al\mbox{.}(2023)]%
        {kai_internal}
\bibfield{author}{\bibinfo{person}{Kai Lukoff}, \bibinfo{person}{Ulrik Lyngs}, \bibinfo{person}{Karina Shirokova}, \bibinfo{person}{Raveena Rao}, \bibinfo{person}{Larry Tian}, \bibinfo{person}{Himanshu Zade}, \bibinfo{person}{Sean~A. Munson}, {and} \bibinfo{person}{Alexis Hiniker}.} \bibinfo{year}{2023}\natexlab{}.
\newblock \showarticletitle{SwitchTube: A Proof-of-Concept System Introducing “Adaptable Commitment Interfaces” as a Tool for Digital Wellbeing}. In \bibinfo{booktitle}{\emph{Proceedings of the 2023 CHI Conference on Human Factors in Computing Systems}} (Hamburg, Germany) \emph{(\bibinfo{series}{CHI '23})}. \bibinfo{publisher}{Association for Computing Machinery}, \bibinfo{address}{New York, NY, USA}, Article \bibinfo{articleno}{197}, \bibinfo{numpages}{22}~pages.
\newblock
\showISBNx{9781450394215}
\href{https://doi.org/10.1145/3544548.3580703}{doi:\nolinkurl{10.1145/3544548.3580703}}


\bibitem[Lukoff et~al\mbox{.}(2018)]%
        {Lukoff_2018}
\bibfield{author}{\bibinfo{person}{Kai Lukoff}, \bibinfo{person}{Cissy Yu}, \bibinfo{person}{Julie Kientz}, {and} \bibinfo{person}{Alexis Hiniker}.} \bibinfo{year}{2018}\natexlab{}.
\newblock \showarticletitle{What Makes Smartphone Use Meaningful or Meaningless?}
\newblock \bibinfo{journal}{\emph{Proc. ACM Interact. Mob. Wearable Ubiquitous Technol.}} \bibinfo{volume}{2}, \bibinfo{number}{1}, Article \bibinfo{articleno}{22} (\bibinfo{date}{mar} \bibinfo{year}{2018}), \bibinfo{numpages}{26}~pages.
\newblock
\href{https://doi.org/10.1145/3191754}{doi:\nolinkurl{10.1145/3191754}}


\bibitem[Lutz and Knop(2020)]%
        {Lutz.2020}
\bibfield{author}{\bibinfo{person}{Sarah Lutz} {and} \bibinfo{person}{Karin Knop}.} \bibinfo{year}{2020}\natexlab{}.
\newblock \showarticletitle{Put down your smartphone -- unless you integrate it into the conversation! An experimental investigation of using smartphones during face to face communication}.
\newblock \bibinfo{journal}{\emph{Studies in Communication and Media}} \bibinfo{volume}{9}, \bibinfo{number}{4} (\bibinfo{year}{2020}), \bibinfo{pages}{516--539}.
\newblock
\showISSN{2192-4007}
\href{https://doi.org/10.5771/2192-4007-2020-4-516}{doi:\nolinkurl{10.5771/2192-4007-2020-4-516}}


\bibitem[Mandi et~al\mbox{.}(2023)]%
        {Mandi.2023}
\bibfield{author}{\bibinfo{person}{Salma Mandi}, \bibinfo{person}{Surjya Ghosh}, \bibinfo{person}{Pradipta De}, {and} \bibinfo{person}{Bivas Mitra}.} \bibinfo{year}{2023}\natexlab{}.
\newblock \showarticletitle{SELFI: Evaluation of Techniques to Reduce Self-report Fatigue by Using Facial Expression of Emotion}.
\newblock In \bibinfo{booktitle}{\emph{Human-Computer Interaction -- INTERACT 2023}}, \bibfield{editor}{\bibinfo{person}{Jos{\'e} {Abdelnour Nocera}}, \bibinfo{person}{Marta {Krist{\'i}n L{\'a}rusd{\'o}ttir}}, \bibinfo{person}{Helen Petrie}, \bibinfo{person}{Antonio Piccinno}, {and} \bibinfo{person}{Marco Winckler}} (Eds.). \bibinfo{series}{Lecture Notes in Computer Science}, Vol.~\bibinfo{volume}{14142}. \bibinfo{publisher}{{Springer Nature Switzerland}}, \bibinfo{address}{Cham}, \bibinfo{pages}{620--640}.
\newblock
\showISBNx{978-3-031-42279-9}
\href{https://doi.org/10.1007/978-3-031-42280-5_39}{doi:\nolinkurl{10.1007/978-3-031-42280-5_39}}


\bibitem[Meinhardt et~al\mbox{.}(2023)]%
        {meinhardt_balancing_2023}
\bibfield{author}{\bibinfo{person}{Luca-Maxim Meinhardt}, \bibinfo{person}{Jan-Henry Belz}, \bibinfo{person}{Michael Rietzler}, {and} \bibinfo{person}{Enrico Rukzio}.} \bibinfo{year}{2023}\natexlab{}.
\newblock \bibinfo{title}{Balancing the {Digital} and the {Physical}: {Discussing} {Push} and {Pull} {Factors} for {Digital} {Well}-being}.
\newblock
\href{https://doi.org/10.48550/ARXIV.2305.12513}{doi:\nolinkurl{10.48550/ARXIV.2305.12513}}
\newblock
\shownote{Version Number: 1}.


\bibitem[Meske and Potthoff(2017)]%
        {meske2017dinu}
\bibfield{author}{\bibinfo{person}{Christian Meske} {and} \bibinfo{person}{Tobias Potthoff}.} \bibinfo{year}{2017}\natexlab{}.
\newblock \showarticletitle{The DINU-model--a process model for the design of nudges}.
\newblock \bibinfo{journal}{\emph{Proceedings of the 25th European Conference on Information Systems (ECIS)}} (\bibinfo{year}{2017}).
\newblock
\showISSN{2587–2597}
\urldef\tempurl%
\url{https://aisel.aisnet.org/ecis2017_rip/11/}
\showURL{%
\tempurl}


\bibitem[Metcalfe(2012)]%
        {Metcalfe2012Behavioural}
\bibfield{author}{\bibinfo{person}{Robert Metcalfe}.} \bibinfo{year}{2012}\natexlab{}.
\newblock \showarticletitle{Behavioural economics and its implications for transport}.
\newblock \bibinfo{journal}{\emph{Journal of Transport Geography}}  \bibinfo{volume}{24} (\bibinfo{year}{2012}), \bibinfo{pages}{503--511}.
\newblock
\href{https://doi.org/10.1016/J.JTRANGEO.2012.01.019}{doi:\nolinkurl{10.1016/J.JTRANGEO.2012.01.019}}


\bibitem[Mildner and Savino(2021)]%
        {Mildner.2021}
\bibfield{author}{\bibinfo{person}{Thomas Mildner} {and} \bibinfo{person}{Gian-Luca Savino}.} \bibinfo{year}{2021}\natexlab{}.
\newblock \showarticletitle{Ethical User Interfaces: Exploring the Effects of Dark Patterns on Facebook}. In \bibinfo{booktitle}{\emph{Extended Abstracts of the 2021 CHI Conference on Human Factors in Computing Systems}}, \bibfield{editor}{\bibinfo{person}{Yoshifumi Kitamura}, \bibinfo{person}{Aaron Quigley}, \bibinfo{person}{Katherine Isbister}, {and} \bibinfo{person}{Takeo Igarashi}} (Eds.). \bibinfo{publisher}{ACM}, \bibinfo{address}{New York, NY, USA}, \bibinfo{pages}{1--7}.
\newblock
\showISBNx{9781450380959}
\href{https://doi.org/10.1145/3411763.3451659}{doi:\nolinkurl{10.1145/3411763.3451659}}


\bibitem[Mildner et~al\mbox{.}(2023)]%
        {Mildner.2023}
\bibfield{author}{\bibinfo{person}{Thomas Mildner}, \bibinfo{person}{Gian-Luca Savino}, \bibinfo{person}{Philip~R. Doyle}, \bibinfo{person}{Benjamin~R. Cowan}, {and} \bibinfo{person}{Rainer Malaka}.} \bibinfo{year}{2023}\natexlab{}.
\newblock \showarticletitle{About Engaging and Governing Strategies: A Thematic Analysis of Dark Patterns in Social Networking Services}. In \bibinfo{booktitle}{\emph{Proceedings of the 2023 CHI Conference on Human Factors in Computing Systems}} (Hamburg, Germany) \emph{(\bibinfo{series}{CHI '23})}. \bibinfo{publisher}{Association for Computing Machinery}, \bibinfo{address}{New York, NY, USA}, Article \bibinfo{articleno}{192}, \bibinfo{numpages}{15}~pages.
\newblock
\showISBNx{9781450394215}
\href{https://doi.org/10.1145/3544548.3580695}{doi:\nolinkurl{10.1145/3544548.3580695}}


\bibitem[Miller-Ott and Kelly(2017)]%
        {Miller-Ott.2017}
\bibfield{author}{\bibinfo{person}{Aimee~E. Miller-Ott} {and} \bibinfo{person}{Lynne Kelly}.} \bibinfo{year}{2017}\natexlab{}.
\newblock \showarticletitle{A Politeness Theory Analysis of Cell-Phone Usage in the Presence of Friends}.
\newblock \bibinfo{journal}{\emph{Communication Studies}} \bibinfo{volume}{68}, \bibinfo{number}{2} (\bibinfo{year}{2017}), \bibinfo{pages}{190--207}.
\newblock
\href{https://doi.org/10.1080/10510974.2017.1299024}{doi:\nolinkurl{10.1080/10510974.2017.1299024}}
\showeprint{https://doi.org/10.1080/10510974.2017.1299024}


\bibitem[Mills(2023)]%
        {MILLS.2023}
\bibfield{author}{\bibinfo{person}{Stuart Mills}.} \bibinfo{year}{2023}\natexlab{}.
\newblock \showarticletitle{Nudge/sludge symmetry: on the relationship between nudge and sludge and the resulting ontological, normative and transparency implications}.
\newblock \bibinfo{journal}{\emph{Behavioural Public Policy}} \bibinfo{volume}{7}, \bibinfo{number}{2} (\bibinfo{year}{2023}), \bibinfo{pages}{309--332}.
\newblock
\showISSN{2398-063X}
\href{https://doi.org/10.1017/bpp.2020.61}{doi:\nolinkurl{10.1017/bpp.2020.61}}


\bibitem[Misra et~al\mbox{.}(2016)]%
        {Misra.2016}
\bibfield{author}{\bibinfo{person}{Shalini Misra}, \bibinfo{person}{Lulu Cheng}, \bibinfo{person}{Jamie Genevie}, {and} \bibinfo{person}{Miao Yuan}.} \bibinfo{year}{2016}\natexlab{}.
\newblock \showarticletitle{The iPhone Effect: The Quality of In-Person Social Interactions in the Presence of Mobile Devices}.
\newblock \bibinfo{journal}{\emph{Environment and Behavior}} \bibinfo{volume}{48}, \bibinfo{number}{2} (\bibinfo{year}{2016}), \bibinfo{pages}{275--298}.
\newblock
\href{https://doi.org/10.1177/0013916514539755}{doi:\nolinkurl{10.1177/0013916514539755}}
\showeprint{https://doi.org/10.1177/0013916514539755}


\bibitem[{Monge Roffarello} and de~Russis(2019)]%
        {mongeroffarello2019race}
\bibfield{author}{\bibinfo{person}{Alberto {Monge Roffarello}} {and} \bibinfo{person}{Luigi de Russis}.} \bibinfo{year}{2019}\natexlab{}.
\newblock \showarticletitle{The Race Towards Digital Wellbeing: Issues and Opportunities}. In \bibinfo{booktitle}{\emph{Proceedings of the 2019 CHI Conference on Human Factors in Computing Systems}} \emph{(\bibinfo{series}{CHI '19})}. \bibinfo{publisher}{{Association for Computing Machinery}}, \bibinfo{address}{New York, NY, USA}, \bibinfo{pages}{1--14}.
\newblock
\showISBNx{9781450359702}
\href{https://doi.org/10.1145/3290605.3300616}{doi:\nolinkurl{10.1145/3290605.3300616}}


\bibitem[{Monge Roffarello} and de~Russis(2022)]%
        {mongeroffarello2022towards}
\bibfield{author}{\bibinfo{person}{Alberto {Monge Roffarello}} {and} \bibinfo{person}{Luigi de Russis}.} \bibinfo{year}{2022}\natexlab{}.
\newblock \showarticletitle{Towards Understanding the Dark Patterns That Steal Our Attention}. In \bibinfo{booktitle}{\emph{Extended Abstracts of the 2022 CHI Conference on Human Factors in Computing Systems}} \emph{(\bibinfo{series}{CHI EA '22})}. \bibinfo{publisher}{{Association for Computing Machinery}}, \bibinfo{address}{New York, NY, USA}.
\newblock
\showISBNx{9781450391566}
\href{https://doi.org/10.1145/3491101.3519829}{doi:\nolinkurl{10.1145/3491101.3519829}}


\bibitem[Nathan and Zeitzer(2013)]%
        {Nathan.2013}
\bibfield{author}{\bibinfo{person}{Nila Nathan} {and} \bibinfo{person}{Jamie Zeitzer}.} \bibinfo{year}{2013}\natexlab{}.
\newblock \showarticletitle{A survey study of the association between mobile phone use and daytime sleepiness in California high school students}.
\newblock \bibinfo{journal}{\emph{BMC public health}}  \bibinfo{volume}{13} (\bibinfo{year}{2013}), \bibinfo{pages}{840}.
\newblock
\href{https://doi.org/10.1186/1471-2458-13-840}{doi:\nolinkurl{10.1186/1471-2458-13-840}}


\bibitem[Nelder(1977)]%
        {Nelder.1977}
\bibfield{author}{\bibinfo{person}{J.~A. Nelder}.} \bibinfo{year}{1977}\natexlab{}.
\newblock \showarticletitle{A Reformulation of Linear Models}.
\newblock \bibinfo{journal}{\emph{Journal of the Royal Statistical Society. Series A (General)}} \bibinfo{volume}{140}, \bibinfo{number}{1} (\bibinfo{year}{1977}), \bibinfo{pages}{48--77}.
\newblock
\showISSN{00359238}
\urldef\tempurl%
\url{http://www.jstor.org/stable/2344517}
\showURL{%
\tempurl}


\bibitem[Okeke et~al\mbox{.}(2018b)]%
        {Okeke.2018}
\bibfield{author}{\bibinfo{person}{Fabian Okeke}, \bibinfo{person}{Michael Sobolev}, \bibinfo{person}{Nicola Dell}, {and} \bibinfo{person}{Deborah Estrin}.} \bibinfo{year}{2018}\natexlab{b}.
\newblock \showarticletitle{Good vibrations}. In \bibinfo{booktitle}{\emph{Proceedings of the 20th International Conference on Human-Computer Interaction with Mobile Devices and Services}}, \bibfield{editor}{\bibinfo{person}{Lynne Bailie} {and} \bibinfo{person}{Nuria Oliver}} (Eds.). \bibinfo{publisher}{ACM}, \bibinfo{address}{New York, NY, USA}, \bibinfo{pages}{1--12}.
\newblock
\showISBNx{9781450358989}
\href{https://doi.org/10.1145/3229434.3229463}{doi:\nolinkurl{10.1145/3229434.3229463}}


\bibitem[Okeke et~al\mbox{.}(2018c)]%
        {okeke2018good}
\bibfield{author}{\bibinfo{person}{Fabian Okeke}, \bibinfo{person}{Michael Sobolev}, \bibinfo{person}{Nicola Dell}, {and} \bibinfo{person}{Deborah Estrin}.} \bibinfo{year}{2018}\natexlab{c}.
\newblock \showarticletitle{Good Vibrations: Can a Digital Nudge Reduce Digital Overload?}. In \bibinfo{booktitle}{\emph{Proceedings of the 20th International Conference on Human-Computer Interaction with Mobile Devices and Services}} \emph{(\bibinfo{series}{MobileHCI '18})}. \bibinfo{publisher}{{Association for Computing Machinery}}, \bibinfo{address}{New York, NY, USA}.
\newblock
\showISBNx{9781450358989}
\href{https://doi.org/10.1145/3229434.3229463}{doi:\nolinkurl{10.1145/3229434.3229463}}


\bibitem[Okeke et~al\mbox{.}(2018a)]%
        {Okeke.2018framework}
\bibfield{author}{\bibinfo{person}{Fabian Okeke}, \bibinfo{person}{Michael Sobolev}, {and} \bibinfo{person}{Deborah Estrin}.} \bibinfo{year}{2018}\natexlab{a}.
\newblock \showarticletitle{Towards A Framework for Mobile Behavior Change Research}. In \bibinfo{booktitle}{\emph{Proceedings of the Technology, Mind, and Society}}. \bibinfo{publisher}{ACM}, \bibinfo{address}{New York, NY, USA}, \bibinfo{pages}{1--6}.
\newblock
\showISBNx{9781450354202}
\href{https://doi.org/10.1145/3183654.3183706}{doi:\nolinkurl{10.1145/3183654.3183706}}


\bibitem[Orzikulova et~al\mbox{.}(2024)]%
        {orzikulova_time2stop_2024}
\bibfield{author}{\bibinfo{person}{Adiba Orzikulova}, \bibinfo{person}{Han Xiao}, \bibinfo{person}{Zhipeng Li}, \bibinfo{person}{Yukang Yan}, \bibinfo{person}{Yuntao Wang}, \bibinfo{person}{Yuanchun Shi}, \bibinfo{person}{Marzyeh Ghassemi}, \bibinfo{person}{Sung-Ju Lee}, \bibinfo{person}{Anind~K Dey}, {and} \bibinfo{person}{Xuhai Xu}.} \bibinfo{year}{2024}\natexlab{}.
\newblock \showarticletitle{{Time2Stop}: {Adaptive} and {Explainable} {Human}-{AI} {Loop} for {Smartphone} {Overuse} {Intervention}}. In \bibinfo{booktitle}{\emph{Proceedings of the {CHI} {Conference} on {Human} {Factors} in {Computing} {Systems}}}. \bibinfo{publisher}{ACM}, \bibinfo{address}{Honolulu HI USA}, \bibinfo{pages}{1--20}.
\newblock
\showISBNx{9798400703300}
\href{https://doi.org/10.1145/3613904.3642747}{doi:\nolinkurl{10.1145/3613904.3642747}}


\bibitem[Panova and Carbonell(2018)]%
        {Panova.2018}
\bibfield{author}{\bibinfo{person}{Tayana Panova} {and} \bibinfo{person}{Xavier Carbonell}.} \bibinfo{year}{2018}\natexlab{}.
\newblock \showarticletitle{Is smartphone addiction really an addiction?}
\newblock \bibinfo{journal}{\emph{Journal of behavioral addictions}} \bibinfo{volume}{7}, \bibinfo{number}{2} (\bibinfo{year}{2018}), \bibinfo{pages}{252--259}.
\newblock
\href{https://doi.org/10.1556/2006.7.2018.49}{doi:\nolinkurl{10.1556/2006.7.2018.49}}


\bibitem[Pinder et~al\mbox{.}(2018)]%
        {Pinder.2018}
\bibfield{author}{\bibinfo{person}{Charlie Pinder}, \bibinfo{person}{Jo Vermeulen}, \bibinfo{person}{Benjamin~R. Cowan}, {and} \bibinfo{person}{Russell Beale}.} \bibinfo{year}{2018}\natexlab{}.
\newblock \showarticletitle{Digital Behaviour Change Interventions to Break and Form Habits}.
\newblock \bibinfo{journal}{\emph{ACM Transactions on Computer-Human Interaction}} \bibinfo{volume}{25}, \bibinfo{number}{3} (\bibinfo{year}{2018}), \bibinfo{pages}{1--66}.
\newblock
\showISSN{1073-0516}
\href{https://doi.org/10.1145/3196830}{doi:\nolinkurl{10.1145/3196830}}


\bibitem[Przybylski and Weinstein(2013)]%
        {Przybylski.2013}
\bibfield{author}{\bibinfo{person}{Andrew~K. Przybylski} {and} \bibinfo{person}{Netta Weinstein}.} \bibinfo{year}{2013}\natexlab{}.
\newblock \showarticletitle{Can you connect with me now? How the presence of mobile communication technology influences face-to-face conversation quality}.
\newblock \bibinfo{journal}{\emph{Journal of Social and Personal Relationships}} \bibinfo{volume}{30}, \bibinfo{number}{3} (\bibinfo{year}{2013}), \bibinfo{pages}{237--246}.
\newblock
\showISSN{0265-4075}
\href{https://doi.org/10.1177/0265407512453827}{doi:\nolinkurl{10.1177/0265407512453827}}


\bibitem[Purohit et~al\mbox{.}(2023a)]%
        {2023cocreation}
\bibfield{author}{\bibinfo{person}{Aditya~Kumar Purohit}, \bibinfo{person}{Torben~Jan Barev}, \bibinfo{person}{Sofia Sch{\"o}bel}, \bibinfo{person}{Andreas Janson}, {and} \bibinfo{person}{Adrian Holzer}.} \bibinfo{year}{2023}\natexlab{a}.
\newblock \showarticletitle{Designing for Digital Wellbeing on a Smartphone: Co-creation of Digital Nudges to Mitigate Instagram Overuse}. In \bibinfo{booktitle}{\emph{Proceedings of the 56th Hawaii International Conference on System Sciences}}. \bibinfo{publisher}{HICSS}, \bibinfo{address}{Hawaii}, \bibinfo{pages}{4087--4096}.
\newblock
\showISBNx{78-0-9981331-6-4}
\href{https://doi.org/10.24251/HICSS.2023.499}{doi:\nolinkurl{10.24251/HICSS.2023.499}}


\bibitem[Purohit et~al\mbox{.}(2023b)]%
        {Purohit.2023}
\bibfield{author}{\bibinfo{person}{Aditya~Kumar Purohit}, \bibinfo{person}{Kristoffer Bergram}, \bibinfo{person}{Louis Barclay}, \bibinfo{person}{Val\'{e}ry Bezen\c{c}on}, {and} \bibinfo{person}{Adrian Holzer}.} \bibinfo{year}{2023}\natexlab{b}.
\newblock \showarticletitle{Starving the Newsfeed for Social Media Detox: Effects of Strict and Self-Regulated Facebook Newsfeed Diets}. In \bibinfo{booktitle}{\emph{Proceedings of the 2023 CHI Conference on Human Factors in Computing Systems}} (Hamburg, Germany) \emph{(\bibinfo{series}{CHI '23})}. \bibinfo{publisher}{Association for Computing Machinery}, \bibinfo{address}{New York, NY, USA}, Article \bibinfo{articleno}{196}, \bibinfo{numpages}{16}~pages.
\newblock
\showISBNx{9781450394215}
\href{https://doi.org/10.1145/3544548.3581187}{doi:\nolinkurl{10.1145/3544548.3581187}}


\bibitem[Purohit and Holzer(2019)]%
        {Purohit.2019}
\bibfield{author}{\bibinfo{person}{Aditya~Kumar Purohit} {and} \bibinfo{person}{Adrian Holzer}.} \bibinfo{year}{2019}\natexlab{}.
\newblock \showarticletitle{Functional Digital Nudges}. In \bibinfo{booktitle}{\emph{Extended Abstracts of the 2019 CHI Conference on Human Factors in Computing Systems}}, \bibfield{editor}{\bibinfo{person}{Stephen Brewster}, \bibinfo{person}{Geraldine Fitzpatrick}, \bibinfo{person}{Anna Cox}, {and} \bibinfo{person}{Vassilis Kostakos}} (Eds.). \bibinfo{publisher}{ACM}, \bibinfo{address}{New York, NY, USA}, \bibinfo{pages}{1--6}.
\newblock
\showISBNx{9781450359719}
\href{https://doi.org/10.1145/3290607.3312876}{doi:\nolinkurl{10.1145/3290607.3312876}}


\bibitem[Purohit and Holzer(2021)]%
        {Purohit2021}
\bibfield{author}{\bibinfo{person}{Aditya~K. Purohit} {and} \bibinfo{person}{Adrian Holzer}.} \bibinfo{year}{2021}\natexlab{}.
\newblock \showarticletitle{Unhooked by Design: Scrolling Mindfully on Social Media by Automating Digital Nudges}. In \bibinfo{booktitle}{\emph{Proceedings of the 27th Americas Conference on Information Systems, AMCIS 2021}}. \bibinfo{publisher}{AIS}, \bibinfo{address}{Virtual Conference}, \bibinfo{pages}{1--10}.
\newblock
\urldef\tempurl%
\url{https://aisel.aisnet.org/amcis2021/sig_hci/sig_hci/7}
\showURL{%
\tempurl}


\bibitem[Rains(2013)]%
        {Rains.2013}
\bibfield{author}{\bibinfo{person}{Stephen~A. Rains}.} \bibinfo{year}{2013}\natexlab{}.
\newblock \showarticletitle{The Nature of Psychological Reactance Revisited: A Meta-Analytic Review}.
\newblock \bibinfo{journal}{\emph{Human Communication Research}} \bibinfo{volume}{39}, \bibinfo{number}{1} (\bibinfo{year}{2013}), \bibinfo{pages}{47--73}.
\newblock
\showISSN{03603989}
\href{https://doi.org/10.1111/j.1468-2958.2012.01443.x}{doi:\nolinkurl{10.1111/j.1468-2958.2012.01443.x}}


\bibitem[Reinecke and Hofmann(2016)]%
        {reinecke_slacking_2016}
\bibfield{author}{\bibinfo{person}{Leonard Reinecke} {and} \bibinfo{person}{Wilhelm Hofmann}.} \bibinfo{year}{2016}\natexlab{}.
\newblock \showarticletitle{Slacking {Off} or {Winding} {Down}? {An} {Experience} {Sampling} {Study} on the {Drivers} and {Consequences} of {Media} {Use} for {Recovery} {Versus} {Procrastination}: {Slacking} {Off} or {Winding} {Down}?}
\newblock \bibinfo{journal}{\emph{Human Communication Research}} \bibinfo{volume}{42}, \bibinfo{number}{3} (\bibinfo{date}{July} \bibinfo{year}{2016}), \bibinfo{pages}{441--461}.
\newblock
\showISSN{03603989}
\href{https://doi.org/10.1111/hcre.12082}{doi:\nolinkurl{10.1111/hcre.12082}}


\bibitem[Reissmann et~al\mbox{.}(2018)]%
        {reissmann2018role}
\bibfield{author}{\bibinfo{person}{Andreas Reissmann}, \bibinfo{person}{Joachim Hauser}, \bibinfo{person}{Ewelina Stollberg}, \bibinfo{person}{Ivo Kaunzinger}, {and} \bibinfo{person}{Klaus~W Lange}.} \bibinfo{year}{2018}\natexlab{}.
\newblock \showarticletitle{The role of loneliness in emerging adults’ everyday use of facebook--An experience sampling approach}.
\newblock \bibinfo{journal}{\emph{Computers in Human Behavior}}  \bibinfo{volume}{88} (\bibinfo{year}{2018}), \bibinfo{pages}{47--60}.
\newblock
\href{https://doi.org/10.1016/j.chb.2018.06.011}{doi:\nolinkurl{10.1016/j.chb.2018.06.011}}


\bibitem[Rixen et~al\mbox{.}(2023)]%
        {RixenIS.2023}
\bibfield{author}{\bibinfo{person}{Jan~Ole Rixen}, \bibinfo{person}{Luca-Maxim Meinhardt}, \bibinfo{person}{Michael Glöckler}, \bibinfo{person}{Marius-Lukas Ziegenbein}, \bibinfo{person}{Anna Schlothauer}, \bibinfo{person}{Mark Colley}, \bibinfo{person}{Enrico Rukzio}, {and} \bibinfo{person}{Jan Guggenheimer}.} \bibinfo{year}{2023}\natexlab{}.
\newblock \showarticletitle{The Loop and Reasons to Break It: Investigating Infinite Scrolling Behaviour in Social Media Applications and Reasons to Stop}.
\newblock \bibinfo{journal}{\emph{Proc. ACM Hum.-Comput. Interact.}} \bibinfo{volume}{7}, \bibinfo{number}{MHCI}, Article \bibinfo{articleno}{1228} (\bibinfo{date}{sep} \bibinfo{year}{2023}), \bibinfo{numpages}{22}~pages.
\newblock
\href{https://doi.org/10.1145/3604275}{doi:\nolinkurl{10.1145/3604275}}


\bibitem[Roffarello and De~Russis(2021)]%
        {alberto_2021}
\bibfield{author}{\bibinfo{person}{Alberto~Monge Roffarello} {and} \bibinfo{person}{Luigi De~Russis}.} \bibinfo{year}{2021}\natexlab{}.
\newblock \showarticletitle{Understanding, Discovering, and Mitigating Habitual Smartphone Use in Young Adults}.
\newblock \bibinfo{journal}{\emph{ACM Trans. Interact. Intell. Syst.}} \bibinfo{volume}{11}, \bibinfo{number}{2}, Article \bibinfo{articleno}{13} (\bibinfo{date}{jul} \bibinfo{year}{2021}), \bibinfo{numpages}{34}~pages.
\newblock
\showISSN{2160-6455}
\href{https://doi.org/10.1145/3447991}{doi:\nolinkurl{10.1145/3447991}}


\bibitem[Roffarello and De~Russis(2023)]%
        {alberto_2023}
\bibfield{author}{\bibinfo{person}{Alberto~Monge Roffarello} {and} \bibinfo{person}{Luigi De~Russis}.} \bibinfo{year}{2023}\natexlab{}.
\newblock \showarticletitle{Achieving Digital Wellbeing Through Digital Self-Control Tools: A Systematic Review and Meta-Analysis}.
\newblock \bibinfo{journal}{\emph{ACM Trans. Comput.-Hum. Interact.}} \bibinfo{volume}{30}, \bibinfo{number}{4}, Article \bibinfo{articleno}{53} (\bibinfo{date}{sep} \bibinfo{year}{2023}), \bibinfo{numpages}{66}~pages.
\newblock
\showISSN{1073-0516}
\href{https://doi.org/10.1145/3571810}{doi:\nolinkurl{10.1145/3571810}}


\bibitem[Ruiz et~al\mbox{.}(2024)]%
        {ruiz_design_2024}
\bibfield{author}{\bibinfo{person}{Nicolas Ruiz}, \bibinfo{person}{Gabriela Molina~León}, {and} \bibinfo{person}{Hendrik Heuer}.} \bibinfo{year}{2024}\natexlab{}.
\newblock \showarticletitle{Design {Frictions} on {Social} {Media}: {Balancing} {Reduced} {Mindless} {Scrolling} and {User} {Satisfaction}}. In \bibinfo{booktitle}{\emph{Proceedings of {Mensch} und {Computer} 2024}}. \bibinfo{publisher}{ACM}, \bibinfo{address}{Karlsruhe Germany}, \bibinfo{pages}{442--447}.
\newblock
\showISBNx{9798400709982}
\href{https://doi.org/10.1145/3670653.3677495}{doi:\nolinkurl{10.1145/3670653.3677495}}


\bibitem[Ryan and Deci(2000)]%
        {ryan_self-determination_2000}
\bibfield{author}{\bibinfo{person}{Richard~M. Ryan} {and} \bibinfo{person}{Edward~L. Deci}.} \bibinfo{year}{2000}\natexlab{}.
\newblock \showarticletitle{Self-determination theory and the facilitation of intrinsic motivation, social development, and well-being.}
\newblock \bibinfo{journal}{\emph{American Psychologist}} \bibinfo{volume}{55}, \bibinfo{number}{1} (\bibinfo{year}{2000}), \bibinfo{pages}{68--78}.
\newblock
\showISSN{1935-990X, 0003-066X}
\href{https://doi.org/10.1037/0003-066X.55.1.68}{doi:\nolinkurl{10.1037/0003-066X.55.1.68}}


\bibitem[Sakel et~al\mbox{.}(2024)]%
        {sakel_social_2024}
\bibfield{author}{\bibinfo{person}{Sophia Sakel}, \bibinfo{person}{Tabea Blenk}, \bibinfo{person}{Albrecht Schmidt}, {and} \bibinfo{person}{Luke Haliburton}.} \bibinfo{year}{2024}\natexlab{}.
\newblock \showarticletitle{The {Social} {Journal}: {Investigating} {Technology} to {Support} and {Reflect} on {Social} {Interactions}}. In \bibinfo{booktitle}{\emph{Proceedings of the {CHI} {Conference} on {Human} {Factors} in {Computing} {Systems}}}. \bibinfo{publisher}{ACM}, \bibinfo{address}{Honolulu HI USA}, \bibinfo{pages}{1--18}.
\newblock
\showISBNx{9798400703300}
\href{https://doi.org/10.1145/3613904.3642411}{doi:\nolinkurl{10.1145/3613904.3642411}}


\bibitem[Samdahl(1991)]%
        {Samdahl.1991}
\bibfield{author}{\bibinfo{person}{Diane~M. Samdahl}.} \bibinfo{year}{1991}\natexlab{}.
\newblock \showarticletitle{Measuring Leisure: Categorical or Interval?}
\newblock \bibinfo{journal}{\emph{Journal of Leisure Research}} \bibinfo{volume}{23}, \bibinfo{number}{1} (\bibinfo{year}{1991}), \bibinfo{pages}{87--93}.
\newblock
\showISSN{0022-2216}
\href{https://doi.org/10.1080/00222216.1991.11969845}{doi:\nolinkurl{10.1080/00222216.1991.11969845}}


\bibitem[Scott et~al\mbox{.}(2024)]%
        {scott2024doing}
\bibfield{author}{\bibinfo{person}{Ava~Elizabeth Scott}, \bibinfo{person}{Leon Reicherts}, {and} \bibinfo{person}{Evropi Stefanidi}.} \bibinfo{year}{2024}\natexlab{}.
\newblock \showarticletitle{Doing CHI together: The benefits of a writing retreat for early-career HCI researchers}.
\newblock \bibinfo{journal}{\emph{Interactions}} \bibinfo{volume}{31}, \bibinfo{number}{3} (\bibinfo{year}{2024}), \bibinfo{pages}{8--9}.
\newblock


\bibitem[Sha and Dong(2021)]%
        {Sha.2021}
\bibfield{author}{\bibinfo{person}{Peng Sha} {and} \bibinfo{person}{Xiaoyu Dong}.} \bibinfo{year}{2021}\natexlab{}.
\newblock \showarticletitle{Research on Adolescents Regarding the Indirect Effect of Depression, Anxiety, and Stress between TikTok Use Disorder and Memory Loss}.
\newblock \bibinfo{journal}{\emph{International journal of environmental research and public health}} \bibinfo{volume}{18}, \bibinfo{number}{16} (\bibinfo{year}{2021}).
\newblock
\href{https://doi.org/10.3390/ijerph18168820}{doi:\nolinkurl{10.3390/ijerph18168820}}


\bibitem[Shahid et~al\mbox{.}(2012)]%
        {Shahid.2012b}
\bibfield{editor}{\bibinfo{person}{Azmeh Shahid}, \bibinfo{person}{Kate Wilkinson}, \bibinfo{person}{Shai Marcu}, {and} \bibinfo{person}{Colin~M. Shapiro}} (Eds.). \bibinfo{year}{2012}\natexlab{}.
\newblock \bibinfo{booktitle}{\emph{STOP, THAT and One Hundred Other Sleep Scales}}.
\newblock \bibinfo{publisher}{{Springer New York}}, \bibinfo{address}{New York, NY}.
\newblock
\showISBNx{978-1-4419-9892-7}
\href{https://doi.org/10.1007/978-1-4419-9893-4}{doi:\nolinkurl{10.1007/978-1-4419-9893-4}}


\bibitem[Shapiro and Wilk(1965)]%
        {Shapiro-Wilk}
\bibfield{author}{\bibinfo{person}{S.~S. Shapiro} {and} \bibinfo{person}{M.~B. Wilk}.} \bibinfo{year}{1965}\natexlab{}.
\newblock \showarticletitle{An Analysis of Variance Test for Normality (Complete Samples)}.
\newblock \bibinfo{journal}{\emph{Biometrika}} \bibinfo{volume}{52}, \bibinfo{number}{3/4} (\bibinfo{year}{1965}), \bibinfo{pages}{591--611}.
\newblock
\showISSN{00063444}
\urldef\tempurl%
\url{http://www.jstor.org/stable/2333709}
\showURL{%
\tempurl}


\bibitem[Sirois and Pychyl(2013)]%
        {sirois_procrastination_2013}
\bibfield{author}{\bibinfo{person}{Fuschia Sirois} {and} \bibinfo{person}{Timothy Pychyl}.} \bibinfo{year}{2013}\natexlab{}.
\newblock \showarticletitle{Procrastination and the {Priority} of {Short}‐{Term} {Mood} {Regulation}: {Consequences} for {Future} {Self}}.
\newblock \bibinfo{journal}{\emph{Social and Personality Psychology Compass}} \bibinfo{volume}{7}, \bibinfo{number}{2} (\bibinfo{date}{Feb.} \bibinfo{year}{2013}), \bibinfo{pages}{115--127}.
\newblock
\showISSN{1751-9004, 1751-9004}
\href{https://doi.org/10.1111/spc3.12011}{doi:\nolinkurl{10.1111/spc3.12011}}


\bibitem[Sobolev et~al\mbox{.}(2021)]%
        {Sobolev.2021}
\bibfield{author}{\bibinfo{person}{Michael Sobolev}, \bibinfo{person}{Rachel Vitale}, \bibinfo{person}{Hongyi Wen}, \bibinfo{person}{James Kizer}, \bibinfo{person}{Robert Leeman}, \bibinfo{person}{J.~P. Pollak}, \bibinfo{person}{Amit Baumel}, \bibinfo{person}{Nehal~P. Vadhan}, \bibinfo{person}{Deborah Estrin}, {and} \bibinfo{person}{Frederick Muench}.} \bibinfo{year}{2021}\natexlab{}.
\newblock \showarticletitle{The Digital Marshmallow Test (DMT) Diagnostic and Monitoring Mobile Health App for Impulsive Behavior: Development and Validation Study}.
\newblock \bibinfo{journal}{\emph{JMIR mHealth and uHealth}} \bibinfo{volume}{9}, \bibinfo{number}{1} (\bibinfo{year}{2021}), \bibinfo{pages}{e25018}.
\newblock
\href{https://doi.org/10.2196/25018}{doi:\nolinkurl{10.2196/25018}}


\bibitem[statcounter(2023)]%
        {AndroidVersionUSA}
\bibfield{author}{\bibinfo{person}{statcounter}.} \bibinfo{year}{2023}\natexlab{}.
\newblock \bibinfo{booktitle}{\emph{Mobile Android Version Market Share United States Of America}}.
\newblock
\urldef\tempurl%
\url{https://gs.statcounter.com/android-version-market-share/mobile/united-states-of-america/#monthly-202208-202310-bar}
\showURL{%
Retrieved Aug 10, 2023 from \tempurl}


\bibitem[{Statista}(2023)]%
        {StatistaSNS}
\bibfield{author}{\bibinfo{person}{{Statista}}.} \bibinfo{year}{2023}\natexlab{}.
\newblock \bibinfo{booktitle}{\emph{Social network usage by brand in the U.S. as of December 2023}}.
\newblock
\urldef\tempurl%
\url{https://www.statista.com/forecasts/997135/social-network-usage-by-brand-in-the-us}
\showURL{%
\tempurl}


\bibitem[Terzimehi{\'c} and Aragon-Hahner(2022)]%
        {Terzimehic.2022b}
\bibfield{author}{\bibinfo{person}{Na{\dj}a Terzimehi{\'c}} {and} \bibinfo{person}{Sarah Aragon-Hahner}.} \bibinfo{year}{2022}\natexlab{}.
\newblock \showarticletitle{I Wish I Had: Desired Real-World Activities Instead of Regretful Smartphone Use}. In \bibinfo{booktitle}{\emph{Proceedings of the 21st International Conference on Mobile and Ubiquitous Multimedia}}, \bibfield{editor}{\bibinfo{person}{Tanja D{\"o}ring}, \bibinfo{person}{Susanne Boll}, \bibinfo{person}{Ashley Colley}, \bibinfo{person}{Augusto Esteves}, {and} \bibinfo{person}{Jo{\~a}o Guerreiro}} (Eds.). \bibinfo{publisher}{ACM}, \bibinfo{address}{New York, NY, USA}, \bibinfo{pages}{47--52}.
\newblock
\showISBNx{9781450398206}
\href{https://doi.org/10.1145/3568444.3568465}{doi:\nolinkurl{10.1145/3568444.3568465}}


\bibitem[Terzimehi{\'c} et~al\mbox{.}(2022)]%
        {Terzimehic.2022}
\bibfield{author}{\bibinfo{person}{Na{\dj}a Terzimehi{\'c}}, \bibinfo{person}{Luke Haliburton}, \bibinfo{person}{Philipp Greiner}, \bibinfo{person}{Albrecht Schmidt}, \bibinfo{person}{Heinrich Hussmann}, {and} \bibinfo{person}{Ville M{\"a}kel{\"a}}.} \bibinfo{year}{2022}\natexlab{}.
\newblock \showarticletitle{MindPhone: Mindful Reflection at Unlock Can Reduce Absentminded Smartphone Use}. In \bibinfo{booktitle}{\emph{Designing Interactive Systems Conference}}, \bibfield{editor}{\bibinfo{person}{Florian~`Floyd' Mueller}, \bibinfo{person}{Stefan Greuter}, \bibinfo{person}{Rohit~Ashok Khot}, \bibinfo{person}{Penny Sweetser}, {and} \bibinfo{person}{Marianna Obrist}} (Eds.). \bibinfo{publisher}{ACM}, \bibinfo{address}{New York, NY, USA}, \bibinfo{pages}{1818--1830}.
\newblock
\showISBNx{9781450393584}
\href{https://doi.org/10.1145/3532106.3533575}{doi:\nolinkurl{10.1145/3532106.3533575}}


\bibitem[Terzimehić et~al\mbox{.}(2023)]%
        {terzimehic_implicit_2023}
\bibfield{author}{\bibinfo{person}{Nađa Terzimehić}, \bibinfo{person}{Fiona Draxler}, \bibinfo{person}{Mariam Ahsanpour}, {and} \bibinfo{person}{Albrecht Schmidt}.} \bibinfo{year}{2023}\natexlab{}.
\newblock \showarticletitle{Implicit {Smartphone} {Use} {Interventions} to {Promote} {Life}-{Technology} {Balance}: {An} {App}-{Market} {Survey}, {Design} {Space} and the {Case} of {Life}-{Relaunched}}. In \bibinfo{booktitle}{\emph{Mensch und {Computer} 2023}}. \bibinfo{publisher}{ACM}, \bibinfo{address}{Rapperswil Switzerland}, \bibinfo{pages}{237--249}.
\newblock
\showISBNx{9798400707711}
\href{https://doi.org/10.1145/3603555.3603578}{doi:\nolinkurl{10.1145/3603555.3603578}}


\bibitem[Thomas~Craig et~al\mbox{.}(2021)]%
        {thomas2021systematic}
\bibfield{author}{\bibinfo{person}{Kelly~J Thomas~Craig}, \bibinfo{person}{Laura~C Morgan}, \bibinfo{person}{Ching-Hua Chen}, \bibinfo{person}{Susan Michie}, \bibinfo{person}{Nicole Fusco}, \bibinfo{person}{Jane~L Snowdon}, \bibinfo{person}{Elisabeth Scheufele}, \bibinfo{person}{Thomas Gagliardi}, {and} \bibinfo{person}{Stewart Sill}.} \bibinfo{year}{2021}\natexlab{}.
\newblock \showarticletitle{Systematic review of context-aware digital behavior change interventions to improve health}.
\newblock \bibinfo{journal}{\emph{Translational behavioral medicine}} \bibinfo{volume}{11}, \bibinfo{number}{5} (\bibinfo{year}{2021}), \bibinfo{pages}{1037--1048}.
\newblock
\href{https://doi.org/10.1093/tbm/ibaa099}{doi:\nolinkurl{10.1093/tbm/ibaa099}}


\bibitem[Thom{\'e}e(2018)]%
        {Thomee.2018}
\bibfield{author}{\bibinfo{person}{Sara Thom{\'e}e}.} \bibinfo{year}{2018}\natexlab{}.
\newblock \showarticletitle{Mobile Phone Use and Mental Health. A Review of the Research That Takes a Psychological Perspective on Exposure}.
\newblock \bibinfo{journal}{\emph{International journal of environmental research and public health}} \bibinfo{volume}{15}, \bibinfo{number}{12} (\bibinfo{year}{2018}).
\newblock
\href{https://doi.org/10.3390/ijerph15122692}{doi:\nolinkurl{10.3390/ijerph15122692}}


\bibitem[{van Berkel} et~al\mbox{.}(2019)]%
        {vanBerkel.2019}
\bibfield{author}{\bibinfo{person}{Niels {van Berkel}}, \bibinfo{person}{Jorge Goncalves}, \bibinfo{person}{Lauri Lov{\'e}n}, \bibinfo{person}{Denzil Ferreira}, \bibinfo{person}{Simo Hosio}, {and} \bibinfo{person}{Vassilis Kostakos}.} \bibinfo{year}{2019}\natexlab{}.
\newblock \showarticletitle{Effect of experience sampling schedules on response rate and recall accuracy of objective self-reports}.
\newblock \bibinfo{journal}{\emph{International Journal of Human-Computer Studies}}  \bibinfo{volume}{125} (\bibinfo{year}{2019}), \bibinfo{pages}{118--128}.
\newblock
\showISSN{10715819}
\href{https://doi.org/10.1016/j.ijhcs.2018.12.002}{doi:\nolinkurl{10.1016/j.ijhcs.2018.12.002}}


\bibitem[{Vanden Abeele}(2021)]%
        {VandenAbeele.2021b}
\bibfield{author}{\bibinfo{person}{Mariek M.~P. {Vanden Abeele}}.} \bibinfo{year}{2021}\natexlab{}.
\newblock \showarticletitle{Digital Wellbeing as a Dynamic Construct}.
\newblock \bibinfo{journal}{\emph{Communication Theory}} \bibinfo{volume}{31}, \bibinfo{number}{4} (\bibinfo{year}{2021}), \bibinfo{pages}{932--955}.
\newblock
\showISSN{1050-3293}
\href{https://doi.org/10.1093/ct/qtaa024}{doi:\nolinkurl{10.1093/ct/qtaa024}}


\bibitem[Vazard(2022)]%
        {Vazard.2022}
\bibfield{author}{\bibinfo{person}{Juliette Vazard}.} \bibinfo{year}{2022}\natexlab{}.
\newblock \showarticletitle{Feeling the Unknown: Emotions of Uncertainty and Their Valence}.
\newblock \bibinfo{journal}{\emph{Erkenntnis}} (\bibinfo{year}{2022}).
\newblock
\showISSN{1572-8420}
\href{https://doi.org/10.1007/s10670-022-00583-1}{doi:\nolinkurl{10.1007/s10670-022-00583-1}}


\bibitem[Venables(1998)]%
        {venables1998exegeses}
\bibfield{author}{\bibinfo{person}{WN Venables}.} \bibinfo{year}{1998}\natexlab{}.
\newblock \showarticletitle{Exegeses on linear models}. In \bibinfo{booktitle}{\emph{S-Plus User’s Conference, Washington DC}}.
\newblock
\urldef\tempurl%
\url{https://citeseerx.ist.psu.edu/document?repid=rep1&type=pdf&doi=7883c3f0f6655248c4df17d09e384028219ef405}
\showURL{%
\tempurl}


\bibitem[Verduyn et~al\mbox{.}(2015a)]%
        {Verduyn.2015}
\bibfield{author}{\bibinfo{person}{Philippe Verduyn}, \bibinfo{person}{David~Seungjae Lee}, \bibinfo{person}{Jiyoung Park}, \bibinfo{person}{Holly Shablack}, \bibinfo{person}{Ariana Orvell}, \bibinfo{person}{Joseph Bayer}, \bibinfo{person}{Oscar Ybarra}, \bibinfo{person}{John Jonides}, {and} \bibinfo{person}{Ethan Kross}.} \bibinfo{year}{2015}\natexlab{a}.
\newblock \showarticletitle{Passive Facebook usage undermines affective well-being: Experimental and longitudinal evidence}.
\newblock \bibinfo{journal}{\emph{Journal of experimental psychology. General}} \bibinfo{volume}{144}, \bibinfo{number}{2} (\bibinfo{year}{2015}), \bibinfo{pages}{480--488}.
\newblock
\href{https://doi.org/10.1037/xge0000057}{doi:\nolinkurl{10.1037/xge0000057}}


\bibitem[Verduyn et~al\mbox{.}(2015b)]%
        {verduyn2015passive}
\bibfield{author}{\bibinfo{person}{Philippe Verduyn}, \bibinfo{person}{David~Seungjae Lee}, \bibinfo{person}{Jiyoung Park}, \bibinfo{person}{Holly Shablack}, \bibinfo{person}{Ariana Orvell}, \bibinfo{person}{Joseph Bayer}, \bibinfo{person}{Oscar Ybarra}, \bibinfo{person}{John Jonides}, {and} \bibinfo{person}{Ethan Kross}.} \bibinfo{year}{2015}\natexlab{b}.
\newblock \showarticletitle{Passive Facebook usage undermines affective well-being: Experimental and longitudinal evidence.}
\newblock \bibinfo{journal}{\emph{Journal of Experimental Psychology: General}} \bibinfo{volume}{144}, \bibinfo{number}{2} (\bibinfo{year}{2015}), \bibinfo{pages}{480}.
\newblock
\href{https://doi.org/10.1037/xge0000057}{doi:\nolinkurl{10.1037/xge0000057}}


\bibitem[Weber et~al\mbox{.}(2020)]%
        {Weber.2020}
\bibfield{author}{\bibinfo{person}{Philip Weber}, \bibinfo{person}{Philip Engelbutzeder}, {and} \bibinfo{person}{Thomas Ludwig}.} \bibinfo{year}{2020}\natexlab{}.
\newblock \showarticletitle{“Always on the Table”: Revealing Smartphone Usages in Everyday Eating Out Situations}. In \bibinfo{booktitle}{\emph{Proceedings of the 11th Nordic Conference on Human-Computer Interaction: Shaping Experiences, Shaping Society}} (Tallinn, Estonia) \emph{(\bibinfo{series}{NordiCHI '20})}. \bibinfo{publisher}{Association for Computing Machinery}, \bibinfo{address}{New York, NY, USA}, Article \bibinfo{articleno}{2}, \bibinfo{numpages}{13}~pages.
\newblock
\showISBNx{9781450375795}
\href{https://doi.org/10.1145/3419249.3420150}{doi:\nolinkurl{10.1145/3419249.3420150}}


\bibitem[Wickens(2002)]%
        {wickens_multiple_2002}
\bibfield{author}{\bibinfo{person}{Christopher~D. Wickens}.} \bibinfo{year}{2002}\natexlab{}.
\newblock \showarticletitle{Multiple resources and performance prediction}.
\newblock \bibinfo{journal}{\emph{Theoretical Issues in Ergonomics Science}} \bibinfo{volume}{3}, \bibinfo{number}{2} (\bibinfo{date}{Jan.} \bibinfo{year}{2002}), \bibinfo{pages}{159--177}.
\newblock
\showISSN{1463-922X, 1464-536X}
\href{https://doi.org/10.1080/14639220210123806}{doi:\nolinkurl{10.1080/14639220210123806}}


\bibitem[Wobbrock and Kientz(2016)]%
        {Wobbrock.2016}
\bibfield{author}{\bibinfo{person}{Jacob~O. Wobbrock} {and} \bibinfo{person}{Julie~A. Kientz}.} \bibinfo{year}{2016}\natexlab{}.
\newblock \showarticletitle{Research Contributions in Human-Computer Interaction}.
\newblock \bibinfo{journal}{\emph{Interactions}} \bibinfo{volume}{23}, \bibinfo{number}{3} (\bibinfo{date}{apr} \bibinfo{year}{2016}), \bibinfo{pages}{38–44}.
\newblock
\showISSN{1072-5520}
\href{https://doi.org/10.1145/2907069}{doi:\nolinkurl{10.1145/2907069}}


\bibitem[Yang et~al\mbox{.}(2020)]%
        {YANG2020112686}
\bibfield{author}{\bibinfo{person}{Jiaxin Yang}, \bibinfo{person}{Xi Fu}, \bibinfo{person}{Xiaoli Liao}, {and} \bibinfo{person}{Yamin Li}.} \bibinfo{year}{2020}\natexlab{}.
\newblock \showarticletitle{Association of problematic smartphone use with poor sleep quality, depression, and anxiety: A systematic review and meta-analysis}.
\newblock \bibinfo{journal}{\emph{Psychiatry Research}}  \bibinfo{volume}{284} (\bibinfo{year}{2020}), \bibinfo{pages}{112686}.
\newblock
\showISSN{0165-1781}
\href{https://doi.org/10.1016/j.psychres.2019.112686}{doi:\nolinkurl{10.1016/j.psychres.2019.112686}}


\end{thebibliography}

\appendix

\newpage

\section{Frequency of Data Points per Participant}\label{app:distribution}

\begin{figure}[ht!]
\centering
 \includegraphics[width=0.4\textwidth]{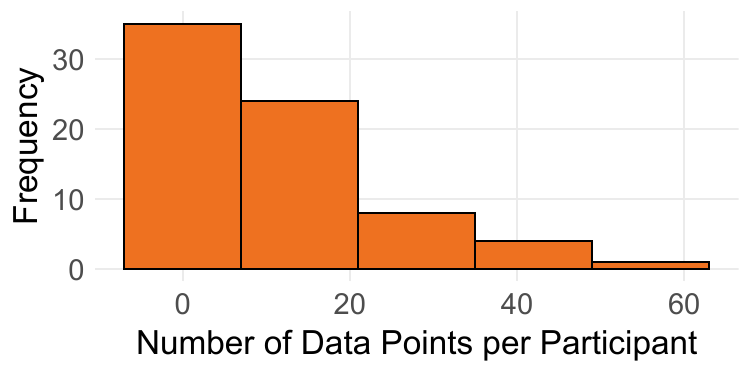}
  \caption{This plot shows the frequency of how many data points were provided per participant (M=$12.88$, SD=$13.02$)}
  \Description{The histogram shows the distribution of data points per participant. The x-axis shows the number of data points, and the y-axis indicates participant frequency. Most participants contributed 20–40 data points, with fewer contributing higher amounts.}
  \label{fig:interface_design}
\end{figure}

\onecolumn

\section{Question Items Used in the User Study}\label{app:usedquestionaires}

\begin{table*}[ht!]
\caption{Question items used in the user study}
\begin{tabularx}{\textwidth}{p{2.3cm}p{6cm}p{4cm}l}

   \textbf{Measurement}& Question Item & Answer Items & Reference   \\
   \hline 
   \addlinespace[2pt]
   \textbf{Reactance} (Threat subscale)
   & 
   I want to be in control, not my phone. \newline
   I like to act independently from my phone. \newline
   I don't want my phone to tell me what to do. \newline
   I don't let my phone impose its will on me. \newline
   I alone determine what to do, not my phone.
   &    5-point Likert scale from\newline
 ``strongly disagree'', to ``strongly agree'' & \cite{Ehrenbrink.2020} \\

\addlinespace[1pt]
   \hline 
   \addlinespace[2pt]
   \textbf{Current Activity}
   & What is your current activity?
   & 7-point Likert scale from (-3), ``definitely leisure'', to (+3), ``definitely not leisure'' & \cite{Samdahl.1991} \\

    \addlinespace[1pt]
   \hline 
   \addlinespace[2pt]
   \textbf{Valence}
   & How do you feel?
   & five images of manikin showing different valence levels & \cite{Bradley.1994}   \\
   
   \addlinespace[1pt]
   \hline 
   \addlinespace[2pt]
   \textbf{Sleepiness}
   & What is your level of sleepiness?
   & 9-point Likert scale from (1), ``extremely alert'', to (9), ``extremely sleepy'' & \cite{Shahid.2012b} \\

\addlinespace[1pt]
   \hline 
   \addlinespace[2pt]
   \textbf{Social Situation}
   & Which one of these best describes people around you?
   & ``alone'', ``with friends/ colleagues/ family members'', ``with strangers'' & \cite{Akpinar.2023} \\

   \addlinespace[1pt]
   \hline 
   \addlinespace[2pt]
   \textbf{Multitasking}
   & Did you do anything else besides being on [app name]?
   & ``yes'', ``no'' & -- \\

    \addlinespace[1pt]
   \hline 
   \addlinespace[2pt]
   \textbf{At Home}
   & Are you currently at home?
   & ``yes'', ``no'' & -- \\

\label{tab:question_items}
\end{tabularx}

\end{table*}

\section{Descriptive Data of the User Study}\label{app:descriptive}
\begin{table*}[ht!]
\caption{Table of the descriptive data of the user study}
\begin{tabularx}{\textwidth}{lrrrrrl}

 \textbf{Contextual Factor} & \textbf{min} & \textbf{max} & \textbf{mean} & \textbf{SD} & \textbf{median} & \textbf{distribution} \\ 
  \hline
  \addlinespace[2pt]
    Sleepiness & 1 & 9 & 4.91 & 2.05 & 5 & \\
    \addlinespace[2pt]
    Current Activity & -3 & 3 & -1.59 & 1.60 & -2 & \\
    \addlinespace[2pt]
    Valence & 1 & 5 & 3.16 & 1.01 & 3 & \\
    \addlinespace[2pt]
    At Home &  &  &  & & & True (86.95\%), False (13.05\%)\\
    \addlinespace[2pt]
    Multitasking &  &  & &  &  & True (37.22\%), False (62.78\%) \\
    \addlinespace[2pt]
    Social Situation &  &  &  &  &  & alone (73.03\%), friends (26.54\%), \\
    \addlinespace[-4pt]
     &  &  &  &  &  & strangers (0.43\%) \\
    
    \\

   \textbf{Dependent Variables} &  &  &  &  &  &  \\
   \hline 
   \addlinespace[2pt]
   Responsiveness & 0s & 67m 20s & 3m 31s & 8m 21s &  8s \\
   \addlinespace[-2pt]
  \qquad \textit{-- log.-trans.} & \textit{0} & \textit{8.30} & \textit{3.07} & \textit{2.14} & \textit{2.20} \\
  \addlinespace[2pt]
   Reactance & 1 & 5 & 3.55 & 1.03 & 3.80 & \\
   \\
   \textbf{App Distribution} &  &  & &  &  &  \\
   \hline
   \addlinespace[3pt]
    
   \multicolumn{7}{l}{TikTok (40.30\%), Reddit (26.48\%), Facebook (13.49\%), Instagram (11.13\%), X (5.56\%), YouTube Shorts (3.03\%)} \\

\label{tab:_descr_stat}
\end{tabularx}
\end{table*}

\end{document}